\pdfoutput=1

\documentclass[twocolumn]{aa}  
\usepackage{graphicx}
\usepackage{rotating}
\usepackage{afterpage}
\usepackage{txfonts}
\usepackage[]{natbib}
\bibpunct{(}{)}{;}{a}{}{,}

\newcommand{\um}{$\mu$m}

\newcommand{\kms}{km\thinspace s$^{-1}$}

\def\degr{\hbox{$^\circ$}}
\def\arcmin{\hbox{$^\prime$}}
\def\arcsec{\hbox{$^{\prime\prime}$}}
\def\utw{\smash{\rlap{\lower5pt\hbox{$\sim$}}}}
\def\udtw{\smash{\rlap{\lower6pt\hbox{$\approx$}}}}

\def\Lsun{\hbox{\it L$_\odot$}}

\def\Teff{\hbox{\it T$_{\rm eff}$}}

\def\Msun{\hbox{\it M$_\odot$}}

\def\Mbol{\hbox{\it M$_{bol}$}}

\def\Teff{\hbox{\it T$_{\rm eff}$}}

\def\K{\hbox{\it K}}

\newcommand{\Ks}{{\it K$_{\rm s}$}}

\newcommand{\Aks}{{\it A$_{\it K_{\rm s}}$}}
\newcommand{\Ak}{{\it A$_{\it K}$}}

\def\BCKs{\hbox{\it BC$_{\it K_{\mathrm s} }$}}

\def\simgr{\mathrel{\hbox{\rlap{\hbox{\lower4pt\hbox{$\sim$}}}\hbox{$>$}}}}
\def\HH{H{\sc ii}}	% HII region

\def\brg{\hbox{ Br$_\gamma$}}
\def\vlsr{\hbox{\it V$_{\rm LSR}$}}

\usepackage{textcomp}

\def\nodata{\hbox{ $..$}}

\begin{document}
\renewcommand{\arraystretch}{0.65}

   \title{ Massive stars in the giant molecular cloud  G23.3$-$0.3  and W41
   \thanks{Based on observations collected at the European Southern Observatory 
   (ESO Programmes 084.D-0769, 085.D-019, 087.D-09609).} 
      \thanks{MM is currently employed by the MPIfR. This works
      was partially carried out at RIT (2009), at ESA (2010), and at the MPIfR.}
      }

   \subtitle{ }

   \author{Maria Messineo
          \inst{1,2,9}
          \and
          Karl~M. Menten
	  \inst{1}
           \and
          Donald~F. Figer
	  \inst{2}
           \and
          Ben Davies
	  \inst{3}
           \and
           J.~Simon Clark
	  \inst{4}
           \and
           Valentin~D. Ivanov
	  \inst{5}
	  \and
	   Rolf-Peter Kudritzki
 	  \inst{6}
	  \and
	   R.~Michael Rich 
	  \inst{7}
	  \and
	   John~W. MacKenty
	  \inst{8}
	  \and
	   Christine Trombley
	  \inst{2}
       }

   \institute{Max-Planck-Institut f\"ur Radioastronomie,
Auf dem H\"ugel 69, D-53121 Bonn, Germany
              \email{messineo@mpifr-bonn.mpg.de}
     \and	      
Center for Detectors, Rochester Institute of Technology, 54 Memorial Drive, Rochester, NY 14623, USA 
     \and
Astrophysics Research Institute, Liverpool John Moores University,
Twelve Quays House, Egerton Wharf, Birkenhead, Wirral.
CH41 1LD, United Kingdom.    	   
     \and
Department of Physics and Astronomy, The Open University, Walton Hall, Milton Keynes, MK7 6AA, UK 	
     \and
European Southern Observatory, Ave. Alonso de CÚrdova 3107, Casilla 19, Santiago, 19001, Chile
     \and
Institute for Astronomy, University of Hawaii, 2680 Woodlawn Drive, Honolulu, HI 96822
     \and
Physics and Astronomy Building, 430 Portola Plaza, Box 951547, Department of Physics 
and Astronomy, University of California, Los Angeles, CA 90095-1547.
     \and
Space Telescope Science Institute, 3700 San Martin Drive, Baltimore, MD 21218
     \and
European Space Agency (ESA), The Astrophysics and Fundamental Physics Missions Division, 
Research and Scientific Support Department, Directorate of Science and Robotic Exploration, 
ESTEC, Postbus 299, 2200 AG Noordwijk, The Netherlands
     }

   \date{Received September 15, 1996; accepted March 16, 1997}

\abstract
{ Young massive stars and stellar clusters continuously form in the Galactic disk,
generating  new \HH\ regions within their natal giant molecular clouds  and subsequently enriching  the interstellar medium via their winds and supernovae.}
{  Massive stars are
among the brightest infrared stars in such regions; their identification  permits 
 the characterisation of the star formation history of the associated cloud as well as constraining the  location 
 of  stellar aggregates and hence their occurrence  
as a function of global environment.  }
{We present a stellar spectroscopic survey  in the direction of 
the giant molecular cloud G23.3$-$0.3. This complex is located  
at a distance of $\sim  4-5$ kpc, and consists of several \HH\ regions and supernova remnants. }
{ We discovered 11 Of$_K^{+}$ stars, one candidate Luminous Blue Variable, several  OB stars, and candidate red supergiants. 
Stars with   $K$-band extinction from $\sim 1.3 - 1.9$ mag appear  to be associated with the GMC G23.3$-$0.3; 
O and B-type{\bf s} satisfying this criterion have spectrophotometric distances 
consistent with that of the giant molecular cloud.
Combining near-IR  spectroscopic  and  photometric data allowed us  
to characterize the multiple sites of star formation within it. 
The O-type stars have masses from $\sim 25 - 45$ \Msun, and  ages of 5-8 Myr.
Two new red supergiants were detected with  interstellar extinction   typical of the cloud; along with the two RSGs within the cluster GLIMPSE9, they trace  an older burst with an age of 20--30 Myr. Massive stars were also  detected in the core of three   supernova remnants -  W41, G22.7$-$0.2, and G22.7583$-$0.4917.}
{A large population  of massive stars appears associated with the GMC G23.3$-$0.3, with the properties inferred for them indicative of an extended history of stars formation.}


   \maketitle
   
\titlerunning{ Massive stars in G23.3-0.3 }
\authorrunning{M. Messineo et al.\ }

%________________________________________________________________

\section{Introduction}

An understanding of the  evolution, and fate of massive
stars ($\ga 8$ \Msun) is of broad astronomical interest, and it is fundamental for studies of galaxies at all redshifts.  
Historically, the majority (70-90\%) of massive stars were thought to be  born in dense clusters, although   recent  observations  also support formation in low-density environments  \citep{lada03,dewit05,wright14}. 
In turn, such star clusters  appear to form in large molecular complexes \citep{clark04,clark09,davies12}, 
and a direct proportionality is often assumed between the cluster masses  
and the masses of the collapsing clouds \citep[e.g.][]{krumholz07,alves07}.
However, observational constraints on the distribution (clusters versus stars in isolation)
and evolution of massive stars are difficult to obtain, 
because of  their  rarity, and   heavy dust obscuration of the richest  star-forming regions of the Galaxy.

The recent completion of multiple  radio and infrared surveys of the Galactic plane\footnote{The Multi-Array Galactic Plane Imaging Survey (MAGPIS)\citep[][]{white05,helfand06}, the Two Micron All Sky Survey  (2MASS) \citep{cutri03}, the Deep Near Infrared Survey of the Southern Sky (DENIS)  \citep{epchtein94}, the UKIRT Infrared Deep Sky Survey  (UKIDSS) \citep[][]{lucas08},
the VISTA Variables in the Via Lactea  survey (VVV) \citep{soto13}, the Midcourse Space Experiment (MSX) \citep{egan03,price01}, 
the  Galactic Legacy Infrared Mid-Plane Survey Extraordinaire (GLIMPSE) \citep{spitzer09} and WISE the  Wide-field Infrared Survey Explore
 (WISE) \citep{wise12}.}  has opened  a golden epoch for studying  the formation,  evolution, 
and environment of massive stars.    
Over the past decade, multi-wavelength analyses of the Galactic plane have revealed several  hundred
new \HH\ regions, and candidate  supernova remnants  \citep[SNRs, e.g.][]{green09,brogan06,helfand06}. 
Moreover, an impressive large number of new candidate  stellar clusters and ionizing stars have been reported;
more than 1800 candidate  clusters were detected with 2MASS data  \citep[e.g.][]{bica03},
more than 90 candidates were found with  GLIMPSE data \citep{mercer05}, 
and  $\sim 100$ candidates with the VVV survey  \citep{borissova11}.

\begin{figure*}
\resizebox{0.45\hsize}{!}{\includegraphics[angle=0]{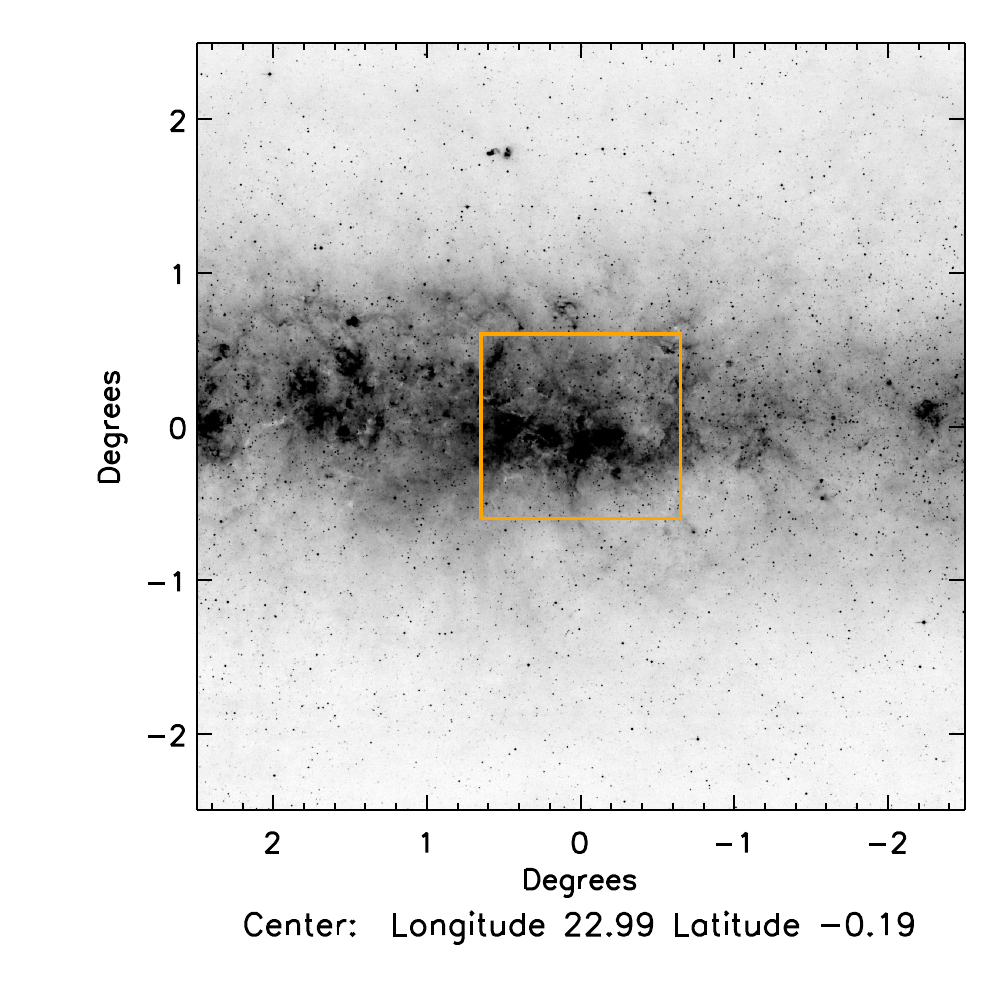}}
\resizebox{0.45\hsize}{!}{\includegraphics[angle=0]{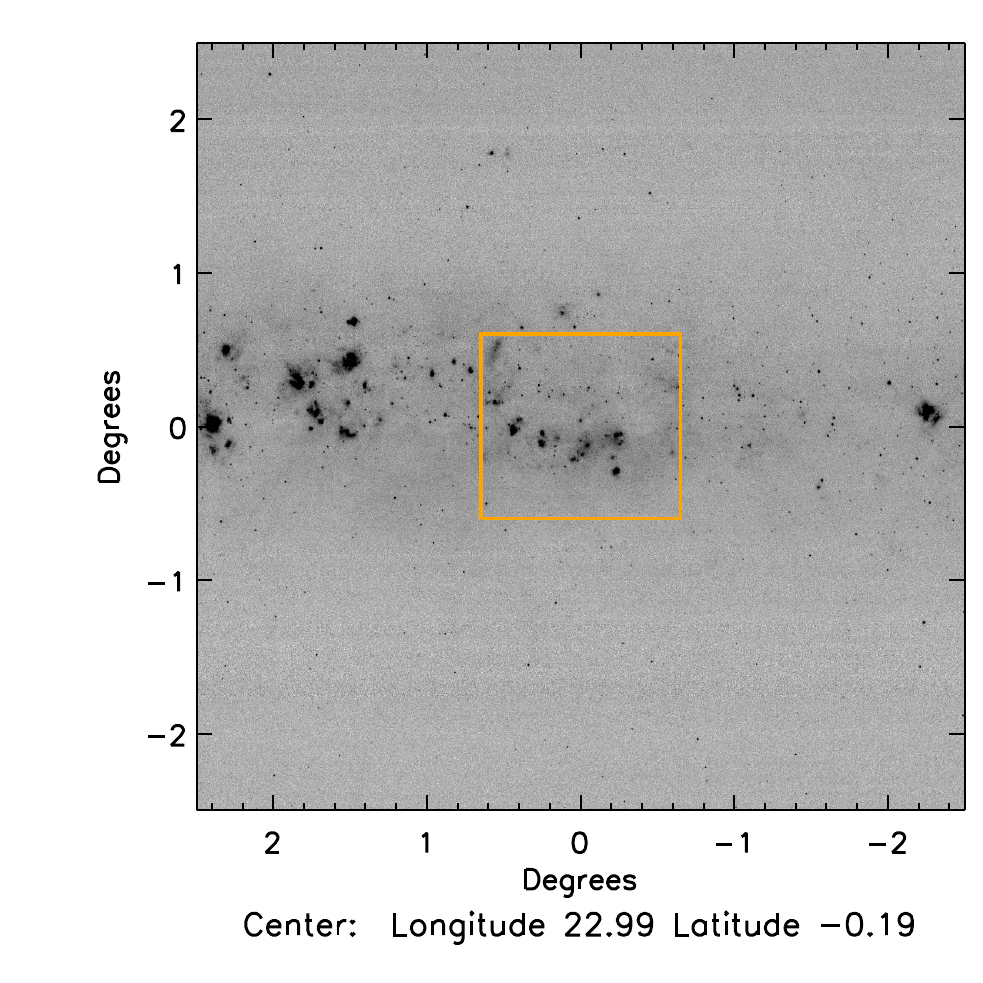}}

\resizebox{0.9\hsize}{!}{\includegraphics[angle=0]{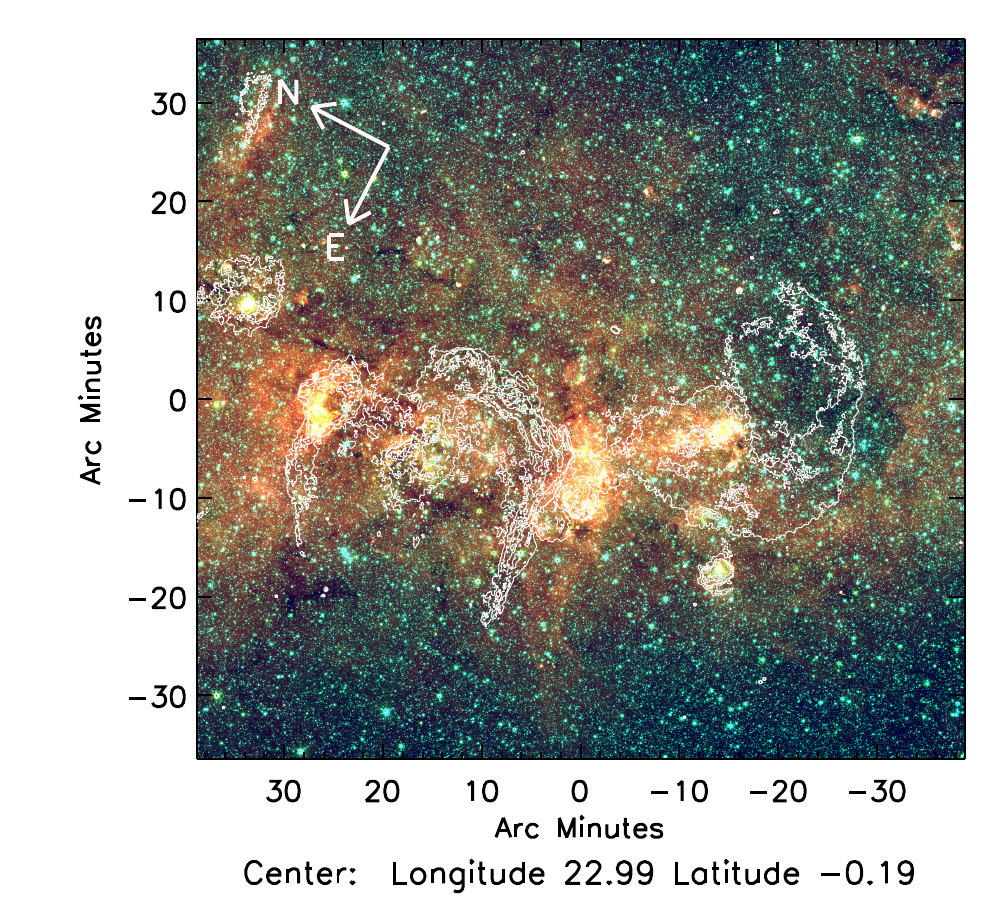}}
\caption{\label{largemap0.fig} 
Top: MSX band A (8 \um; left) and band E (20 \um; right) images of G23.3--0.3. 
The square indicates the area displayed in the composite image at the bottom.
Bottom: False-color composite image of the G23.3--0.3 complex, 
which was created with GLIMPSE data: 3.6 \um\ (blue), 4.5 \um\ (green), and 8.0 \um\ (red). 
Contours of the 20 cm emission detected  by MAGPIS \citep{white05,helfand06} 
at 0.002, 0.003, 0.004   Jy beam$^{-1}$ levels are super-imposed.  
Galactic longitude is along the x-axis, and
Galactic latitude is along the y-axis.} 
\end{figure*}

The Galactic giant molecular cloud (GMC) GMC G23.3$-$0.3 (object "[23,78]" in \citet{dame86})
is found at  a  distance  of 4--5  kpc \citep{albert06}. A remarkable number of candidate
stellar clusters appear associated with this region  \citep[e.g.][]{messineo10}, and
four SNRs \citep[G$22.9917-0.3583$, G022.7$-$00.2, W41, and G22.7583$-$0.4917,][]{green09,helfand06,leahy08}  
are projected against it 
\citep[as shown by][]{messineo10}. The presence of SNRs  suggests that massive star formation has been active in 
multiple sites of this GMC, as do the stellar cluster number 9 in   \citet[][]{mercer05} \citep[hereafter GLIMPSE9,][]{messineo10}, 
cluster number 10 in  \citet[]{mercer05} (hereafter GLIMPSE10), 
$[BDS2003]$117, and $[BDS2003]$118   \citep{bica03}.
Additional regions with massive stars were identified by \citet{messineo10}.

Given this, G23.3$-$0.3 appears to be an ideal laboratory for the  investigation of   massive stars and multi-seeded star
formation.  The rich star clusters associated with  the complex allow us to study 
the mode and progression  of  star formation in this region and to sample rare evolutionary 
phases 
of massive stars, such as Wolf-Rayets (WRs), red supergiants (RSGs), and  luminous blue variables (LBVs).
The presence of supernova  remnants (SNRs)  indicates that  star formation has been progressing for some time, with the current 
stellar population providing  information  on  the initial masses of the 
supernova progenitors, and  on the fate of massive stars. 

In this paper, we present the result of a   spectroscopic survey of selected bright stars
in the direction of   GMC G23.3$-$0.3. 
In Sect. \ref{obs}, the  spectroscopic 
observations and data reduction are presented, along with available photometric data.
In Sect. \ref{analizza}, we describe the spectral types, the reddening properties,
and the selection of massive stars likely associated with the GMC.
Luminosities of the massive stars are derived.
Eventually, in Sect. \ref{discussion}, we summarize the results, and briefly discuss the 
spatial distribution of the detected massive stars, their ages, and
their connection with the supernova remnants.

\section{Observations and data reduction}
\label{obs}

\begin{table*}
\caption{\label{regions} Surveyed regions  \citep[see Table 4 of][]{messineo10} and 
supernova remnants.}
\begin{tabular}{@{\extracolsep{-.08in}}llllll} 
\hline
\hline 
Overdensity &  RA[J2000]  & DEC[J2000]   & Rad (\arcmin) & SNR        &         Reference   \\
\hline
REG1/ [BDS2003]118  &18 34 15.1 & $-$08 20 42   & 1.2& G23.5667$-$0.0333 (SNR5)            &1                         \\
GLIMPSE9Large       &18 34 09.6 & $-$09 13 53   & 3.0$^a$& border of G22.7$-$0.2 (SNR2)    &   new                                 \\
                    &           &               &    & near G22.7583$-$0.4917 (SNR3)       &                                       \\ 
GLIMPSE9 (cluster)  &18 34 09.6 & $-$09 13 53   & 0.3&                                     & 2, 3   \\
REG2                &18 34 41.1 & $-$08 34 22   & 4.0& border of W41                       & 2                    \\ 
REG4/GLIMPSE10      &18 34 31.6 & $-$08 46 47   & 5.0& core of W41                         & 2, 3   \\ 
REG5                &18 34 20.0 & $-$08 59 48   & 5.0& G22.9917$-$0.3583 (SNR4)            & 2                     \\ 
REG7/[BDS2003]117   &18 34 27.7 & $-$09 15 52   & 2.0$^b$& core of G22.7583$-$0.4917 (SNR3)& 2, 1                         \\ 
RSGCX1              &18 33 08.9 & $-$09 09 14   & 4.5& core of G22.7$-$0.2 (SNR2)          &    new                                 \\
\hline
\end{tabular}
\begin{list}{}{}
\item[ ] {\bf Notes.} $^a$A larger region enclosing the GLIMPSE9 cluster was surveyed. ~
$^b$ The quoted radius encloses only the bulk of the nebulosity seen at 3.6 \um.
\item[ ] {\bf References.}
(1) \citet{bica03};~
(2) \citet{messineo10};~
(3) \citet{mercer05}.
\end{list}
\end{table*}

\subsection{SINFONI data}
The observations were made with 
the Spectrograph for INtegral Field Observations in the Near Infrared
(SINFONI)  \citep{eisenhauer03} on  the Yepun Very Large Telescope,
under the ESO programs 084.D-0769 and 085.D-0192 (P.I. Messineo).
We observed  $\sim 100$  stars with $0.6 < H-$\Ks$ < 1.4$  mag
and $11 < $\Ks$ < 6$ mag  from selected fields (see Table \ref{regions});  
their color-color distribution is shown in 
Sect.\ \ref{section.ext}.
A total number of 89 data-cubes were obtained, and a total number of 104 stellar
spectra were extracted  from these  cubes.

We used  SINFONI in non-AO mode, with a pixel scale
of $0\rlap{.}^{\prime\prime} 250$ pix$^{-1}$,   the
$K$-grating (1.95-2.45 \um), and a resolving power R $\approx 4400$.

Exposures were taken in a  target-sky-sky-target sequence, using a
fixed  sky position. Integration times ($DIT \times NDIT$)
ranged from 1 s to 53 s in period 84,
and from 1 s to 93 s in period 85.  Two exposures  were taken for each position.
Telluric standard stars of  B-type were observed at an airmass 
within 0.2 dex from the airmass of the science observations, and immediately 
before or after the science observation.

Data reduction was performed as described in \citet{messineo07}.
The construction of a wavelength-calibrated  data-cube, along with the removal of the instrumental signatures,
was performed with version 3.9.0 of the ESO SINFONI  pipeline  
\citep{schreiber04,modigliani07}. Each science frame was sky-subtracted, and  flat-fielded. 
Dead/hot pixels were removed by interpolation;  geometric distortions 
were corrected. A wavelength-calibration map was obtained using daytime arc-lamp lines. 
Possible shifts  in wavelengths (up to 0.4 pixels) were checked, and corrected with observed  
 OH sky lines \citep{oliva92,rousselot00} by  cross-correlating  the OH line positions
with a template spectrum with OH lines at zero velocity.

Stellar traces were extracted from the cubes, and corrected for atmospheric 
and instrumental responses
by dividing the spectra of the targets by the spectra  of B-type stars.
The \brg\ and \ion{He}{I} lines   were removed from the spectra of the standards with a linear interpolation,
and the resulting spectra were multiplied by  a black body curve, F$(_\lambda)$, with the effective 
temperature of the  star. 
Some  spectra  with low signal-to-noise displayed residuals of OH sky lines;  
in these seven stars, we  removed the residuals of the OH sky lines
at 2.0008 \um, 2.0276 \um, 2.0413 \um, 2.0563 \um, 2.0729 \um, 2.1506 \um, 2.1802 \um, 2.1955 \um, 
2.2126 \um, and 2.2312 \um\ with a linear interpolation.  
The absolute coordinates of the SINFONI fields generally agree  with the 2MASS coordinates within 1\arcsec\ or 2\arcsec.
The astrometry of each field was  aligned with a 2MASS image or UKIDSS image.

We examined stellar traces with a signal-to-noise ratio above   20-40.

Table \ref{table.obspectra} lists the early-type stars,
and  Tables \ref{table.crsgspectra} and \ref{table.giantspectra}
list the late-type stars. Finding charts are provided in Appendix C.

\begin{table*}
{\tiny
\caption{\label{table.obspectra} List of detected early-type stars.  }
\begin{tabular}{@{\extracolsep{-.04in}}lll|lll|rr|l }
\hline 
\hline 
 {\rm ID}   & \multicolumn{2}{c}{\rm Coordinates}  &  &\multicolumn{2}{c}{\rm Spectral Detection}  & ($J-$\Ks)$_o$ & ($H-$\Ks)$_o$ &{\rm Comment } \\ 
\hline 
           & {\rm RA(J2000)} & {\rm DEC(J2000)} &{\rm Instr.}  &  {\rm Spectrum} & {\Teff} &   \\
\hline 
 &{\rm [hh mm ss]}   & {\rm [deg mm ss]}        &                  &               & [K]     & \\ 
\hline 

  1 & 18 33 18.14 & $-09$~24~09.9 &          SofI &             OBe &24300$\pm$ 8800 & -0.12 & -0.06 &                     \\
  2 & 18 33 52.19 & $-09$~10~38.2 &         SofI &         O9-9.5e &29300$\pm$ 1800 & -0.16 & -0.07 &           BD$-09$4766 \\
  3 & 18 34 00.86 & $-09$~15~41.5 &      SINFONI &          O6-7f$_K$+ &35700$\pm$ 1000 & -0.21 & -0.10 &                     \\
  4 & 18 34 05.74 & $-09$~16~00.6 &      SINFONI &        O7-8.5f$_K$+ &31800$\pm$ 1500 & -0.21 & -0.10 &                     \\
  5 & 18 34 06.25 & $-09$~15~17.9 &      SINFONI &          O6-7f$_K$+ &35700$\pm$ 1000 & -0.21 & -0.10 &                     \\
  6 & 18 34 08.75 & $-09$~13~59.9 &      SINFONI &            B0-3 &23800$\pm$ 6700 & -0.16 & -0.08 &                     \\
  7 & 18 34 09.25 & $-09$~03~06.0 &      SINFONI &           B4-A2 &12700$\pm$ 3600 & -0.02 &  0.00 &                     \\
  8 & 18 34 10.50 & $-09$~14~04.4 &      SINFONI &            B0-3 &23800$\pm$ 6700 & -0.16 & -0.08 &                     \\
  9 & 18 34 10.59 & $-09$~13~43.9 &      SINFONI &        O7-8.5f$_K$+ &33100$\pm$ 1500 & -0.21 & -0.10 &                     \\
 10 & 18 34 10.70 & $-09$~13~58.7 &      SINFONI &             OF &\nodata   & -0.06 & -0.01 &                     \\
 11 & 18 34 11.30 & $-09$~13~56.4 &      SINFONI &             OF &\nodata   & -0.06 & -0.01 &                     \\
 12 & 18 34 11.81 & $-08$~55~44.9 &      SINFONI &           B4-A2 &12700$\pm$ 3600 & -0.02 &  0.00 &                     \\
 13 & 18 34 12.14 & $-09$~00~23.6 &      SINFONI &           B4-A2 &12700$\pm$ 3600 & -0.02 &  0.00 &                     \\
 14 & 18 34 12.17 & $-09$~12~29.9 &      SINFONI &          O6-7f$_K$+ &34500$\pm$ 1200 & -0.21 & -0.10 &                     \\
 15 & 18 34 13.47 & $-09$~14~31.9 &      SINFONI &          O6-7f$_K$+ &34500$\pm$ 1200 & -0.21 & -0.10 &                     \\
 16 & 18 34 14.47 & $-08$~44~22.9 &      SINFONI &         O9-9.5e &29300$\pm$ 1800 & -0.16 & -0.07 &                     \\
 17 & 18 34 15.88 & $-08$~45~45.2 &      SINFONI &        O9-9.5f$_K$+ &31400$\pm$ 1100 & -0.19 & -0.09 &                     \\
 18 & 18 34 17.26 & $-08$~46~50.0 &      SINFONI &          O6-7f$_K$+ &34500$\pm$ 1200 & -0.21 & -0.10 &                     \\
 19 & 18 34 18.14 & $-08$~57~18.4 &      SINFONI &           B4-A2 &12700$\pm$ 3600 & -0.02 &  0.00 &                     \\
 20 & 18 34 18.85 & $-08$~45~32.9 &      SINFONI &           B4-A2 &12700$\pm$ 3600 & -0.02 &  0.00 &                     \\
 21 & 18 34 19.19 & $-08$~46~17.6 &      SINFONI &         B7.5-A2 &12900$\pm$ 3900 & -0.02 &  0.00 &                     \\
 22 & 18 34 21.70 & $-08$~28~20.9 &         SofI &            cLBV &13200$\pm$ 2300 &  0.01 & -0.01 &                     \\
 23 & 18 34 23.79 & $-08$~49~18.1 &      SINFONI &          O6-7f$_K$+ &35700$\pm$ 1000 & -0.21 & -0.10 &                     \\
 24 & 18 34 26.38 & $-09$~00~49.1 &      SINFONI &             OF &\nodata   & -0.06 & -0.01 &                     \\
 25 & 18 34 27.67 & $-09$~15~51.1 &      SINFONI &            O4f$_K$+ &38200$\pm$ 2500 & -0.21 & -0.10 &                     \\
 26 & 18 34 28.48 & $-08$~59~31.1 &      SINFONI &           B4-A2 &12700$\pm$ 3600 & -0.02 &  0.00 &                     \\
 27 & 18 34 30.15 & $-08$~44~40.6 &      SINFONI &             OF &\nodata   & -0.06 & -0.01 &                     \\
 28 & 18 34 30.84 & $-08$~58~40.1 &      SINFONI &           B4-A2 &12700$\pm$ 3600 & -0.02 &  0.00 &                     \\
 29 & 18 34 30.95 & $-08$~58~37.8 &      SINFONI &             OF &\nodata   & -0.06 & -0.01 &                     \\
 30 & 18 34 33.83 & $-08$~32~57.9 &      SINFONI &             OF &\nodata   & -0.06 & -0.01 &                     \\
 31 & 18 34 33.92 & $-08$~32~59.6 &      SINFONI &            B0-3 &23800$\pm$ 6700 & -0.16 & -0.08 &                     \\
 32 & 18 34 35.17 & $-09$~00~39.9 &      SINFONI &           B4-A2 &12700$\pm$ 3600 & -0.02 &  0.00 &                     \\
 33 & 18 34 35.74 & $-09$~01~27.6 &      SINFONI &             OF &\nodata   & -0.06 & -0.01 &                     \\
 34 & 18 34 36.94 & $-08$~47~54.7 &      SINFONI &             OF &\nodata   & -0.06 & -0.01 &                     \\
 35 & 18 34 38.36 & $-08$~50~49.7 &      SINFONI &             OF &\nodata   & -0.06 & -0.01 &                     \\
 36 & 18 34 42.63 & $-08$~45~01.9 &      SINFONI &          O6-7f$_K$+ &34500$\pm$ 1200 & -0.21 & -0.10 &                     \\
 37 & 18 34 42.86 & $-08$~45~02.9 &      SINFONI &             OF &\nodata   & -0.06 & -0.01 &                     \\
 38 & 18 34 50.71 & $-08$~46~16.0 &      SINFONI &            B0-3 &23800$\pm$ 6700 & -0.16 & -0.08 &                     \\
  $[$MFD2010$]$ 3 & 18 34 08.68 & $-09$~14~11.1 &              &            B0-3 &21500$\pm$ 6000 & -0.08 & -0.04 &     $[$MFD2010$]$ 3$^a$ \\
  $[$MFD2010$]$ 4 & 18 34 08.54 & $-09$~14~11.8 &              &            B0-3 &21500$\pm$ 6000 & -0.08 & -0.04 &     $[$MFD2010$]$ 4$^a$ \\
$[$MVM2011$]$ 39 & 18 33 47.64 & $-09$~23~07.7 &              &             WC8 &65000$\pm$ 5000 &  0.43 &  0.38 & $[$MVM2011$]$ 39$^b$  \\

\hline
\end{tabular}

\begin{list}{}{}
\item[] 
{\bf Notes. }  Identification numbers 
are followed by
celestial coordinates,  instrument,  spectral types, 
estimated effective temperatures, \Teff, intrinsic near-infrared colors,  and comments. 
Two B supergiants detected by \citet{messineo10}, and a WR discovered by \citet{mauerhan11} 
are appended  to the table.~
We used  the collection of infrared colors and temperatures per spectral types as listed in the Appendix of \citet{messineo11}.
For every star (for example a O6-7 star), we assumed the mean temperature of the range considered, 
and as error  half range.
($^a$) \citet{messineo10}.~
($^b$) \citet{mauerhan11}.
\end{list}
}
\end{table*}

\subsection{SofI data}
An additional 47  objects 
were detected with the Son of Isaac (SofI) spectrograph
on the ESO New Technology Telescope (NTT) on La Silla during the ESO program 087.D-09609 
(P.I, Messineo), on the nights of June 10, 11, and 12, 2011. 

Observations with  SofI on the NTT
were performed with the medium resolution grism, a slit-width of 1\arcsec, 
and the \Ks\ filter. 
A coverage  from 2.0 \um\ to 2.3 \um\ at a resolving power of $\sim 1900$ was obtained.
Medium resolution spectra in $H-$band 
were taken only for one target,  a candidate LBV; 
a  slit with a width of 1\arcsec\ was used, which provided a coverage from 1.5 \um\ to 1.8 \um\
at a resolving power of $\sim 1250$.
The objects were nodded along the slit to obtain pairs of frames, which were subtracted and flat-fielded.
In a few observations, the stellar traces did not move (no nodding, no jitter), 
and we subtracted each frame with darks.
The two-dimensional frames were rectified with a bilinear interpolation of  stellar traces and 
arc lines. 
Stellar traces were extracted from   individual frames, aligned  in wavelength, 
and co-added. Correction for atmospheric and instrumental responses were performed with spectra of 
B-type standards (taken in the same manner as for the targets, and with linearly 
interpolated  \brg\ and \ion{He}{I} lines). We multiplied the results by a black body curve, 
F$_\lambda$.

\begin{table*} 
\caption{\label{table.crsgspectra} Spectra of late-type stars that are potential RSGs 
 ($L > 4 \times 10^4 $\Lsun\  for a distance of 4.6 kpc).
} 
{\tiny
\begin{tabular}{@{\extracolsep{-.06in}}lll|llllllr|lrrr|rr| l}
\hline 
\hline 
 {\rm ID}   & {\rm RA(J2000)} & {\rm DEC(J2000)}   & \multicolumn{7}{c}{\rm Spectral Type} & Comment\\ 
\hline 
 &                 &                      &{\rm Instr.}  &  {\rm EW(CO)} &{\rm Sp[RGB]}  & {\rm \Teff[RGB]$^*$}   & {\rm Sp[RSG]}  &  {\rm \Teff[RSG]$^*$}  & {\rm  H$_2$O}$^+$ \\ 
 &{\rm [hh mm ss]}  & {\rm [deg mm ss]}   &             &   {\rm[AA]}    &               & {\rm [K]}              &                &  {\rm [K]}             &     [\%]       &\\ 
\hline 
 39 &  18 32 36.02 &  $-$9~08~03.5 &            SofI &   29 &       M5 &    3450$\pm$  203&       K5 &    3869$\pm$  137&    8 &                               \\
 40 &  18 33 08.89 &  $-$9~08~32.6 &            SofI &   33 &     $..$ &    3223$\pm$  226&       M0 &    3790$\pm$  124&   11 &               IRAS18303$$-$$0910\\
 41 &  18 33 13.90 &  $-$9~06~23.2 &            SofI &   23 &       M1 &    3745$\pm$  130&       K3 &    3985$\pm$  121&    0 &                               \\
 42 &  18 33 15.02 &  $-$9~08~32.2 &            SofI &   23 &       M1 &    3745$\pm$  130&       K3 &    3985$\pm$  121&   10 &                               \\
 43 &  18 33 35.24 &  $-$8~47~57.7 &            SofI &   32 &     $..$ &    3223$\pm$  226&       M0 &    3790$\pm$  124&   18 &                             BG$^a$\\
 44 &  18 33 37.80 &  $-$9~21~38.1 &            SofI &   21 &       M0 &    3790$\pm$  124&       K2 &    4049$\pm$  131&    8 &                               \\
 45 &  18 33 40.98 &  $-$9~03~25.2 &            SofI &   26 &       M3 &    3605$\pm$  120&       K3 &    3985$\pm$  121&  $-$16 &                               \\
 46 &  18 34 10.36 &  $-$9~13~52.9 &         SINFONI &   64 &     $..$ &    3223$\pm$  226&       M3 &    3605$\pm$  120&  $-$76 &                     [MFD2010]8$^b$\\
 47 &  18 34 23.17 &  $-$8~48~38.6 &         SINFONI &   61 &     $..$ &    3223$\pm$  226&       M2 &    3660$\pm$  140&   $-$6 &                               \\
 48 &  18 34 33.86 &  $-$8~44~21.2 &         SINFONI &   47 &       M6 &    3336$\pm$  226&       K5 &    3869$\pm$  137&  $-$16 &                               \\
 $[$MFD2010$]$ 5&  18 34 09.86 &  $-$9~14~23.8 &         SINFONI &    $..$ &     $..$ &    3223$\pm$  226&     M1.5 &    3710$\pm$  152&  $..$ &    $[$MFD2010$]$ 5$^b$   \\
 BD$-$08 4635 &  18 34 51.88 &  $-$8~36~40.8 &         SINFONI &    $..$ &     $..$ &    3223$\pm$  226&       M2 &    3660$\pm$  140&  $..$&                     BD$-$08 4635$^c$\\
 BD$-$08 4639 &  18 35 31.06 &  $-$8~41~23.4 &         SINFONI &    $..$ &     $..$ &    3223$\pm$  226&       K2 &    4049$\pm$  131&  $..$ &                     BD$-$08 4639$^c$\\
 BD$-$08 4645 &  18 36 21.66 &  $-$8~52~40.0 &         SINFONI &    $..$ &     $..$ &    3223$\pm$  226&       M2 &    3660$\pm$  140&  $..$ &                     BD$-$08 4645$^c$\\
\hline
\end{tabular}
\begin{list}{}{}
\item[] {\bf Notes.}
Identification numbers are followed by celestial coordinates, instrument,  EW(CO)s, 
spectral types, \Teff, H$_2$O indexes, and comments.
Two spectral types are reported; the first was obtained using the
relation for red giants (Sp[RGB]), the latter using that for red supergiants (Sp[RSG]).  
We appended to the table  RSG [MFD2010]5
\citep{messineo10},   RSG BD$-08$ 4645, 
 BD$-08$ 4635, and  BD$-08$ 4639  \citep{skiff13}. ~
($^*$) Temperature errors account for accuracy in spectral types of $\pm2$.~
($^+$) The H$_2$O index depends on the correction for \Aks; a variation of 10\% in \Aks\ typically affects the H$_2$O by 20\%. 
($^a$) BG= object in the background of the cloud.~
($^b$) \citet{messineo10}.~
($^c$) \citet{skiff13}.
\end{list}
}
\end{table*}

\subsection{Infrared photometry}

We searched for counterparts of the observed stars in the 
2MASS Catalog of Point Sources \citep{cutri03}, in the third release of 
DENIS data  at CDS (catalog  B/denis)
\citep{epchtein94}, in the GLIMPSE catalog \citep{spitzer09}, and in  
the WISE catalog \citep{wise12}; 
we used the closest match within a search radius of 2\arcsec.
We searched in the UKIDSS catalog \citep{lucas08} with a search radius of 1\arcsec,
and retained only counterparts in the linear regime ($K \ga 10.2$ mag). 
The  II/293 (GLIMPSE) catalog  from CDS is a combination of the original GLIMPSE-I (v2.0),
GLIMPSE-II (v2.0), and GLIMPSE-3D catalogs. 
We also searched  for counterparts in the Version 2.3 of the MSX
Point Source Catalog   \citep{egan03,price01} with a search radius of 5\arcsec.
MSX upper limits were removed. 
WISE counterparts were retained only if their signal-to-noise ratio was larger than 2.0.
Near-infrared and GLIMPSE counterparts were visually checked 
with 2MASS/UKIDSS and GLIMPSE charts.
For most of sources, WISE  band-3 and band-4 provided  upper limit magnitudes, due
to confusion.

In addition, we searched for possible $B$, $V$, $R$-band matches in 
The Naval Observatory Merged Astrometric Dataset (NOMAD) by \citet{zacharias04}.
The photometric data are listed in Table \ref{table.phot}.
For a few  targets (missing in both 2MASS and UKIDSS), \Ks\ counterparts 
were estimated  from the SINFONI cubes (with a typical  uncertainty of $\sim 0.3$ mag).
For stars [MFD2010]3, [MFD2010]4, and [MFD2010]5, 
$H$ and $K$-band measurements were  obtained with the Near Infrared Camera 
and Multi-Object Spectrometer \citep[NICMOS,][]{skinner98} \citep{messineo10}.

{\tiny
\begin{table*}[u] \renewcommand{\arraystretch}{0.8}
\caption{ \label{table.phot} Infrared measurements  of the spectroscopically
detected early-type stars and candidate RSGs. } 
{\tiny
\begin{tabular}{@{\extracolsep{-.10in}} l|rrr|rrr|rrr|rrrr|r|rrrr|rrrrr}
\hline 
    &  \multicolumn{3}{c}{\rm 2MASS$^a$}   &\multicolumn{3}{c}{\rm DENIS} &\multicolumn{3}{c}{\rm UKIDSS$^b$} &   \multicolumn{4}{c}{\rm GLIMPSE}   &  \multicolumn{1}{c}{\rm MSX}& \multicolumn{4}{c}{\rm WISE}  & \multicolumn{1}{c}{\rm NOMAD} &\\ 
\hline 
 {\rm ID$^f$} & {\it J} & {\it H} &  $K_{\mathrm s}$  &
 {\it I} & {\it J} &  $K_{\mathrm s}$ & 
 {\it J} & {\it H} &  $K$ & 
 {\rm [3.6]} & {\rm [4.5]} & {\rm [5.8]} & {\rm [8.0]} &
 {\it A}  &{\it W1} &{\it W2} & {\it W3} &{\it W4} & {\it R} & \\ 
\hline 
 &{\rm [mag]}   &	{\rm [mag]}    & {\rm [mag]}     & {\rm [mag]} &{\rm [mag]}  &  {\rm [mag]} &{\rm [mag]} &{\rm [mag]}  & {\rm [mag]}&{\rm [mag]}&{\rm [mag]}&{\rm [mag]}&{\rm [mag]}&{\rm [mag]}& {\rm [mag]}& {\rm [mag]}&{\rm [mag]}& {\rm [mag]}& {\rm [mag]}& \\ 
\hline 
                             1  &   9.66 &  9.35 &  9.17 & 10.49 &  9.77 &  9.20 & \nodata & \nodata & \nodata &  9.07 &  8.89 &  8.70 &  8.44 & \nodata &  8.98 &  8.85 & \nodata &  7.06 &  10.82 &  \\
                             2  &   7.35 &  6.84 &  6.61 &  9.13 &  6.90 &  6.58 & \nodata & \nodata & \nodata &  6.86 &  6.46 &  6.32 &  6.38 & \nodata &  6.39 &  6.33 &  6.25 &  3.72 &   9.85 &  \\
                             3  &  14.01 & 11.85 & 10.75 & \nodata & 13.98 & 10.83 & 13.99 & 11.88 & 10.76 & 10.05 &  9.82 &  9.65 &  9.97 & \nodata & 10.14 &  9.86 &  8.58 &  5.66 & \nodata &  \\
                             4  &  12.73 & 10.91 &  9.88 & \nodata & 12.75 &  9.87 & \nodata & \nodata & \nodata &  9.33 &  9.06 &  8.96 &  8.98 & \nodata &  9.23 &  8.89 &  7.52 &  5.46 & \nodata &  \\
                             5  &  13.30 & 11.35 & 10.38 & \nodata & 13.23 & 10.30 & 13.25 & 11.40 & 10.35 &  9.63 &  9.51 &  9.31 &  9.40 & \nodata &  9.59 &  9.30 &  8.48 &  5.85 & \nodata &  \\
                             6  &  13.15 & 11.43 & 10.53 & \nodata & 13.09 & 10.26 & 13.47 & 12.47 & 10.74 & \nodata & \nodata & \nodata & \nodata & \nodata & \nodata & \nodata & \nodata & \nodata & \nodata &  \\
                             7  &  11.71 & 11.22 & 10.96 & 13.28 & 11.66 & 10.90 & 11.75 & 11.60 & 10.97 & 10.81 & 10.71 & 10.61 & 10.29 & \nodata & 10.80 & 10.88 & \nodata &  4.04 &  14.16 &  \\
                             8$^c$  & \nodata&\nodata&\nodata& \nodata & 14.91 & 12.11 & 14.91 & 13.01 & 12.01 & \nodata & \nodata & \nodata & \nodata & \nodata & \nodata & \nodata & \nodata & \nodata & \nodata&  \\
                             9  &  13.43 & 11.45 & 10.40 & \nodata & 13.34 & 10.39 & 13.48 & 11.55 & 10.50 &  9.77 & \nodata &  9.48 &  9.50 & \nodata & \nodata & \nodata & \nodata & \nodata & \nodata &  \\
                            10$^c$   & \nodata&\nodata&\nodata& \nodata & \nodata & \nodata & 16.31 & 14.16 & 13.12 & \nodata & \nodata & \nodata & \nodata & \nodata & \nodata & \nodata & \nodata & \nodata & \nodata&  \\
                            11$^c$   & \nodata&\nodata&\nodata& \nodata & \nodata & \nodata & 17.70 & 15.86 & 15.06 & \nodata & \nodata & \nodata & \nodata & \nodata & \nodata & \nodata & \nodata & \nodata & \nodata&  \\
                            12  &  10.93 & 10.57 & 10.39 & \nodata & \nodata & \nodata & 11.15 & 12.00 & 10.48 & 10.40 & 10.31 & 10.29 & 10.24 & \nodata & 10.05 &  9.97 & \nodata &  4.60 &  12.62 &  \\
                            13  &  10.84 & 10.51 & 10.38 & 11.74 & 10.81 & 10.34 & 10.98 & 10.85 & 10.41 & 10.28 & 10.25 & 10.23 & 10.20 & \nodata & 10.20 & 10.28 & \nodata &  2.22 &  12.72 &  \\
                            14  &  13.36 & 11.25 & 10.18 & \nodata & 13.22 & 10.11 & \nodata & \nodata & \nodata &  9.51 &  9.11 &  8.89 &  8.89 & \nodata &  9.47 &  8.87 &  6.98 &  2.36 & \nodata &  \\
                            15  &  12.21 & 10.42 &  9.52 & \nodata & 12.10 &  9.44 & \nodata & \nodata & \nodata &  8.93 &  8.73 &  8.63 &  8.75 & \nodata &  8.96 &  8.73 & \nodata &  4.66 & \nodata&  \\
                            16  &  12.07 & 10.15 &  9.19 & \nodata & 12.03 &  9.10 & \nodata & \nodata & \nodata &  8.38 &  8.22 &  8.02 &  7.99 & \nodata &  8.46 &  8.15 &  6.25 &  3.96 & \nodata &  \\
                            17  &  12.90 & 11.37 & 10.58 & 18.13 & 12.90 & 10.47 & 12.90 & 11.53 & 10.61 & 10.00 &  9.89 &  9.87 &  9.76 & \nodata &  9.89 &  9.66 & \nodata & \nodata & \nodata &  \\
                            18  &  12.46 & 10.72 &  9.96 & \nodata & 12.36 &  9.80 & \nodata & \nodata & \nodata &  9.27 &  9.08 &  9.02 &  9.07 & \nodata &  9.33 &  9.08 & \nodata &  4.66 & \nodata &  \\
                            19  &  11.06 & 10.58 & 10.36 & 12.32 & 11.12 & 10.27 & 11.17 & 11.09 & 10.36 & 10.22 & 10.17 & 10.15 & \nodata & \nodata & 10.25 &  9.98 &  7.02 & \nodata &  13.41 &  \\
                            20$^c$   & \nodata&\nodata&\nodata& 14.50 & 13.29 & 11.93 & 13.06 & 12.52 & 12.42 & \nodata & \nodata & \nodata & \nodata & \nodata & \nodata & \nodata & \nodata & \nodata & \nodata &  \\
                            21  &  12.73 & 11.42 & 10.76 & 17.20 & 12.61 & 10.64 & 12.72 & 11.59 & 10.74 & 10.29 & 10.15 & 10.02 & 10.02 & \nodata & 10.37 & 10.18 & \nodata & \nodata & \nodata &  \\
                            22  &   9.78 &  8.42 &  7.63 & 13.92 &  9.67 &  7.51 & \nodata & \nodata & \nodata &  6.89 &  6.51 &  6.17 &  5.93 &  5.95 &  6.85 &  6.40 &  5.74 &  4.41 &  16.75 &  \\
                            23  &  12.75 & 11.20 & 10.43 & 17.90 & 12.77 & 10.38 & 12.69 & 11.20 & 10.38 &  9.83 &  9.66 &  9.55 &  9.93 & \nodata & 10.05 &  9.87 & \nodata &  2.19 & \nodata &  \\
                            24$^d$  & \nodata&\nodata& 14.66 & \nodata & \nodata & \nodata & \nodata & \nodata & \nodata & \nodata & \nodata & \nodata & \nodata & \nodata & \nodata & \nodata & \nodata & \nodata & \nodata&  \\
                            25  &  12.16 & 10.67 &  9.90 & \nodata & 12.34 & 10.16 & \nodata & \nodata & \nodata &  9.44 &  9.12 &  9.02 & \nodata & \nodata & \nodata & \nodata & \nodata & \nodata & \nodata &  \\
                            26  &  11.27 & 10.92 & 10.73 & 12.63 & 11.57 & 10.98 & 11.54 & 11.36 & 10.74 & 10.60 & 10.50 & 10.25 &  9.74 & \nodata & 10.76 & 10.76 & \nodata &  3.22 &  13.29 &  \\
                            27$^d$  & \nodata&\nodata& 14.16 & \nodata & \nodata & \nodata & 15.01 & 13.94 & \nodata & \nodata & \nodata & \nodata & \nodata & \nodata & \nodata & \nodata & \nodata & \nodata &  16.22 &  \\
                            28  &  11.38 & 11.04 & 10.78 & 12.80 & 11.67 & 11.06 & 11.60 & 11.43 & 10.84 & 10.69 & 10.67 & 10.12 & \nodata & \nodata & 10.40 & 10.05 &  5.81 &  2.95 &  13.81 &  \\
                            29$^c$   & \nodata&\nodata&\nodata& \nodata & \nodata & \nodata & 15.26 & 14.53 & 14.30 & \nodata & \nodata & \nodata & \nodata & \nodata & \nodata & \nodata & \nodata & \nodata & \nodata&  \\
                            30$^c$   & \nodata&\nodata&\nodata& \nodata & \nodata & \nodata & 15.17 & 13.59 & 12.85 & \nodata & \nodata & \nodata & \nodata & \nodata & \nodata & \nodata & \nodata & \nodata & \nodata&  \\
                            31  &  12.50 & 11.07 & 10.32 & 17.75 & 12.77 & 10.67 & 12.49 & 11.31 & 10.39 &  9.80 &  9.81 &  9.79 & \nodata & \nodata &  9.52 &  9.31 &  5.93 &  3.32 & \nodata &  \\
                            32  &  11.22 & 10.79 & 10.57 & 12.87 & 11.58 & 10.77 & 11.50 & 11.34 & 10.61 & 10.40 & 10.31 & \nodata & \nodata & \nodata & 10.29 & 10.14 & \nodata &  3.83 &  13.38 &  \\
                            33$^d$  & \nodata&\nodata& 13.25 & 15.02 & \nodata & \nodata & 13.66 & 12.79 & \nodata & \nodata & \nodata & \nodata & \nodata & \nodata & \nodata & \nodata & \nodata & \nodata &  15.66 &  \\
                            34$^d$  & \nodata&\nodata& 12.79 & \nodata & \nodata & \nodata & \nodata & \nodata & \nodata & \nodata & \nodata & \nodata & \nodata & \nodata & \nodata & \nodata & \nodata & \nodata &  18.15 &  \\
                            35$^c$   &  13.71 &\nodata&\nodata& \nodata & \nodata & \nodata & 13.81 & 13.18 & 13.01 & \nodata & \nodata & \nodata & \nodata & \nodata & \nodata & \nodata & \nodata & \nodata &  15.26 &  \\
                            36  &  13.37 & 11.18 & 10.06 & \nodata & 13.64 & 10.33 & \nodata & \nodata & \nodata &  9.26 &  9.04 &  8.92 &  9.00 & \nodata &  9.27 &  8.90 &  6.07 &  0.90 & \nodata &  \\
                            37$^c$   & \nodata&\nodata&\nodata& \nodata & \nodata & \nodata & 15.88 & 13.52 & 12.40 & \nodata & \nodata & \nodata & \nodata & \nodata & \nodata & \nodata & \nodata & \nodata & \nodata&  \\
                            38  &  12.40 & 11.13 & 10.50 & 17.13 & 12.71 & 10.81 & 12.35 & 11.18 & 10.52 & 10.07 &  9.92 &  9.63 & 10.24 & \nodata &  9.88 &  9.68 &  7.12 &  3.96 & \nodata &  \\
                             $[$MFD2010$]$ 3$^e$  &  10.66 &  8.93 &  7.96 & 17.06 & 10.50 &  7.63 & \nodata & \nodata & \nodata &  6.46 &  6.81 &  6.47 &  6.63 & \nodata &  6.59 &  6.31 &  6.65 &  4.71 & \nodata &  \\
                             $[$MFD2010$]$ 4$^e$  & \nodata& 10.21 &  9.14 & \nodata & \nodata & \nodata & \nodata & \nodata & \nodata & \nodata & \nodata & \nodata & \nodata & \nodata &  \nodata &  \nodata &  \nodata &  \nodata & \nodata&  \\
                            $[$MVM2011$]$ 39   &  12.18 & 10.52 &  9.36 & 17.77 & 12.20 &  9.42 & \nodata & \nodata & \nodata &  8.53 &  8.03 &  7.78 &  7.51 & \nodata &  8.67 &  8.06 &  7.57 & \nodata & \nodata &  \\
                            39  &   7.36 &  5.62 &  4.84 & \nodata & \nodata &  4.11 &\nodata&\nodata&\nodata&  4.31 &  4.67 &  4.22 &  4.26 &  4.15 &  4.54 &  4.06 &  4.28 &  3.78 & \nodata &  \\
                            40  &   9.71 &  6.66 &  5.08 & \nodata &  9.61 &  4.27 &\nodata&\nodata&\nodata&  4.00 &  4.31 &  3.62 &  3.22 &  3.15 &  4.95 &  3.40 &  2.32 &  0.97 & \nodata &  \\
                            41  &   9.60 &  7.45 &  6.52 & 15.73 &  9.48 &  6.49 &\nodata&\nodata&\nodata&  6.73 &  6.05 &  5.75 &  5.68 &  5.68 &  6.06 &  5.92 &  5.62 &  3.94 & \nodata&  \\
                            42  &   8.79 &  6.93 &  6.14 & 13.59 &  8.72 &  5.90 &\nodata&\nodata&\nodata&  6.66 &  5.95 &  5.56 &  5.54 &  5.58 &  5.63 &  5.57 &  5.86 & \nodata &  17.22 &  \\
                            43  & \nodata& 10.49 &  7.91 & 17.14 & 14.97 &  7.95 &\nodata&\nodata&\nodata&  6.77 &  6.41 &  5.54 &  5.61 & \nodata &  6.47 &  5.98 &  6.32 & \nodata & \nodata &  \\
                            44  &   8.69 &  6.73 &  5.84 & 14.33 &  8.64 &  5.07 &\nodata&\nodata&\nodata&  7.40 &  5.97 &  5.16 &  5.16 &  5.07 &  5.42 &  5.29 &  5.21 &  4.04 & \nodata &  \\
                            45  &   9.41 &  7.25 &  6.21 & 16.20 &  9.30 &  6.16 &\nodata&\nodata&\nodata&  5.60 &  6.18 &  5.33 &  5.26 &  5.15 &  5.66 &  5.41 &  4.85 &  3.74 & \nodata &  \\
                            46  &  10.22 &  7.59 &  6.29 & \nodata & 10.02 &  6.25 &\nodata&\nodata&\nodata&  5.38 & \nodata &  4.89 &  4.78 & \nodata &  5.05 &  4.91 &  4.14 &  2.47 & \nodata&  \\
                            47  &   9.59 &  7.32 &  6.19 & 16.63 &  9.59 &  6.23 &\nodata&\nodata&\nodata&  5.52 &  6.12 &  5.30 &  5.30 &  4.95 &  5.56 &  5.37 &  4.91 &  2.95 & \nodata &  \\
                            48  &   9.03 &  7.00 &  6.06 & 15.90 &  9.44 &  6.30 &\nodata&\nodata&\nodata& \nodata & \nodata &  5.24 &  5.25 &  5.16 & \nodata & \nodata & \nodata & \nodata & \nodata &  \\
                             $[$MFD2010$]$ 5$^e$  &  11.41 &  8.43 &  7.05 & \nodata & 11.32 &  6.97 &\nodata&\nodata&\nodata&  6.75 &  7.39 &  5.70 &  5.84 & \nodata &  6.09 &  5.97 &  6.26 &  4.34 & \nodata &  \\
                            BD$-$08 4635  &   4.79 &  3.45 &  3.05 &  8.57 & \nodata &  3.73 &\nodata&\nodata&\nodata& \nodata &  3.91 & \nodata & \nodata &  2.75 &  3.04 &  2.51 &  2.92 &  2.87 &   9.90 &  \\
                            BD$-$08 4639  &   4.05 &  3.06 &  2.77 &  8.69 & \nodata &  3.85 &\nodata&\nodata&\nodata& \nodata & \nodata & \nodata & \nodata &  2.84 & \nodata & \nodata & \nodata & \nodata &   8.05 &  \\
                            BD$-$08 4645  &   3.92 &  2.73 &  2.29 &  8.97 & \nodata &  3.89 &\nodata&\nodata&\nodata& \nodata & \nodata & \nodata & \nodata &  2.40 & \nodata & \nodata & \nodata & \nodata &  10.69 &  \\

\hline
\end{tabular}
\begin{list}{}{}
\item[]{\bf Notes.} (${\mathrm{a}}$) 2MASS upper limits and confused stars were removed all, but star \#4.~
(${\mathrm{b}}$) Small corrections ($J \approx +0.1$ mag, $H \approx -0.1$ mag, 
$K \approx 0.0$ mag)  were applied to match the 2MASS photometric system.~
(${\mathrm{c}}$) UKIDSS values were used.~
(${\mathrm{d}}$) \Ks\ was estimated from the SINFONI data-cube.~
(${\mathrm{e}}$) $H$ and \Ks\ were taken from \citet{messineo10}.~
(${\mathrm{f}}$) Identification numbers are taken from Table \ref{table.obspectra},
\ref{table.crsgspectra}, and \ref{table.giantspectra}.
\end{list}
}
\end{table*} 
}

{\tiny
\addtocounter{table}{-1}
\begin{table*}[u] \renewcommand{\arraystretch}{0.8} 
\caption{ continuation  of Table \ref{table.phot}. Associated errors.} 
{\tiny
\begin{tabular}{@{\extracolsep{-.10in}} l|rrr|rrr|rrr|rrrr|r|rrrr|rrrrr}
\hline 
    &  \multicolumn{3}{c}{\rm 2MASS}   &\multicolumn{3}{c}{\rm DENIS} &\multicolumn{3}{c}{\rm UKIDSS} &   \multicolumn{4}{c}{\rm GLIMPSE}   &  \multicolumn{1}{c}{\rm MSX}& \multicolumn{4}{c}{\rm WISE}  & \\ 
\hline 
 {\rm ID} & $ J_{\mathrm err}$ & $H_{\mathrm err}$ & $K_{\mathrm serr}$  &
 $I_{\mathrm err}$ & $J_{\mathrm err}$ & $K_{\mathrm serr}$ & 
 $ J_{\mathrm err}$ & $H_{\mathrm err}$ & $K_{\mathrm err}$ & 
 {[3.6$_{\mathrm err}$]} &  [4.5$_{\mathrm err}$] &  [5.8$_{\mathrm err}$] &  [8.0$_{\mathrm err}$] &
 {\it A$_{\mathrm err}$}  &{\it W1$_{\mathrm err}$} &{\it  W2$_{\mathrm err}$ }&{\it  W3$_{\mathrm err}$}&{\it  W4$_{\mathrm err}$}&  \\ 
\hline 
 &{\rm [mag]}   &	{\rm [mag]}    & {\rm [mag]}     & {\rm [mag]} &{\rm [mag]}  &  {\rm [mag]} &{\rm [mag]}  &{\rm [mag]}  & {\rm [mag]}&{\rm [mag]}&{\rm [mag]}&{\rm [mag]}&{\rm [mag]}&{\rm [mag]}& {\rm [mag]}& {\rm [mag]}&{\rm [mag]}& {\rm [mag]}&  \\ 
\hline 
                              1  &   0.02 &  0.02 &  0.02 &  0.03 &  0.07 &  0.07 &\nodata&\nodata&\nodata&  0.04 &  0.05 &  0.05 &  0.03 &\nodata&  0.03 &  0.02 &\nodata&  0.48 &  \\
                             2  &   0.02 &  0.03 &  0.02 &  0.06 &  0.16 &  0.09 &\nodata&\nodata&\nodata&  0.05 &  0.07 &  0.03 &  0.03 &\nodata&  0.04 &  0.02 &  0.05 &  0.35 &  \\
                             3  &   0.03 &  0.02 &  0.02 &\nodata&  0.12 &  0.08 &  0.002 &  0.001 &  0.001 &  0.05 &  0.06 &  0.06 &  0.08 &\nodata&  0.04 &  0.03 &  0.31 &  0.23 &  \\
                             4  & \nodata&\nodata&\nodata&\nodata&  0.10 &  0.08 &\nodata&\nodata&\nodata&  0.07 &  0.05 &  0.05 &  0.05 &\nodata&  0.03 &  0.03 &  0.18 &  0.31 &  \\
                             5  &   0.02 &  0.02 &  0.03 &\nodata&  0.10 &  0.08 &  0.001 &\nodata&  0.001 &  0.03 &  0.05 &  0.05 &  0.05 &\nodata&  0.03 &  0.03 &  0.35 &  0.15 &  \\
                             6  &   0.12 &  0.02 &  0.02 &\nodata&  0.10 &  0.08 &  0.001 &  0.001 &  0.001 &\nodata&\nodata&\nodata&\nodata&\nodata&\nodata&\nodata&\nodata&\nodata&  \\
                             7  &   0.03 &  0.03 &  0.02 &  0.03 &  0.09 &  0.08 &  0.001 &  0.001 &  0.001 &  0.05 &  0.06 &  0.08 &  0.10 &\nodata&  0.04 &  0.06 &\nodata&  0.07 &  \\
                             8  & \nodata&\nodata&\nodata&\nodata&  0.14 &  0.10 &  0.003 &  0.001 &  0.001 &\nodata&\nodata&\nodata&\nodata&\nodata&\nodata&\nodata&\nodata&\nodata&  \\
                             9  &   0.05 &  0.06 &  0.09 &\nodata&  0.10 &  0.08 &  0.001 &\nodata&  0.001 &  0.07 &\nodata&  0.05 &  0.06 &\nodata&\nodata&\nodata&\nodata&\nodata&  \\
                            10  & \nodata&\nodata&\nodata&\nodata&\nodata&\nodata&  0.009 &  0.003 &  0.003 &\nodata&\nodata&\nodata&\nodata&\nodata&\nodata&\nodata&\nodata&\nodata&  \\
                            11  & \nodata&\nodata&\nodata&\nodata&\nodata&\nodata&  0.030 &  0.014 &  0.016 &\nodata&\nodata&\nodata&\nodata&\nodata&\nodata&\nodata&\nodata&\nodata&  \\
                            12  &   0.03 &  0.03 &  0.03 &\nodata&\nodata&\nodata&\nodata&  0.001 &  0.001 &  0.06 &  0.05 &  0.08 &  0.09 &\nodata&  0.03 &  0.04 &\nodata&  0.08 &  \\
                            13  &   0.02 &  0.02 &  0.02 &  0.03 &  0.08 &  0.08 &\nodata&\nodata&  0.001 &  0.05 &  0.06 &  0.07 &  0.16 &\nodata&  0.04 &  0.09 &\nodata&  0.03 &  \\
                            14  &   0.03 &  0.03 &  0.02 &\nodata&  0.10 &  0.08 &\nodata&\nodata&\nodata&  0.06 &  0.06 &  0.04 &  0.08 &\nodata&  0.03 &  0.02 &  0.06 &  0.04 &  \\
                            15  &   0.03 &  0.03 &  0.02 &\nodata&  0.09 &  0.07 &\nodata&\nodata&\nodata&  0.04 &  0.05 &  0.04 &  0.05 &\nodata&  0.03 &  0.03 &\nodata&  0.07 &  \\
                            16  &   0.03 &  0.02 &  0.02 &\nodata&  0.09 &  0.07 &\nodata&\nodata&\nodata&  0.03 &  0.05 &  0.04 &  0.06 &\nodata&  0.03 &  0.03 &  0.06 &  0.14 &  \\
                            17  &   0.02 &  0.03 &  0.03 &  0.19 &  0.10 &  0.08 &  0.001 &\nodata&  0.001 &  0.07 &  0.06 &  0.07 &  0.07 &\nodata&  0.03 &  0.03 &\nodata&\nodata&  \\
                            18  &   0.02 &  0.02 &  0.02 &\nodata&  0.09 &  0.08 &\nodata&\nodata&\nodata&  0.05 &  0.05 &  0.03 &  0.06 &\nodata&  0.03 &  0.03 &\nodata&  0.29 &  \\
                            19  &   0.02 &  0.02 &  0.02 &  0.03 &  0.08 &  0.08 &\nodata&\nodata&\nodata&  0.05 &  0.06 &  0.08 &\nodata&\nodata&  0.05 &  0.05 &  0.14 &\nodata&  \\
                            20  & \nodata&\nodata&\nodata&  0.06 &  0.09 &  0.10 &  0.001 &  0.001 &  0.002 &\nodata&\nodata&\nodata&\nodata&\nodata&\nodata&\nodata&\nodata&\nodata&  \\
                            21  &   0.03 &  0.02 &  0.02 &  0.12 &  0.10 &  0.08 &  0.001 &  0.001 &  0.001 &  0.06 &  0.06 &  0.06 &  0.09 &\nodata&  0.05 &  0.05 &\nodata&\nodata&  \\
                            22  &   0.03 &  0.04 &  0.03 &  0.06 &  0.05 &  0.06 &\nodata&\nodata&\nodata&  0.03 &  0.06 &  0.03 &  0.03 &  0.05 &  0.03 &  0.02 &  0.06 &  0.17 &  \\
                            23  &   0.03 &  0.03 &  0.02 &  0.17 &  0.10 &  0.08 &  0.001 &\nodata&  0.001 &  0.04 &  0.06 &  0.05 &  0.09 &\nodata&  0.04 &  0.04 &\nodata&  0.06 &  \\
                            24  & \nodata&\nodata&  0.30 &\nodata&\nodata&\nodata&\nodata&\nodata&\nodata&\nodata&\nodata&\nodata&\nodata&\nodata&\nodata&\nodata&\nodata&\nodata&  \\
                            25  &   0.03 &  0.03 &  0.02 &\nodata&  0.09 &  0.08 &\nodata&\nodata&\nodata&  0.06 &  0.10 &  0.11 &\nodata&\nodata&\nodata&\nodata&\nodata&\nodata&  \\
                            26  &   0.02 &  0.02 &  0.02 &  0.04 &  0.08 &  0.08 &\nodata&\nodata&  0.001 &  0.04 &  0.08 &  0.08 &  0.15 &\nodata&  0.07 &  0.11 &\nodata&  0.11 &  \\
                            27  & \nodata&\nodata&  0.30 &\nodata&\nodata&\nodata&  0.003 &  0.002 &\nodata&\nodata&\nodata&\nodata&\nodata&\nodata&\nodata&\nodata&\nodata&\nodata&  \\
                            28  &   0.03 &  0.03 &  0.03 &  0.04 &  0.08 &  0.08 &\nodata&\nodata&  0.001 &  0.08 &  0.09 &  0.10 &\nodata&\nodata&  0.05 &  0.08 &  0.10 &  0.04 &  \\
                            29  & \nodata&\nodata&\nodata&\nodata&\nodata&\nodata&  0.004 &  0.004 &  0.008 &\nodata&\nodata&\nodata&\nodata&\nodata&\nodata&\nodata&\nodata&\nodata&  \\
                            30  & \nodata&\nodata&\nodata&\nodata&\nodata&\nodata&  0.003 &  0.002 &  0.002 &\nodata&\nodata&\nodata&\nodata&\nodata&\nodata&\nodata&\nodata&\nodata&  \\
                            31  &   0.04 &  0.05 &  0.03 &  0.22 &  0.09 &  0.08 &  0.001 &\nodata&  0.001 &  0.07 &  0.09 &  0.12 &\nodata&\nodata&  0.03 &  0.05 &  0.05 &  0.16 &  \\
                            32  &   0.02 &  0.02 &  0.04 &  0.04 &  0.08 &  0.08 &\nodata&\nodata&  0.001 &  0.07 &  0.10 &\nodata&\nodata&\nodata&  0.08 &  0.09 &\nodata&  0.38 &  \\
                            33  & \nodata&\nodata&  0.30 &  0.07 &\nodata&\nodata&  0.001 &  0.001 &\nodata&\nodata&\nodata&\nodata&\nodata&\nodata&\nodata&\nodata&\nodata&\nodata&  \\
                            34  & \nodata&\nodata&  0.30 &\nodata&\nodata&\nodata&\nodata&\nodata&\nodata&\nodata&\nodata&\nodata&\nodata&\nodata&\nodata&\nodata&\nodata&\nodata&  \\
                            35  &   0.05 &\nodata&\nodata&\nodata&\nodata&\nodata&  0.001 &  0.001 &  0.003 &\nodata&\nodata&\nodata&\nodata&\nodata&\nodata&\nodata&\nodata&\nodata&  \\
                            36  &   0.02 &  0.03 &  0.02 &\nodata&  0.10 &  0.08 &\nodata&\nodata&\nodata&  0.05 &  0.05 &  0.06 &  0.04 &\nodata&  0.03 &  0.03 &  0.06 &  0.05 &  \\
                            37  & \nodata&\nodata&\nodata&\nodata&\nodata&\nodata&  0.006 &  0.002 &  0.002 &\nodata&\nodata&\nodata&\nodata&\nodata&\nodata&\nodata&\nodata&\nodata&  \\
                            38  &   0.03 &  0.02 &  0.02 &  0.16 &  0.09 &  0.08 &  0.001 &\nodata&  0.001 &  0.07 &  0.06 &  0.06 &  0.16 &\nodata&  0.03 &  0.04 &  0.07 &  0.10 &  \\
                             $[$MFD2010$]$ 3  &   0.04 &  0.02 &  0.30 &  0.11 &  0.08 &  0.07 &\nodata&\nodata&\nodata&  0.15 &  0.17 &  0.04 &  0.05 &\nodata&  0.04 &  0.02 &  0.09 &  0.08 &  \\
                             $[$MFD2010$]$ 4  & \nodata&  0.02 &  0.02 &\nodata&\nodata&\nodata&\nodata&\nodata&\nodata&\nodata&\nodata&\nodata&\nodata&\nodata&  0.04 &  0.02 &  0.09 &  0.08 &  \\
                            $[$MVM2011$]$ 39   &   0.02 &  0.02 &  0.02 &  0.16 &  0.09 &  0.07 &\nodata&\nodata&\nodata&  0.04 &  0.05 &  0.03 &  0.03 &\nodata&  0.03 &  0.02 &  0.10 &\nodata&  \\
                            39  &   0.02 &  0.03 &  0.02 &\nodata&\nodata&  0.21 &\nodata&\nodata&\nodata&  0.06 &  0.04 &  0.02 &  0.03 &  0.05 &  0.10 &  0.05 &  0.02 &  0.07 &  \\
                            40  &   0.02 &  0.04 &  0.02 &\nodata&  0.07 &  0.21 &\nodata&\nodata&\nodata&  0.06 &  0.05 &  0.03 &  0.02 &  0.05 &  0.07 &  0.07 &  0.02 &  0.02 &  \\
                            41  &   0.02 &  0.04 &  0.02 &  0.06 &  0.06 &  0.11 &\nodata&\nodata&\nodata&  0.12 &  0.05 &  0.03 &  0.02 &  0.05 &  0.04 &  0.02 &  0.03 &  0.04 &  \\
                            42  &   0.03 &  0.04 &  0.02 &  0.03 &  0.07 &  0.16 &\nodata&\nodata&\nodata&  0.10 &  0.05 &  0.04 &  0.03 &  0.05 &  0.05 &  0.03 &  0.04 &\nodata&  \\
                            43  & \nodata&  0.03 &  0.03 &  0.12 &  0.15 &  0.06 &\nodata&\nodata&\nodata&  0.11 &  0.08 &  0.03 &  0.02 &\nodata&  0.04 &  0.02 &  0.08 &\nodata&  \\
                            44  &   0.04 &  0.04 &  0.02 &  0.04 &  0.07 &  0.20 &\nodata&\nodata&\nodata&  0.30 &  0.11 &  0.02 &  0.03 &  0.05 &  0.05 &  0.03 &  0.02 &  0.05 &  \\
                            45  &   0.02 &  0.04 &  0.02 &  0.08 &  0.07 &  0.11 &\nodata&\nodata&\nodata&  0.05 &  0.09 &  0.03 &  0.03 &  0.05 &  0.05 &  0.03 &  0.03 &  0.06 &  \\
                            46  &   0.03 &  0.05 &  0.02 &\nodata&  0.07 &  0.10 &\nodata&\nodata&\nodata&  0.06 &\nodata&  0.03 &  0.03 &\nodata&  0.06 &  0.03 &  0.02 &  0.03 &  \\
                            47  &   0.02 &  0.03 &  0.02 &  0.09 &  0.07 &  0.11 &\nodata&\nodata&\nodata&  0.10 &  0.10 &  0.03 &  0.03 &  0.05 &  0.05 &  0.03 &  0.03 &  0.06 &  \\
                            48  &   0.02 &  0.03 &  0.03 &  0.10 &  0.07 &  0.12 &\nodata&\nodata&\nodata&\nodata&\nodata&  0.03 &  0.02 &  0.05 &\nodata&\nodata&\nodata&\nodata&  \\
                             $[$MFD2010$]$ 5  &   0.02 &  0.03 &  0.03 &\nodata&  0.08 &  0.08 &\nodata&\nodata&\nodata&  0.08 &  0.23 &  0.03 &  0.02 &\nodata&  0.06 &  0.02 &  0.07 &  0.05 &  \\
                            BD$-$08 4635  &   0.24 &  0.22 &  0.26 &  0.05 &\nodata&  0.15 &\nodata&\nodata&\nodata&\nodata&  0.18 &\nodata&\nodata&  0.05 &  0.11 &  0.06 &  0.02 &  0.03 &  \\
                            BD$-$08 4639  &   0.22 &  0.18 &  0.22 &  0.02 &\nodata&  0.16 &\nodata&\nodata&\nodata&\nodata&\nodata&\nodata&\nodata&  0.05 &  0.12 &  0.08 &  0.02 &  0.05 &  \\
                            BD$-$08 4645  &   0.21 &  0.18 &  0.19 &  0.04 &\nodata&  0.17 &\nodata&\nodata&\nodata&\nodata&\nodata&\nodata&\nodata&  0.05 &  0.31 &  0.14 &  0.03 &  0.03 &  \\

\hline
\end{tabular}
}
\end{table*} 
}

\subsection{Previously known massive stars in the direction of the complex}

In the SIMBAD astronomical archive, we found  matches  for 11 out of 151 observed stars.
The alias names are provided in Tables \ref{table.obspectra}, 
\ref{table.crsgspectra}, and \ref{table.giantspectra}.

\citet{messineo10} reported the detections of a few massive stars
in the direction of the GLIMPSE9 cluster;  [MFD2010]3  and [MFD2010]4 
are  two B0-5 supergiants;  [MFD2010]5 and [MFD2010]8  
are two RSG stars.  Our detection number \#46 coincides with  star [MFD2010]8.

We searched the lists of known WRs presented by 
 \citet{vanderhucht01}, \citet{mauerhan11}, and \citet{shara11}. 
The WR number 39 (WC8) in \citet{mauerhan11} (thereafter, we call it [MVM2011]39)
is projected onto  SNR  G22.07$-$0.3.

We searched in the Galactic spectroscopic database by \citet{skiff13} for known RSGs.
BD$-08$ 4645 (EIC 685) is reported as a M2 I by 
\citet{whitney83} and \citet{sylvester98}.
BD$-08$ 4635 and BD$-08$ 4639 are two  bright sources with IR colors 
similar to that of RSG BD$-08$ 4645. \citet{skiff13} lists them 
as M2 and K2 types, respectively.

These massive stars and candidate massive stars were added to the list of newly 
detected stars, and their photometric properties were re-investigated.

\section{Results}

\label{analizza}

\subsection{Spectral classification}

\subsubsection{Early-type stars}
\label{earlyclass}

\begin{figure*}
\begin{center}
\resizebox{0.4\hsize}{!}{\includegraphics[angle=0]{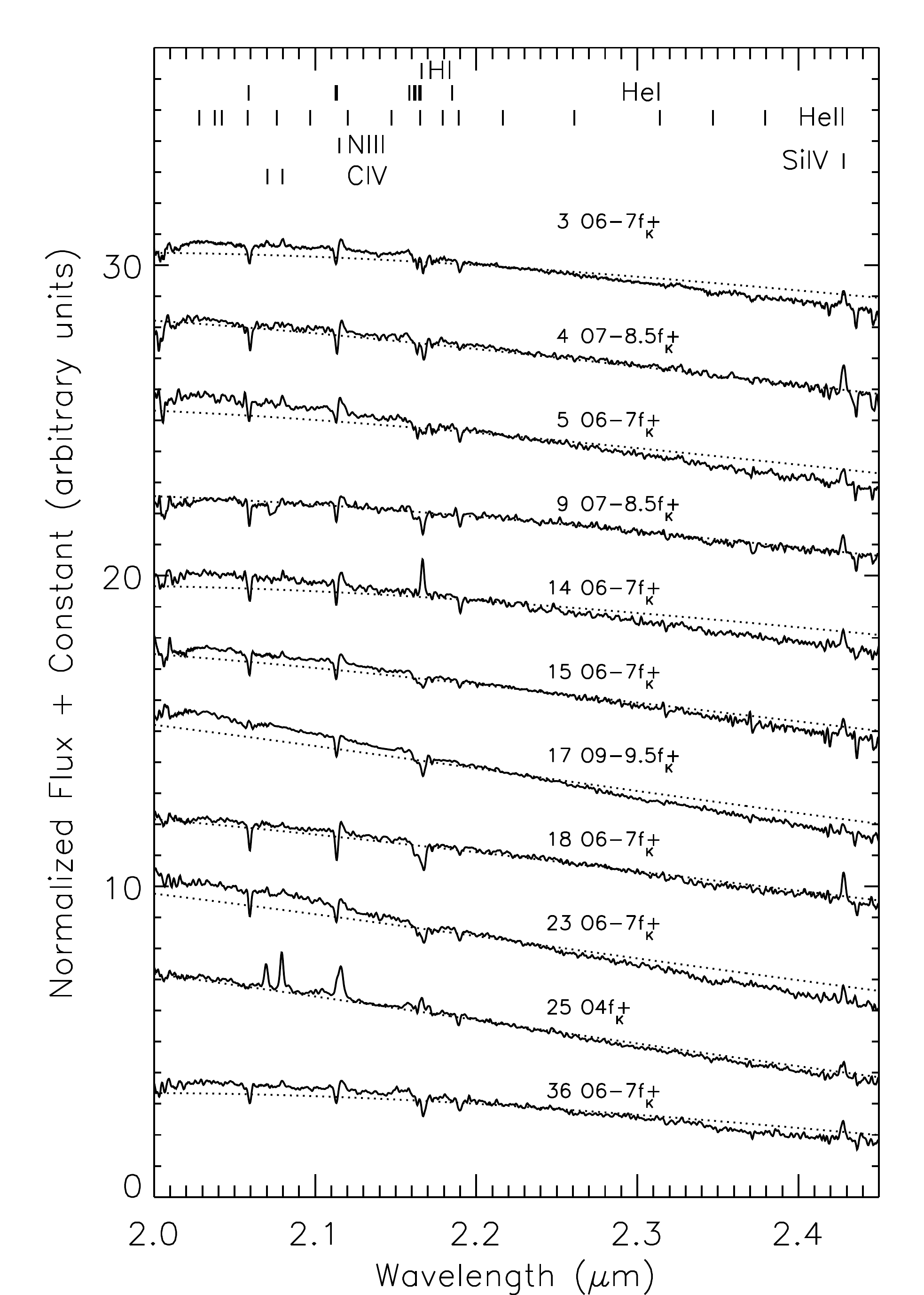}}
\resizebox{0.4\hsize}{!}{\includegraphics[angle=0]{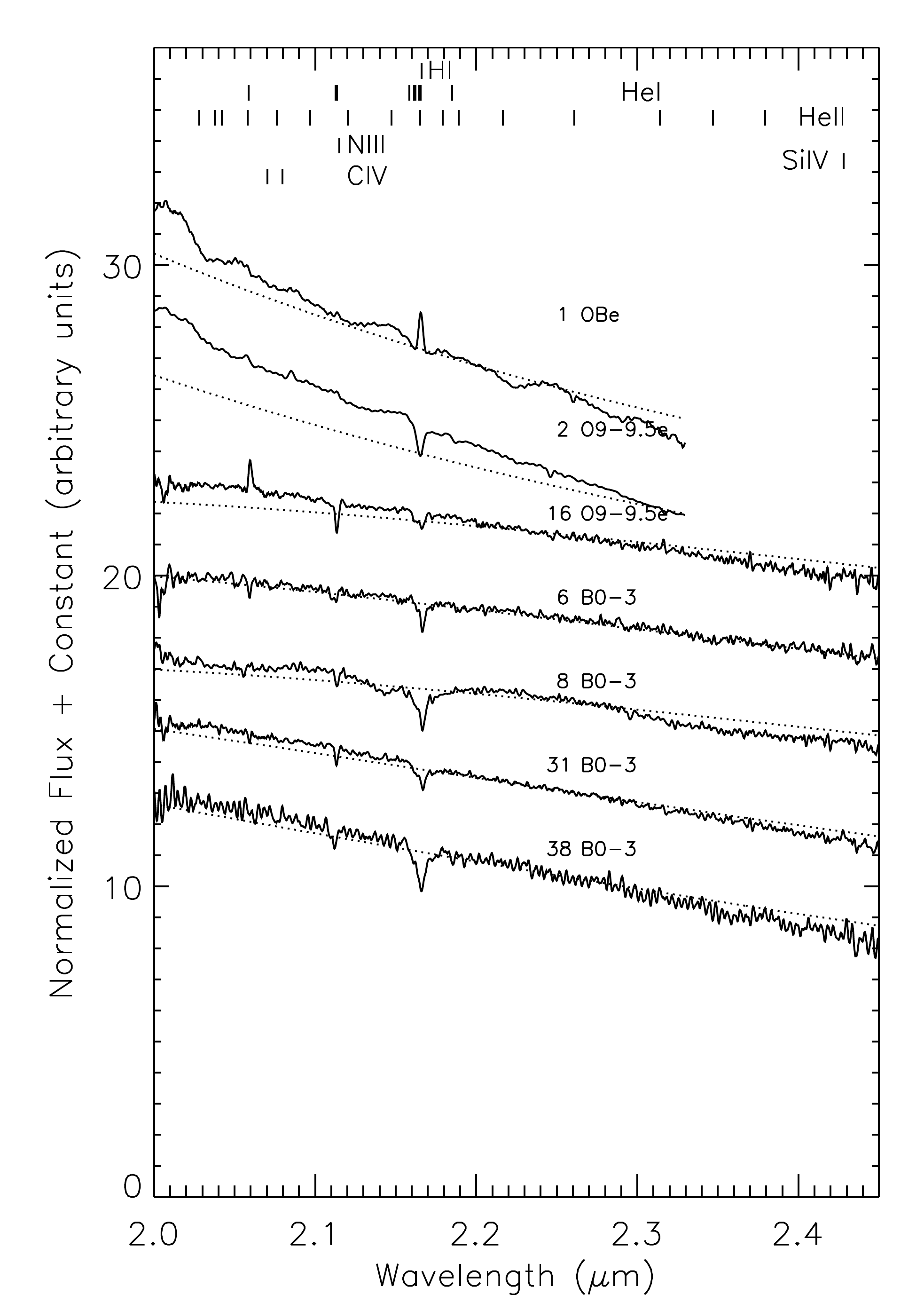}}
\end{center}
\begin{center}
\resizebox{0.4\hsize}{!}{\includegraphics[angle=0]{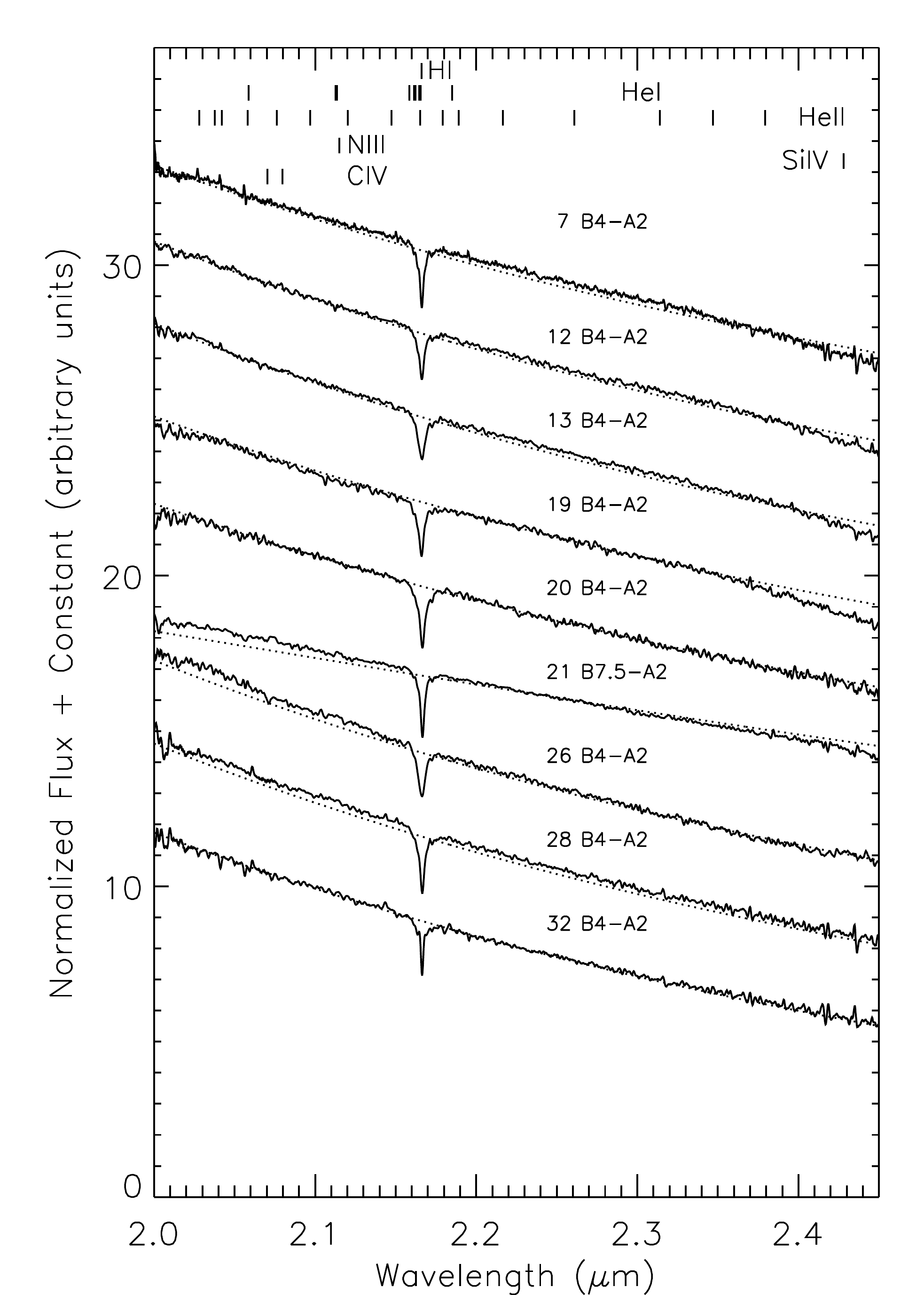}}
\resizebox{0.4\hsize}{!}{\includegraphics[angle=0]{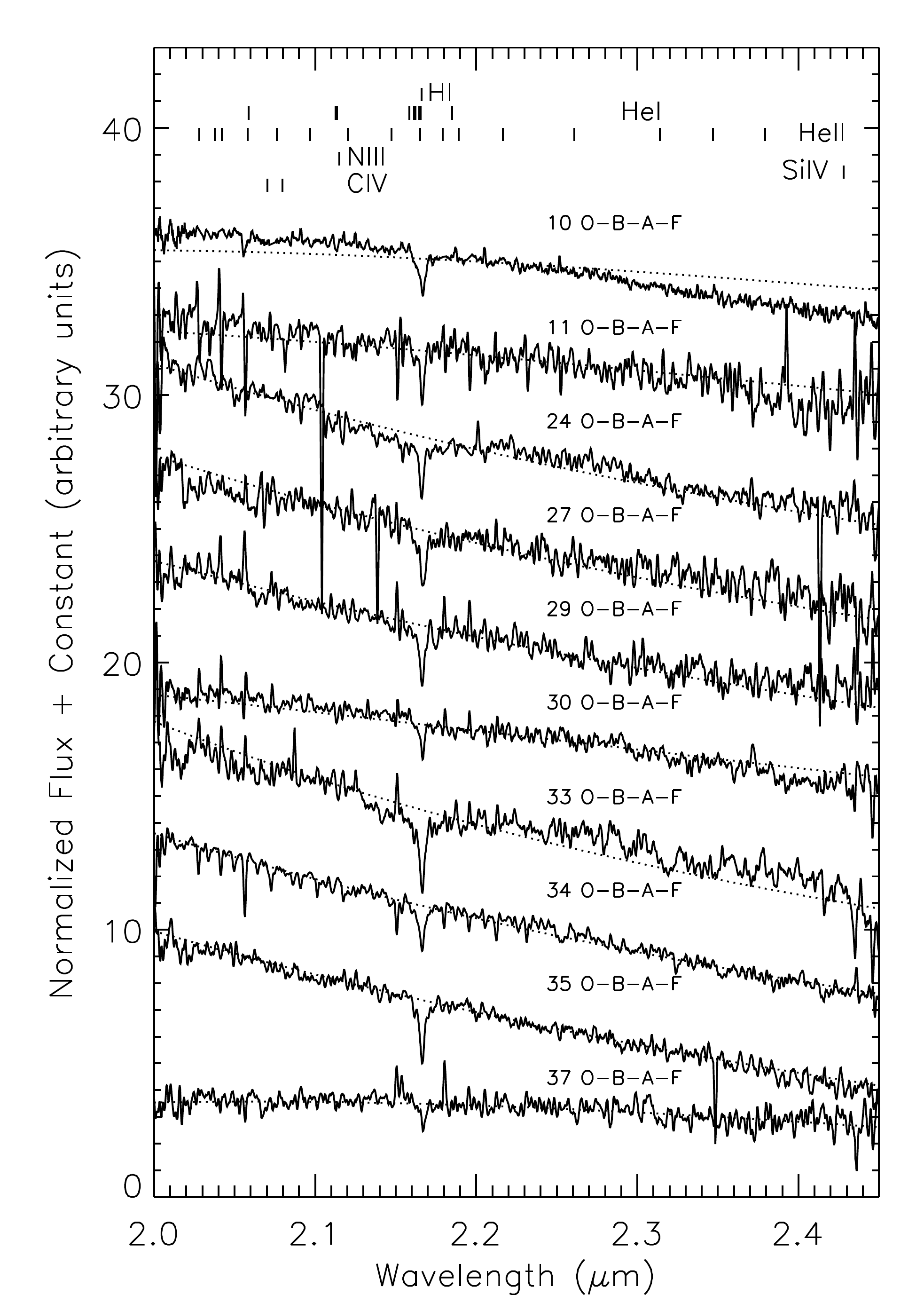}}
\end{center}
\caption{ \label{early.fig} Normalized spectra (arbitrarily shifted for clarity) 
of detected early-type stars.   The positions of lines from
\ion{H}{I}, \ion{He}{I}, and \ion{He}{II}  are marked and labeled at the top, along with  those of other
detected lines from \ion{Si}{IV}, \ion{N}{III}, and \ion{C}{IV}. 
The  spectra were multiplied by the black body of the standard star.
Dotted lines show  black bodies with the temperatures inferred from the observed stellar 
spectral types; these black bodies were reddened with individual 
\Aks\ values (for details see Section \ref{section.ext}).}
\end{figure*}

A total number of 38 early-type stars were detected
(see Figs.\ \ref{early.fig} and \ref{fig:splbv}).
We classified them by comparison with infrared spectroscopic atlases
\citep[e.g.][]{hanson96,morris96,figer97,hanson05}, 
by using  \ion{H}{I}, \ion{He}{I},  \ion{He}{II}, \ion{N}{III}, and \ion{C}{IV} lines.
\ion{C}{IV} lines are typical of O4-7 types, more rarely appear in O8 type;
the \ion{N}{III} complex at 2.115 \um\ disappears in stars later than  O8.5-O9 type;
the \ion{He}{II} line at 2.189 \um\ is present in O-type stars down to O9-type; 
the  \ion{He}{I} line at 2.112 \um\  is observed from O4-type  down to B8 (for supergiants), 
or B3  (for dwarfs), and the strengths of the \ion{He}{I} absorption line  at 2.112 \um\
increases from early-O to late-O. 
The \ion{He}{I} line at 2.058 \um\ is usually seen down to $\sim$B3 \citep{davies12}.

We used the prefix f$_K$+    to denote a spectral classification in  $K$-band 
similar to that given in the optical window by \citet{apellaniz07} and \citet{farina09}. 
We, thereby, defined an Of$_K$+ stars as a star with
a $K$-band spectrum that shows the \ion{N}{III}/\ion{C}{III} complex at 2.115 \um\  
in emission, and \ion{Si}{IV} at 2.428 \um\ in emission. 
There are only a few previous reports on 
the \ion{Si}{IV} line at 2.428 \um; the line was identified  in some
WRs and O supergiants of  the Arches cluster \citep{martins08}, and transitional objects
(e.g.\ cLBVs) in the vicinity of the Galactic center \citep{martins07}. 
The detected   O-type stars are all Of$_K$+.

\vspace{+1em}
{\it {\rm Of$_K$+} type stars ( panel 1 of Fig.\ \ref{early.fig}):} \\
The spectrum of star \#25  shows  strong \ion{C}{IV}  lines at 2.0705 \um\  
and  at 2.0796 \um\ in emission, and the  broad \ion{N}{III}/\ion{C}{III} complex  at  2.115 \um\ in emission, 
the \brg\ line at 2.1661 \um\ in absorption with a wind signature in emission, 
the \ion{He}{II} line at 2.1891 \um\ in absorption, 
and the \ion{Si}{IV}  line at 2.428 \um\ in emission.
These lines are typically detected in O stars with types from  4 to 6. 
In \citet{hanson96} and \citet{hanson05}, the  strength of the carbon  lines  appears 
to increase  with earlier types;
therefore, star \#25  is likely a O4-5f$_K$+ supergiant, 
similar to HD15570 \citep[see spectrum in ][]{hanson05}.

\clearpage

The spectra of stars \#3, \#5, \#14, \#15, \#18, \#23, and \#36  
display signatures of O6-7f$_K$+ stars;  they are
characterized by  the \ion{He}{I}  line at
2.058 \um,  a weak \ion{C}{IV} line at 2.0796 \um\ in emission,
a prominent \ion{He}{I}  line at 2.112 \um\ in absorption,
the \ion{N}{III} complex  at 2.115 \um\ in emission,
the \brg\  (mostly in absorption),  the \ion{He}{II} line at 2.189 \um\ in absorption, 
and the  \ion{Si}{IV} line at 2.248 \um. 
The spectra of stars \#3 and \#23 have the additional detection of  a \ion{C}{IV} line 
at 2.0705 \um. 
The spectrum of star \#14  has the \brg\ line in emission 
(O6-7f$_K$+);  the \brg\ lines of stars \#3 and \#5  display a wind signature.

The spectra of stars \#4  and \#9  have the \ion{He}{I} 
lines at 2.058 \um\ and 2.112 \um\ in absorption, the \ion{N}{III} at 2.115 \um\ in emission, the \brg\ line, 
the \ion{He}{II} line at 2.189 \um\ in absorption, and the \ion{Si}{IV} line at 2.428 \um. 
Star \#4  has a \brg\ line in absorption with a signature of wind in emission.
The non-detection of \ion{C}{IV} lines, the presence of \ion{N}{III} and \ion{He}{II} lines, and \ion{Si}{IV}
suggest a later Of$_K$+ (O7-O8.5$^{+}_K$).

The spectrum of star \#17  displays a \ion{He}{I} line at 2.112 \um\ in absorption, 
a weak \ion{N}{III} complex at  2.115 \um\  in emission, the \brg\ line in absorption, 
and the  \ion{Si}{IV} line at 2.248 \um\ in emission.
Since there is not \ion{He}{II} at 2.189 \um, but \ion{N}{III} emission is still detected, this star appears 
a (O9-O9.5)f$_K$+.\\

\vspace{+1em}
{\it Late-O and B type stars ( panel 2 of Fig.\ \ref{early.fig}):} 

The spectrum of star \#1  presents the \brg\ line  in emission.

The spectrum of star \#2 has the \brg\ line in absorption, and a hint  for the \ion{He}{I} line at  2.058 \um\
in emission, and for the \ion{He}{II} line at 2.189 \um\ in absorption. The lack of \ion{N}{III} at 2.115 \um, and the hint 
for \ion{He}{I} and   \ion{He}{II}, suggest a O9-O9.5e.

The spectrum of star \#16  shows the \ion{He}{I} line at 2.058 \um\ in emission, 
the \ion{He}{I} line at 2.112 \um\ in absorption, the \ion{N}{III} line at 2.115 \um\  in emission, 
and the \brg\ line in absorption.
The absence of \ion{He}{II} and  presence of \ion{N}{III} 
suggest a O9-9.5 type.
The 2.058 \um\ emission  indicates a supergiant luminosity class \citep{hanson96}.

The \ion{He}{I} line at 2.112 \um\  and the  \brg\ line in absorption are detected in the spectra of 
stars \#6, \#8,  \#31, and \#38.
The detection of \ion{He}{I} lines and the absence of \ion{N}{III} emission at 2.115 \um\ 
and of the \ion{He}{II} line at 2.189 \um\ suggest a B0-8I or a B0-3V. 
There is a hint for \ion{He}{I} at 2.058 \um\ in the spectra of 
stars  \#6 and \#8 (B0-3);  there is a hint for \ion{Si}{IV} at 2.248 \um\ in  the  spectrum of star \#31.

\vspace{+1em}
{\it B-A type stars ( panel 3 of Fig.\ \ref{early.fig}):} \\
We assigned a  B4-A2 type (dwarfs), or B7.5-A2 type  (supergiants)
to stars with  only a detected \brg\ line in absorption: \#7,    \#12, \#13, \#19, \#20, \#21,
\#26,  \#28,  and \#32.\\

\vspace{+1em}
{\it O-B-A-F type stars ( panel 4 of Fig.\ \ref{early.fig}):} \\
Stars with noisy spectra and  marginal detections of  \brg\ lines are  labeled  O-B-A-F
(stars \#10, \#11, \#24, \#27,\#29, \#30, \#33, \#34, \#35, and \#37).\\ 

The noisy structures  around 2.00 \um\ are due to a poor atmospheric correction.

\subsubsection{A candidate Luminous Blue Variable.}
\label{sec.lbv}

\begin{table}
\caption{ \label{splbv} List of  lines detected in the new spectra of the cLBV (\#22).}
\begin{tabular}{@{\extracolsep{-.08in}}lllrr}
Line & Vacuum $\lambda$  &   Obs. $\lambda$$^*$        &  EW$^+$  \\
     & [$\mu$m]          &    [$\mu$m]                             & [$\AA$]      \\
\ion{H}{I} 18-4                                     & 1.53460$^{e,f}$   & 1.53483  & $1.1\pm0.2$  \\ 
+$[\rm{\ion{Fe}{II}}]$ a$^4$F$_{9/2}-$a$^4$D$_{5/2}$& 1.53389$^{a,f}$   &   blended \\
\ion{H}{I} 17-4                                     & 1.54432$^{e,f}$   & 1.54473  & $1.0\pm0.2$ \\
\ion{H}{I} 16-4                                     & 1.55607$^{e,f}$   & 1.55624  & $2.3\pm0.5$ \\
\ion{H}{I} 15-4                                     & 1.57049$^{e,f}$   & 1.57089  & $1.7\pm0.3$ \\
\ion{Fe}{II}  $z^2$ I$_{11/2}-$3 d$^5$ 4 s$^2$ I$_{11/2}$ &   1.5776$^d$& 1.57663  & $1.6\pm0.3$ \\ 
\ion{H}{I} 14-4                                     & 1.58849$^{e,f}$   & 1.58885  & $2.0\pm0.5$ \\
\ion{H}{I} 13-4                                     & 1.61137$^{e,f}$   & 1.61177  & $1.5\pm1.0$ \\ 
\ion{H}{I} 12-4                                     & 1.64117$^{e,f}$   & 1.64151  & $1.9\pm0.4$ \\ 
$[\rm{\ion{Fe}{II}}]$ a$^4$F$_{9/2}-$a$^4$D$_{7/2}$ & 1.64400$^{a,f}$   & 1.64457  & $0.6\pm0.1$ \\ 
$[\rm{\ion{Fe}{II}}]$ a$^4$F$_{5/2}-$a$^4$D$_{1/2}$ & 1.66422$^{a,f}$   & 1.66510  & $0.5\pm0.1$ \\ 
\ion{H}{I} 11-4                                     & 1.68111$^{e,f}$   & 1.68136  & $3.7\pm0.8$ \\
\ion{Fe}{II}  $z^4$ F$_{9/2}-$ c$^4$  F$_{9/2}$     & 1.68778$^{d,f}$   & 1.68814  & $1.5\pm0.4$ \\ 
$[\rm{\ion{Fe}{II}}]$ a$^4$F$_{5/2}-$a$^4$D$_{3/2}$ & 1.71159$^{a,f}$   & 1.71151  & $0.4\pm0.3$ \\ 
\ion{H}{I} 10-4                                     & 1.73669$^{e,f}$   & 1.73700  & $3.1\pm0.3$ \\
\ion{He}{I}                                         & 2.05869$^{d,f}$   & 2.05950  & $7.8\pm0.5$\\
\ion{Fe}{II} $z^4$ F$^0_{3/2}-$ c$^4$  F$_{3/2}$    & 2.091$^d$         & 2.09009  & $0.9\pm0.3$  \\
\ion{Mg}{II}                                        & 2.13748$^{d,f}$   & 2.13808  & $0.6\pm0.5$ $^g$\\
\ion{Mg}{II}                                        & 2.14380$^{d,f}$   & 2.14453  & $0.3\pm0.3$ $^h$ \\
\ion{H}{I}  7-4                                     & 2.16612$^{e,f}$   & 2.16691  & $15.8\pm0.4$ \\
\ion{Na}{I}                                         & 2.206$^{d,f}$     &  2.2082  & $4.6\pm0.6$ \\
\ion{Na}{I}                                         & 2.20897$^{d,f}$   & blended  &\\

\end{tabular}

\begin{list}{}{}
\item[{\bf Notes.}]($^a$)  \citet{morris96,reunanen07}.~
%($^b$) from the  Atomic Line List v2.04.~
%($^c$) \citet{morris96}.~
($^d$) \citet{morris96,clark99}.~
($^e$) \citet{storey95}.~
($^f$) from the NIST line list.~
($^+$) Errors are calculated with the formula number 7 of \citet{zorro}.
Only lines with a significance of 1 sigma are listed.~  
($^*$) Absolute wavelength accuracy of each single frame is within 1.6\AA\
(based on OH lines).~
($^g$) The line peak is at 3$\sigma$.~
($^h$) The line peak is at 2$\sigma$.

\end{list}

\end{table}

\begin{figure*}
\begin{center}
\resizebox{0.6\hsize}{!}{\includegraphics[angle=0]{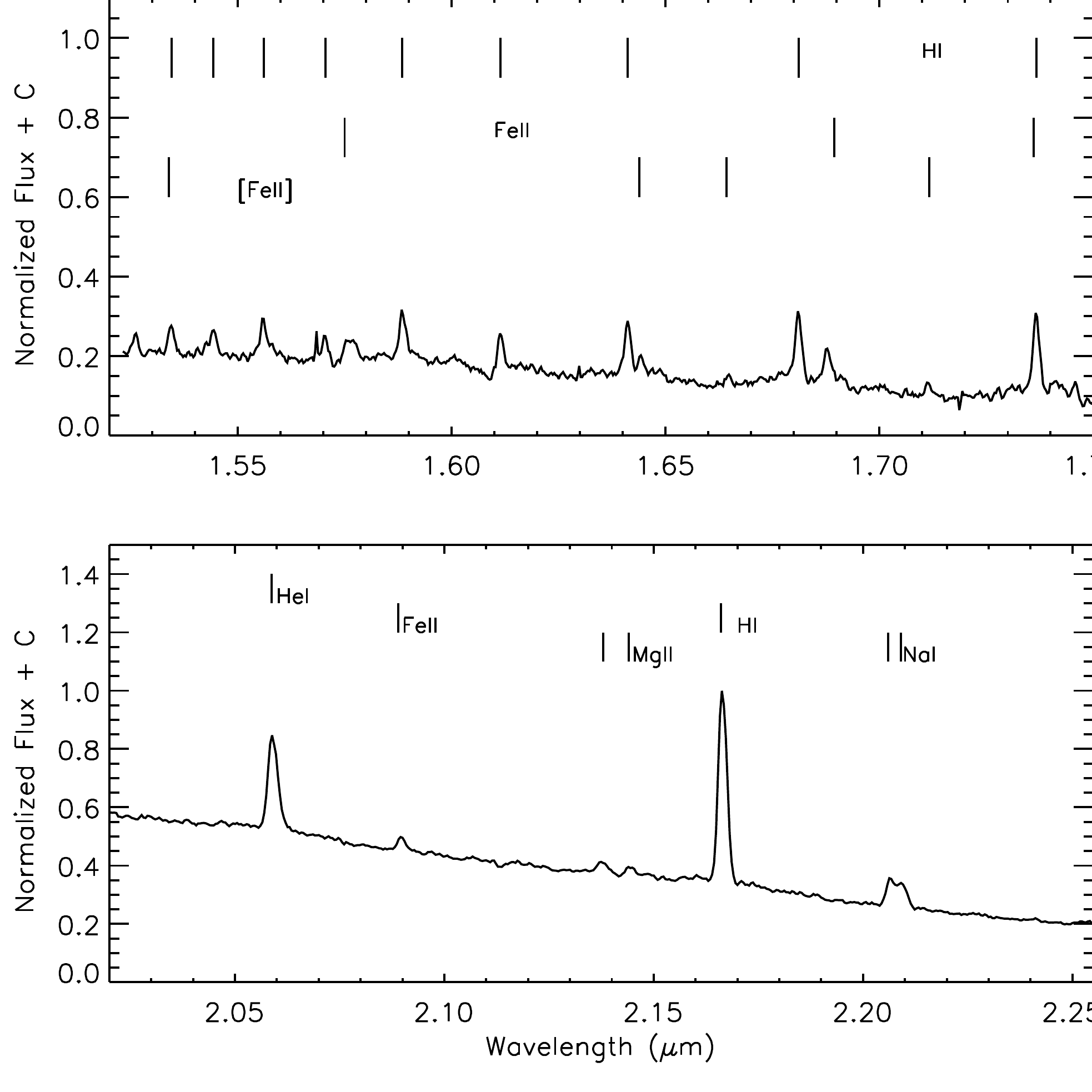} }
\end{center}
\caption{ \label{fig:splbv}  Spectra of the cLBV \#22. 
The positions of \ion{H}{I} and \ion{He}{I} lines are marked and labeled at the top, along with
those of other detected lines (\ion{Na}{I}, \ion{Mg}{II}, \ion{Fe}{II}). 
} 
\end{figure*}

In Figure \ref{fig:splbv} and Table \ref{splbv}, 
the spectral features of star \#22 are shown.
The $H$-band spectrum of star \#22   is characterized  by  \ion{H}{I} lines in emission 
and by a  number of iron lines (\ion{Fe}{II}), which are mostly forbidden ([\ion{Fe}{II}]). 
The  $K$-band spectrum shows emission lines from  \ion{He}{I}, \ion{H}{I}, 
\ion{Mg}{II}, \ion{Na}{I}, and \ion{Fe}{II}.

These lines are typical of massive  objects (for example  B[e]s, LBVs) in transition 
from the blue supergiant phase to the more evolved Wolf-Rayet stage, 
with cold envelopes or disks \citep[e.g.][]{morris96}.  
The possible evolutionary link between the disk-bearing B[e]s and the multi-wind LBVs is unclear, 
and this is a current topic of ongoing discussions \citep[e.g][]{crowther95,clark12b}.
LBVs display a large variety of stellar spectra; their definition is actually
based on their    variability and   sporadic strong  outbursts 
\citep[e.g.][]{thackeray74,humphreys78}.  

The $H$-band spectrum of star \#22  presents   \ion{H}{I} lines in emission (as in the spectrum of S Dor)
  and several Fe lines, which  recall   the rich spectrum of LBV WRA 751  \citep{morris96,smith02}.
The $K$-band spectra of  the stars Pistol,  Wra17$-$96, $G26.47+0.02$,  $G24.73+0.69$,   
and  HR Car exhibit  the same emission  lines as  those of star \#22 
\citep{figer95,morris96,clark03,egan02}. 
These impressive similarities with  other LBV spectra suggest that star \#22  
is a candidate LBV (cLBV\footnote{the prefix "c" (candidate) indicates that a photometric
monitoring is not available yet.}).

\begin{figure}
\resizebox{0.99\hsize}{!}{\includegraphics[angle=0]{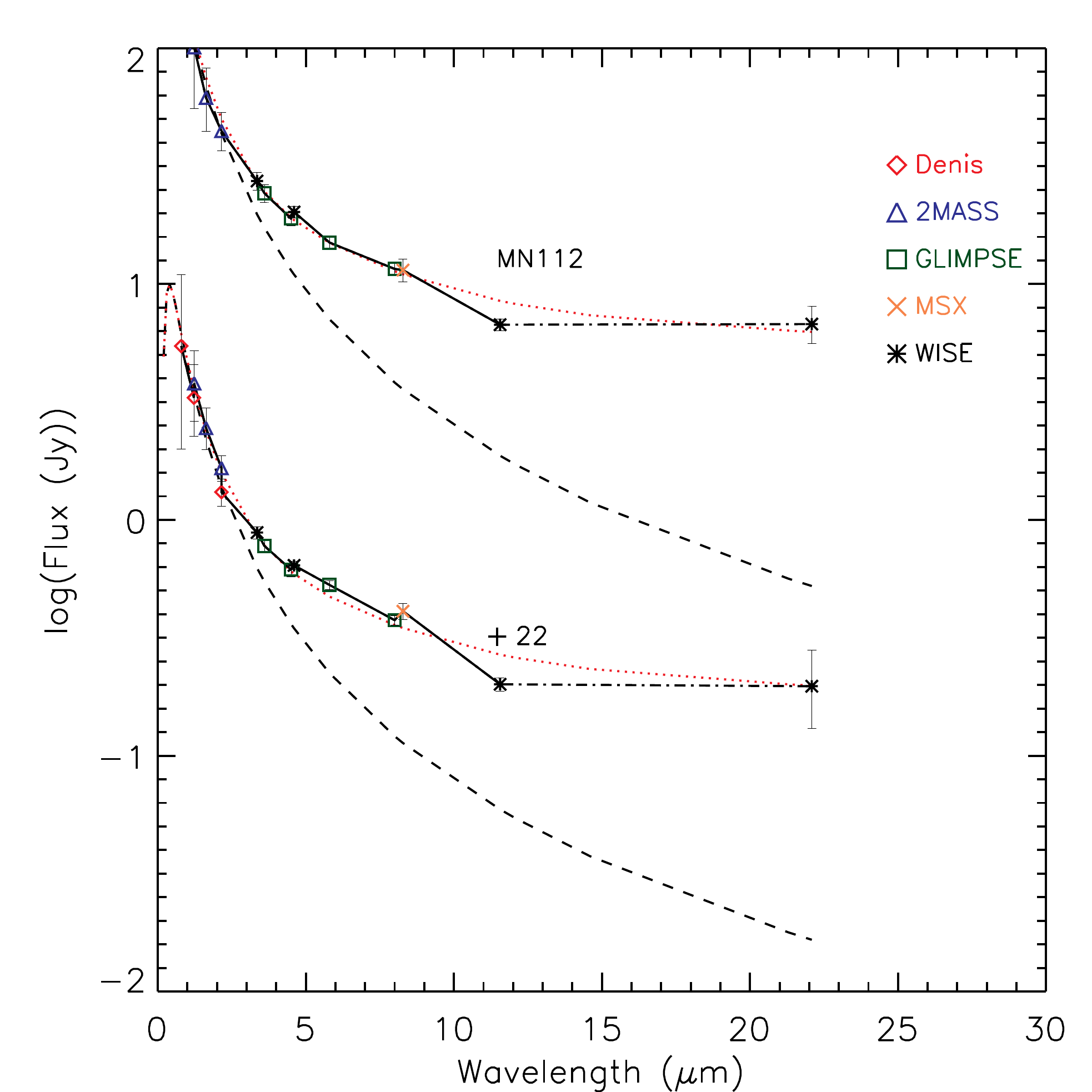}}
\caption{ \label{fig.sed} SEDs of star \#22 (this work) and of the cLBV MN112 discovered
by \citet{gvaramadze10}. Flux densities in the DENIS, 2MASS, MSX, 
GLIMPSE, and WISE bands are plotted  with diamonds, triangles, squares,
crosses,  and asterisks, respectively. 
 The WISE 3 (11 \um) and 4 (20 \um) measurements were marked as affected by confusion.
The long-dashed curves
are black-bodies with the stellar effective temperatures.
The dotted curves are modified black-bodies, which we created by adding 
to the continuum a 10-15\% of free-free emission ( $\propto \lambda^{-0.6}$),
a warm dust component at 650 k, and a cold dust component at 150 k.}
\end{figure}

The cLBV  has been detected as a point-source up to 20 \um\ (W4 band of the WISE survey).
With a GLIMPSE  [3.6]$-$[5.8] = 0.72 mag and a [3.6]$-$[8.0]= 0.96 mag, star \#22  well 
fits in the GLIMPSE color distribution  found for known Galactic LBV stars \citep{messineo12}.
The SED  of cLBV \#22  resembles that of cLBV MN112 \citep{gvaramadze10}, with
an  excess  at several mid-infrared wavelengths (see Fig.\ \ref{fig.sed}); 
however,  in contrast to MN112,
an extended circumstellar nebulae is not detected. 
We did not find significant photometric variations in the $J$- and \Ks-band of DENIS 
and 2MASS (Table \ref{callbv}).
Nevertheless, high probability of being a variable point source is reported in band $W3$ (11.6 \um) by the 
WISE catalog.

\begin{table}
\caption{\label{callbv}  Near-infrared measurements of cLBV \#22.} 
\begin{tabular}{lrrr}
\hline
     &DENIS1& DENIS2&  2MASS\\
Date &  $22-05-1999$  & $29-08-2000$ &$29-04-1999$  \\
\hline
$I$  &  $13.88\pm0.06$& $13.92\pm0.04$&    \\
$J$  &  $9.70\pm0.05$ &  $9.67\pm0.07$&  $9.78\pm0.02$\\
$H$  &  $..$              &  $..$             &  $8.42\pm0.04$\\
\Ks  &  $7.65\pm0.06$ &  $7.51\pm0.07$&  $7.63\pm0.02$\\
\hline
\end{tabular}
\begin{list}{}{}
\item[{\bf Notes.}]  
Two epochs of DENIS simultaneous
$IJ$\Ks\ measurements were available. 
We re-calibrated the DENIS measurements by 
using   point sources within 1\arcmin;
a significant offset was found for epoch one.
\end{list}

\end{table}

\subsubsection{Late-type stars}
\label{late.class}

\begin{figure}
\resizebox{\hsize}{!}{\includegraphics[angle=0]{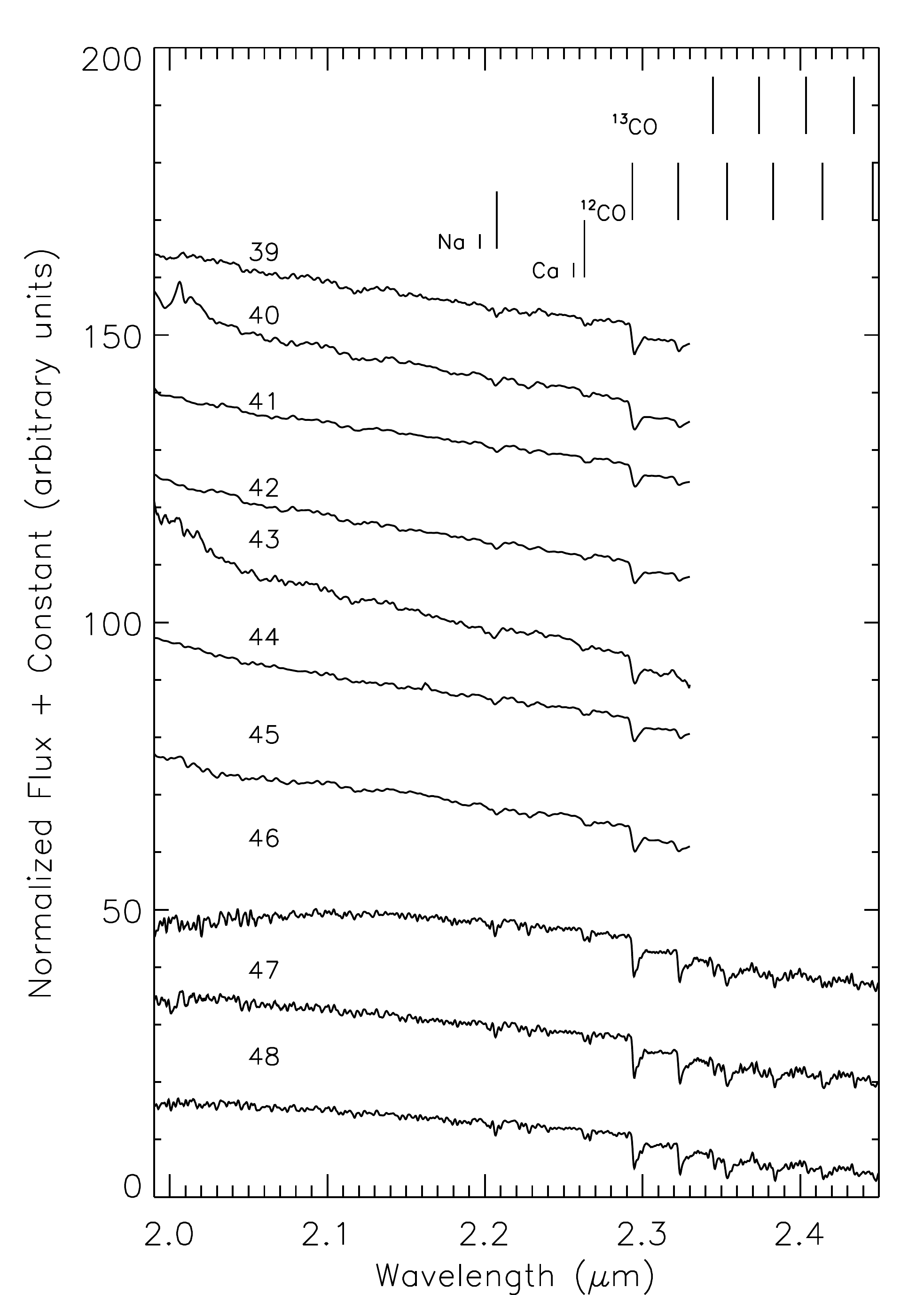}}
\caption{ \label{rsg.fig} Normalized spectra (arbitrarily shifted for clarity) 
of detected candidate red supergiants.  
The  spectra were multiplied by the black body of the 
standard star, and de-reddened.
The spectra with shorter coverage were taken with  SofI.
} 
\end{figure}

The equivalent width of the CO band-head, EW(CO), at 2.29 \um\ linearly correlates with the stellar 
temperature (\Teff).  CO absorption also strengthens with increasing luminosity.
Therefore, the EW(CO) and \Teff\  values of giants and RSGs follow two distinct 
relations  \citep{blum03,figer06,davies07}; 
the sequence of RSGs extends to larger  values of EW(CO).

The EWs are based on the \citet{kleinmann86} spectra.
We smoothed the reference spectra of \citet{kleinmann86} to the resolution 
of the observed ones;  we de-reddened each  target spectrum 
with the extinction law by \citet{messineo05} and 
the E($J-$\Ks) color excess (see  Sect.\ \ref{section.ext}).
The continuum was taken from 2.285 \um\ to  2.290 \um.
The EW(CO)s in unit of Angstroms were obtained 
by integrating the line strength of the CO feature,
1-Flux(CO)/Flux(continuum), in wavelengths \citep[from 2.290 \um\ to 2.320 \um, e.g.][]{figer06}. 
EW(CO)s from medium-resolution spectra taken with SofI
were measured in a narrower region, from 2.285 \um\ to 2.307 \um. 
Typical uncertainties of the  estimated spectral-types  are within a factor of two, 
as estimated  by slightly shifting the continuum region and the reddening.

Stars with EW(CO)s larger than that of a M7 giant were
classified as candidate RSGs or  variable AGB stars.
A detailed discussion on the identification of AGB stars,
which contaminate both red giant and RSG sequences, 
is provided in Appendix \ref{agbsel}.
After having excluded one  AGB star (\#56), we found that four other stars show 
EWs larger that that of an M7III star: \#40, \#43, \#46, and \#47.

Spectral types for the 113 detected late-type stars are  listed in Tables 
\ref{table.crsgspectra} and \ref{table.giantspectra}.  
Each list is sorted by coordinates.
Some spectra of bright late-type stars are displayed in Fig.\
\ref{rsg.fig}.

\subsection{Determination of \Aks}
\label{section.ext}

In the near-infrared, the attenuation of a star's light by interstellar dust 
absorption is wavelength-dependent, and may be expressed by a power 
law $A_\lambda$ $\propto \lambda^{-\alpha}$.

For every star, we estimated the effective extinction in \Ks-band, \Aks, 
by measuring the near-infrared color-excess, and
by using   $\alpha=-1.9$ \citep{messineo05}.  
We adopted the intrinsic infrared colors per spectral type tabulated by \citet{messineo11}; 
they were taken from \citet{martins06} (O-stars in the Bessell system), 
\citet{wegner94} (B-A stars in the Johnson system), \citet{johnson66} 
(B-A dwarfs in the Johnson system), 
\citet{koornneef83}  (B-A supergiants and late-types in the Koornneef system),
\citet{lejeune01} (colors of dwarfs from O3 to A5 in the Bessell system).
The used compilation uses data in the Johnson, Bessell, and Koornneef filter systems.
Color transformations were not applied, but no significant deviations were found.
There is no significant difference between the SAAO and the Johnson system 
\citep{carter90,blum00}. \citet{carpenter01} found differences between the SAAO system 
(or Koornneff system) and the 2mass system well within 0.1 mag.
Table \ref{table.obspectra} lists the adopted intrinsic $(J-K)_o$ and 
$(H-K)_o$ colors of early-types.

\begin{figure}
\begin{center}
\resizebox{0.7\hsize}{!}{\includegraphics[angle=0]{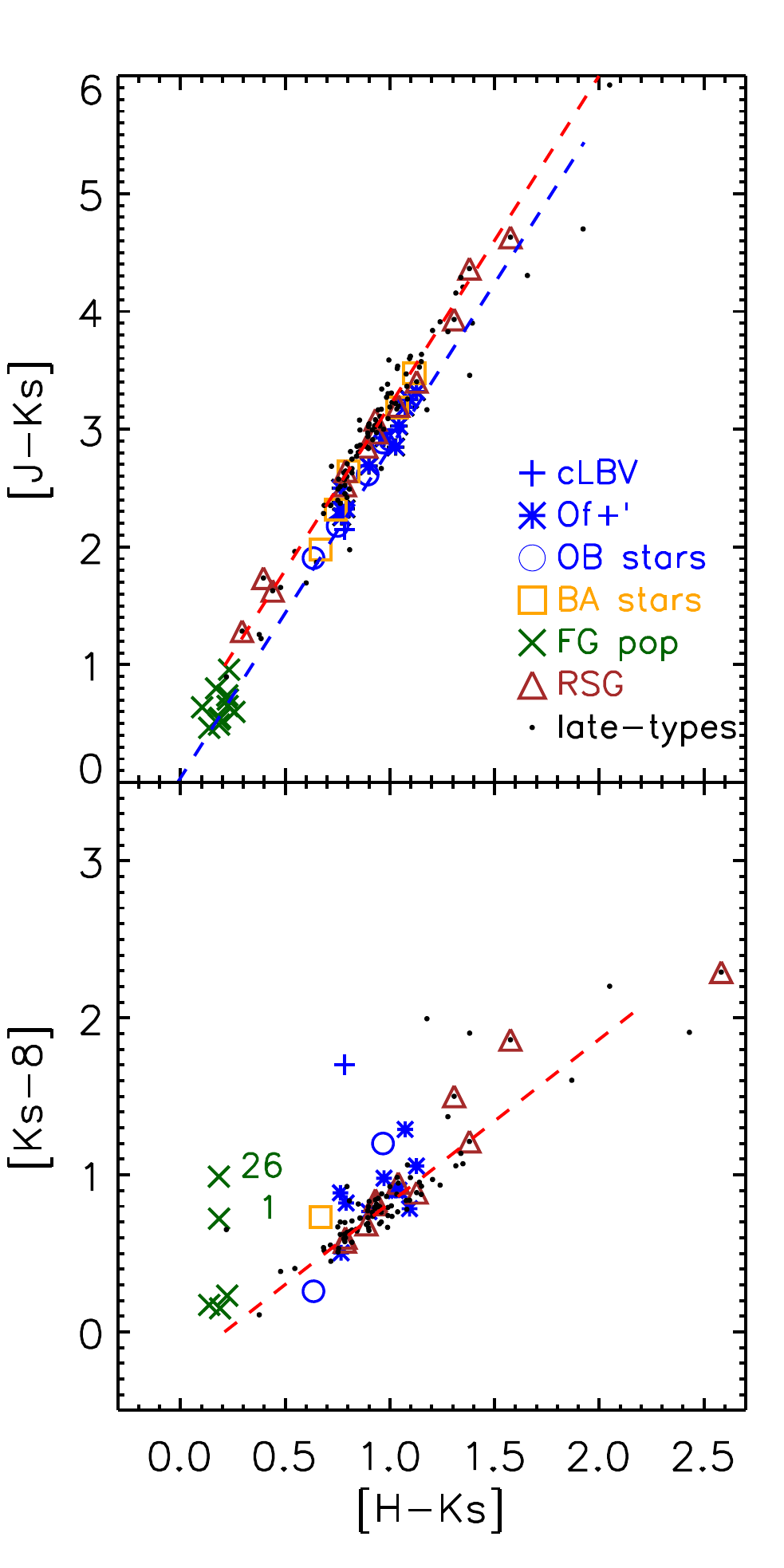}}
\end{center}
\caption{ \label{colcol}  {\it Top panel:} the $J-$\Ks\ versus $H-$\Ks\ diagram of 
the observed stars. Spectral-types 
are marked as shown in  the figure legend; the two dashed curves indicate the 
reddening curves of  naked M1 and O9 stars.
{\it Bottom panel:}   \Ks$-8$ versus $H-$\Ks\ diagram;
a reddening curve for an M1 star is shown with a dashed line.
Star \#22 (cLBV) shows infrared excess at 8 \um. 
The two foreground stars \#1  (OBe) and \#26 (B4-A4) have  notable infrared excess.
Star \#7 was not plotted (8 \um\ confused). 
} 
\end{figure}

\begin{figure}
\resizebox{0.99\hsize}{!}{\includegraphics[angle=0]{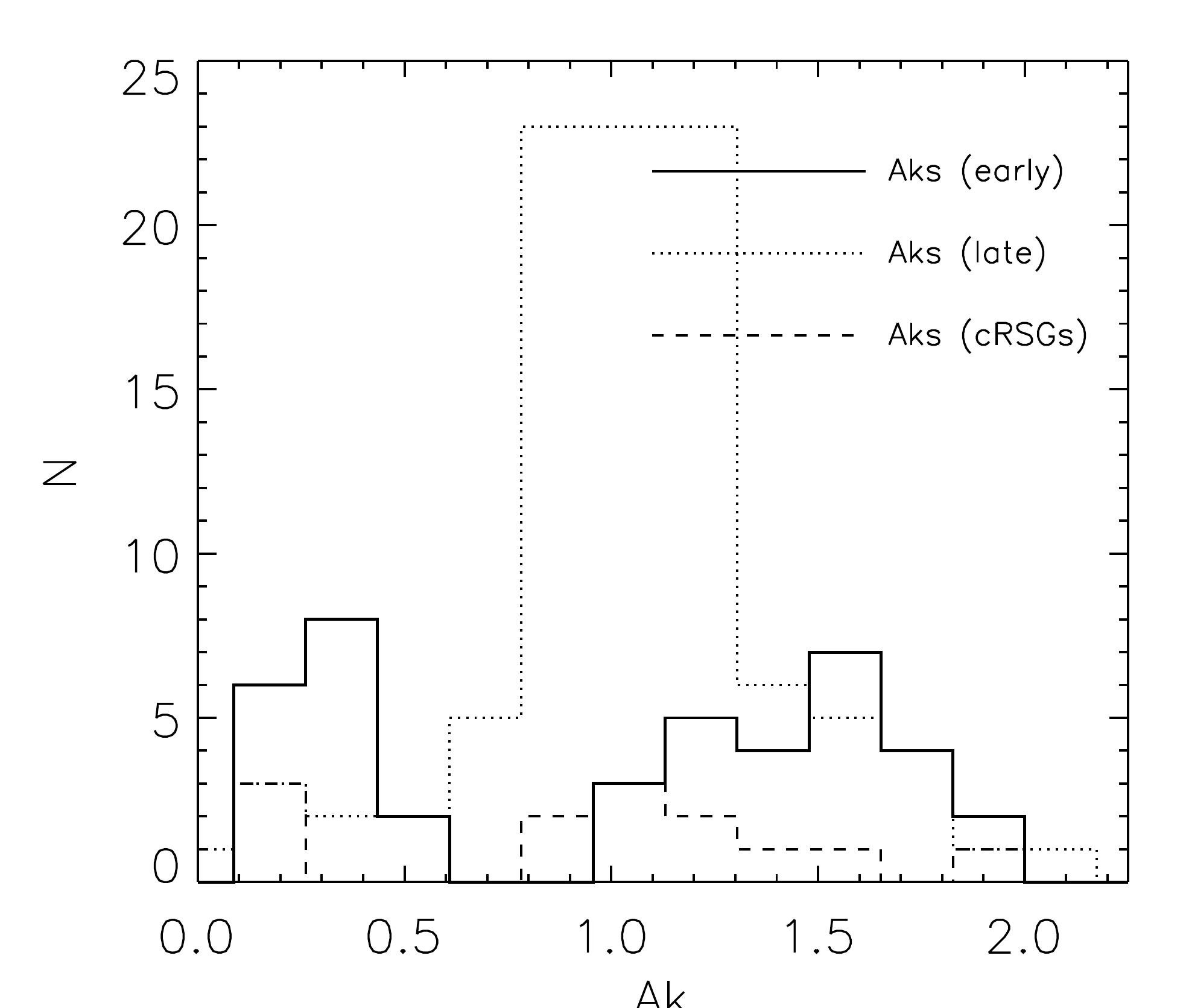}}
\caption{ \label{early.extinction} 
Histograms of the \Aks\ values for early-type stars
(solid line), late-type stars (dotted line), and candidate red supergiants  
from Table \ref{table.mbolrsg} (dashed line).
} 
\end{figure}

We  assumed as interstellar extinction individual \Aks\ values.
For the  detected early-type stars, we preferred the total interstellar extinction
\Aks\ from the shortest color $E(J-H)$;  for late-type stars, we used individual 
\Aks\ from $J-$\Ks\ (or $H-$\Ks) \citep{koornneef83}.  

A $J-$\Ks\ versus $H-$Ks diagram of the observed sources is shown in Fig.\ \ref{colcol}.
The cLBV displays an infrared excess  long-ward of 2 \um;
the O-type stars nicely follow the reddening vectors.
The bulk of  detected late-type stars (with exclusion of  a few AGBs)  
lacks   strong dust excess (Fig.\ \ref{colcol}), 
as  also inferred from the $Q1$  parameter (see Appendix \ref{q1q2sec}).

Figure \ref{early.extinction} shows the distribution of \Aks\ for early- 
and late-type stars. Two distinct populations of early-type stars are found; 
there is a group of bluer objects with
\Aks  $< 0.8$ mag, and a group with \Aks\ from 0.9 mag to 2.0 mag.
The distribution of  \Aks\ of late-type stars peaks around 0.9 mag, and 
appears unrelated to that of early-types.

At infrared wavelength, Galactic interstellar extinction has been best modeled
using a power-law with  an index $\alpha$ from $-1.61$   
\citep[for example, ][]{rieke85, indebetouw05} to  about $-2.1$  
\citep{nishiyama06} and \citet{stead09}.
For a reddening,  E($H-$\Ks), of 0.7 and 1.3 mag, $\alpha = -1.9$ 
yields \Aks=1 and 2 mag, while $\alpha =-1.61$ would yield \Aks=1.19 and 2.37 mag, and
$\alpha =-2.2$ would yield \Aks=0.86 and 1.61 mag.
Therefore, Rieke's law would brighten the de-reddened \Ks\ and \Mbol\ of $-0.19$ and $-0.37$ mag;
an index  of $-2.2$  would dim the de-reddened \Ks\ and \Mbol\ of $+0.14$ and $+0.39$ mag.
An index of $-1.9$ provides consistent values of interstellar extinction  
from  multicolor reddenings ( e.g.  $E(J-H)$, $E(H-K)$), and $E(J-K)$).

\subsection{Spectro-photometric distances}
\label{section.dist}

\begin{table*}\renewcommand{\arraystretch}{0.8}
\caption{\label{table.distance} Spectrophotometric distances of individual early-type stars. 
}
\begin{tabular}{@{\extracolsep{-.09in}}lrrll l l l rrrlr}
\hline
  ID & \Ks$_o$ & \Aks &     Spectral&     Class &  {\it M$_{\rm K}$}(I)$^b$  &{\it  M$_{\rm K}$}(III)$^b$ &{\it M$_{\rm K}$}(V)$^b$ &    DM I$^a$ &  DM IIII$^a$   & DM V$^a$ & Region  \\
     & [mag]   & [mag] &             &           &  [mag]           & [mag]            & [mag]         &  [mag]      &  [mag]     & [mag]& \\   
\hline
  16 &   7.50 &   1.68 &     O9-9.5e &           I &  -5.39$\pm$  0.83 & -4.47$\pm$  0.65 & -3.30$\pm$  0.76 &  {\bf 12.89 $\pm$   0.83 } &      { 11.97 $\pm$   0.65 } &      { 10.80 $\pm$   0.76 } &          REG4 &  \\
  15 &   7.93 &   1.59 &      O6-7f$_K$+ &           I &  -5.28$\pm$  0.66 & -4.84$\pm$  0.57 & -3.99$\pm$  0.65 &  {\bf 13.21 $\pm$   0.66 } &      { 12.77 $\pm$   0.57 } &      { 11.92 $\pm$   0.65 } &      GLIMPSE9 &  \\
  36 &   8.13 &   1.93 &      O6-7f$_K$+ &           I &  -5.28$\pm$  0.66 & -4.84$\pm$  0.57 & -3.99$\pm$  0.65 &  {\bf 13.41 $\pm$   0.66 } &      { 12.97 $\pm$   0.57 } &      { 12.12 $\pm$   0.65 } &          REG4 &  \\
   4 &   8.26 &   1.62 &    O7-8.5f$_K$+ &           I &  -5.39$\pm$  0.83 & -4.66$\pm$  0.61 & -3.63$\pm$  0.71 &  {\bf 13.65 $\pm$   1.17 } &      { 12.92 $\pm$   1.02 } &      { 11.89 $\pm$   1.08 } &      GLIMPSE9 &  \\
  14 &   8.32 &   1.86 &      O6-7f$_K$+ &           I &  -5.28$\pm$  0.66 & -4.84$\pm$  0.57 & -3.99$\pm$  0.65 &  {\bf 13.60 $\pm$   0.66 } &      { 13.16 $\pm$   0.57 } &      { 12.31 $\pm$   0.65 } &      GLIMPSE9 &  \\
  18 &   8.41 &   1.55 &      O6-7f$_K$+ &           I &  -5.28$\pm$  0.66 & -4.84$\pm$  0.57 & -3.99$\pm$  0.65 &  {\bf 13.69 $\pm$   0.66 } &      { 13.25 $\pm$   0.57 } &      { 12.40 $\pm$   0.65 } &          REG4 &  \\
  25 &   8.57 &   1.34 &        O4f$_K$+ &           I &  -5.16$\pm$  0.63 & -5.05$\pm$  0.63 & -4.41$\pm$  0.78 &  {\bf 13.73 $\pm$   0.63 } &      { 13.62 $\pm$   0.63 } &      { 12.98 $\pm$   0.78 } &          REG7&  \\
   5 &   8.65 &   1.73 &      O6-7f$_K$+ &         III &  -5.28$\pm$  0.66 & -4.84$\pm$  0.57 & -3.99$\pm$  0.65 &     { 13.93 $\pm$   0.66 } &   {\bf 13.49 $\pm$   0.57 } &      { 12.64 $\pm$   0.65 } &      GLIMPSE9 &  \\
   9 &   8.65 &   1.75 &    O7-8.5f$_K$+ &         III &  -5.39$\pm$  0.83 & -4.66$\pm$  0.61 & -3.63$\pm$  0.71 &     { 14.04 $\pm$   0.84 } &   {\bf 13.31 $\pm$   0.62 } &      { 12.28 $\pm$   0.72 } &      GLIMPSE9 &  \\
   3 &   8.85 &   1.90 &      O6-7f$_K$+ &         III &  -5.28$\pm$  0.66 & -4.84$\pm$  0.57 & -3.99$\pm$  0.65 &     { 14.13 $\pm$   0.66 } &   {\bf 13.69 $\pm$   0.57 } &      { 12.84 $\pm$   0.65 } &      GLIMPSE9 &  \\
  23 &   9.04 &   1.39 &      O6-7f$_K$+ &         III &  -5.28$\pm$  0.66 & -4.84$\pm$  0.57 & -3.99$\pm$  0.65 &     { 14.32 $\pm$   0.66 } &   {\bf 13.88 $\pm$   0.57 } &      { 13.03 $\pm$   0.65 } &          REG4 &  \\
  17 &   9.21 &   1.37 &    O9-9.5f$_K$+ &         III &  -5.39$\pm$  0.83 & -4.47$\pm$  0.65 & -3.30$\pm$  0.76 &     { 14.60 $\pm$   0.83 } &   {\bf 13.68 $\pm$   0.65 } &      { 12.51 $\pm$   0.76 } &          REG4 &  \\
\hline
   $[$MFD2010$]$ 3 &   6.48 &   1.48 &        B0-3 &           I &  -6.27$\pm$  0.92 & -3.47$\pm$  1.43 & -2.38$\pm$  1.27 &  {\bf 12.75 $\pm$   0.97 } &      {  9.95 $\pm$   1.46 } &      {  8.86 $\pm$   1.30 } &      GLIMPSE9 &  \\
   $[$MFD2010$]$ 4 &   7.48 &   1.66 &        B0-3 &           I &  -6.27$\pm$  0.92 & -3.47$\pm$  1.43 & -2.38$\pm$  1.27 &  {\bf 13.75 $\pm$   0.92 } &      { 10.95 $\pm$   1.43 } &      {  9.86 $\pm$   1.27 } &      GLIMPSE9 &  \\
   6 &   9.02 &   1.51 &        B0-3 &         III &  -6.49$\pm$  1.14 & -3.47$\pm$  1.43 & -2.38$\pm$  1.27 &     { 15.51 $\pm$   1.15 } &   {\bf 12.49 $\pm$   1.43 } &      { 11.40 $\pm$   1.27 } &      GLIMPSE9 &  \\
  31 &   9.05 &   1.26 &        B0-3 &         III &  -6.49$\pm$  1.14 & -3.47$\pm$  1.43 & -2.38$\pm$  1.27 &     { 15.54 $\pm$   1.14 } &   {\bf 12.52 $\pm$   1.43 } &      { 11.43 $\pm$   1.27 } &          REG2 &  \\
  38 &   9.37 &   1.13 &        B0-3 &         III &  -6.49$\pm$  1.14 & -3.47$\pm$  1.43 & -2.38$\pm$  1.27 &     { 15.86 $\pm$   1.14 } &   {\bf 12.84 $\pm$   1.43 } &      { 11.75 $\pm$   1.27 } &          REG4 &  \\
  21 &   9.64 &   1.11 &     B7.5-A2 &         III &  -7.04$\pm$  0.65 &  \nodata    & -0.41$\pm$  1.25 &     { 16.68 $\pm$   0.65 } &   $..$ &      { 10.05 $\pm$   1.25 } &          REG4 &  \\
   8 &  10.34 &   1.67 &        B0-3 &         III &  -6.49$\pm$  1.14 & -3.47$\pm$  1.43 & -2.38$\pm$  1.27 &     { 16.83 $\pm$   1.14 } &   {\bf 13.81 $\pm$   1.43 } &      { 12.72 $\pm$   1.27 } &      GLIMPSE9 &  \\
\hline
   2 &   6.10 &   0.51 &     O9-9.5e &           I &  -5.39$\pm$  0.83 & -4.47$\pm$  0.65 & -3.30$\pm$  0.76 &  {\bf 11.49 $\pm$   0.83 } &      { 10.57 $\pm$   0.65 } &      {  9.40 $\pm$   0.76 } &          \nodata &  \\
  19 &   9.94 &   0.42 &       B4-A2 &           V &  -7.04$\pm$  0.65 &  \nodata    & -0.41$\pm$  1.25 &     { 16.98 $\pm$   0.65 } &      { $..$ } &   {\bf 10.35 $\pm$   1.25 } &          REG5 &  \\
  12 &  10.07 &   0.32 &       B4-A2 &           V &  -7.04$\pm$  0.65 &  \nodata    & -0.41$\pm$  1.25 &     { 17.11 $\pm$   0.65 } &      { $..$ } &   {\bf 10.48 $\pm$   1.25 } &          REG5 &  \\
  13 &  10.08 &   0.29 &       B4-A2 &           V &  -7.04$\pm$  0.65 &  \nodata    & -0.41$\pm$  1.25 &     { 17.12 $\pm$   0.65 } &      { $..$ } &   {\bf 10.49 $\pm$   1.25 } &          REG5 &  \\
  32 &  10.19 &   0.37 &       B4-A2 &           V &  -7.04$\pm$  0.65 &  \nodata    & -0.41$\pm$  1.25 &     { 17.23 $\pm$   0.65 } &      { $..$ } &   {\bf 10.60 $\pm$   1.25 } &          REG5 &  \\
  26 &  10.42 &   0.31 &       B4-A2 &           V &  -7.04$\pm$  0.65 &  \nodata    & -0.41$\pm$  1.25 &     { 17.46 $\pm$   0.65 } &      { $..$ } &   {\bf 10.83 $\pm$   1.25 } &          REG5 &  \\
  28 &  10.48 &   0.30 &       B4-A2 &           V &  -7.04$\pm$  0.65 &  \nodata    & -0.41$\pm$  1.25 &     { 17.52 $\pm$   0.65 } &      { $..$ } &   {\bf 10.89 $\pm$   1.25 } &          REG5 &  \\
   7 &  10.54 &   0.42 &       B4-A2 &           V &  -7.04$\pm$  0.65 &  \nodata    & -0.41$\pm$  1.25 &     { 17.58 $\pm$   0.65 } &      { $..$ } &   {\bf 10.95 $\pm$   1.25 } &          REG5 &  \\
  20 &  11.95 &   0.47 &       B4-A2 &           V &  -7.04$\pm$  0.65 &  \nodata    & -0.41$\pm$  1.25 &     { 18.99 $\pm$   0.65 } &      { $..$ } &   {\bf 12.36 $\pm$   1.25 } &          REG4 &  \\
\hline
\end{tabular}

\begin{list}{}{}
\item[]{\bf Notes.}
Identification numbers from Table \ref{table.obspectra} are followed
by de-reddened \Ks, (\Ks$_o$), \Aks, spectral types, estimated luminosity classes,  
absolute magnitudes in the \Ks-band
for class I,  III, V,  distances, and regions.
The table lists  O-type  stars with \Aks$>0.8$ mag,  then  B-stars with \Aks$>0.8$ mag, 
and finally  star with \Aks$<0.8$ mag.~ Each block is ordered by de-reddened \Ks.~
($^a$) Distance modulus for the  estimated luminosity classes  are marked in bold (see text).~
($^b$) Quoted errors on  M$_{\rm K}$ are calculated by assuming an error of 
0.5 mag on a single type \citep[e.g.][]{bibby08, humphreys84}. Note that the quoted M$_{\rm K}$
values are used to derive spectrophotometric DMs. Later, in Table \ref{table.mbolearly},
we will assume a common distance and recalculate M$_{\rm K}$ and \Mbol.~
\end{list}
\end{table*}

\begin{table*}
\caption{ \label{table.distsum}   Average spectro-photometric distance of stars
with \Aks$>$0.8 mag. 
}
\begin{tabular}{@{\extracolsep{-.09in}}llrrrrlrr}
\hline
Spec.     &  Lum Class& Nstar & $<$\Aks$>$ & {\it M}$_{\rm K}$ & $<$DM$>$  &  REF. \\
          &      &    & [mag]  &  [mag]    & [mag]          &    \\
      O4-6&     I&  1 &   1.34 &   -5.16&  13.73$\pm$  0.63 &  1,2    \\
      O6-7&     I&  4 &   1.73 &   -5.28&  13.48$\pm$  0.21 &  1,2,3   \\
    O7-8.5&     I&  1 &   1.62 &   -5.39&  13.65$\pm$  1.17 &  3  \\
    O9-9.5&     I&  1 &   1.68 &   -5.39&  12.89$\pm$  0.83 &  3   \\
      O6-7&   III&  3 &   1.67 &   -4.84&  13.69$\pm$  0.20 &  4  \\
    O7-8.5&   III&  1 &   1.75 &   -4.66&  13.31$\pm$  0.62 &  4  \\
    O9-9.5&   III&  1 &   1.37 &   -4.47&  13.68$\pm$  0.65 &  4  \\
      B0-3&     I&  2 &   1.57 &   -6.27&  13.25$\pm$  0.71 &  5   \\
\hline
\end{tabular}
\begin{list}{}{}
\item[] {\bf Notes.}
Classes were assigned by assuming  similar distances; only 
supergiants yielded independent estimates.
For each group (see Table \ref{table.distance} ) and luminosity class, we report the number of
stars (Nstars), average \Aks, average distance modulus, and standard deviation.  ~
\item[]
{\bf References.}
(1) \citet{figer02};~
(2) \citet{martins08};~
(3) \citet{messineo11};~
(4) \citet{martins06};~
(5) \citet{bibby08}.
\end{list}
\end{table*}

\begin{table*}
\caption{\label{table.mbolearly} List of estimated stellar parameters for the sample of early-type 
stars with \Aks $>0.8$ mag.  
}
\begin{tabular}{@{\extracolsep{-.09in}}llrrrrrrrrll  }
\hline 
{\rm ID} &Sp. Type & \Ks$_o$ & \Aks($JH$) &\Aks($J$\Ks) &\Aks($H$\Ks) & $Q1$& \BCKs & DM & \Mbol &  \\ 
         &         &   [mag] &[mag]       &[mag]        & [mag]       &[mag] & [mag] &[mag]&[mag]\\
\hline 

  3 &    O6$-$7f$_K$+& 8.85 $\pm$ 0.04 &  1.90 $\pm$ 0.03 & 1.86 $\pm$ 0.02 & 1.79 $\pm$ 0.05 &  0.19 $\pm$  0.08 & $-$4.28 $\pm$  0.09 &  13.31 $\pm$    0.17& $-$8.74$\pm$  0.20 \\
  4 &  O7$-$8.5f$_K$+& 8.26 $\pm$ 0.82 &  1.62 $\pm$ 0.20 & 1.64 $\pm$ 0.20 & 1.69 $\pm$ 0.20 & $-$0.03 $\pm$  0.80 & $-$3.97 $\pm$  0.12 &  13.31 $\pm$    0.17& $-$9.02$\pm$  0.85 \\
  5 &    O6$-$7f$_K$+& 8.65 $\pm$ 0.04 &  1.73 $\pm$ 0.03 & 1.68 $\pm$ 0.02 & 1.60 $\pm$ 0.05 &  0.20 $\pm$  0.08 & $-$4.28 $\pm$  0.09 &  13.31 $\pm$    0.17& $-$8.94$\pm$  0.20 \\
  6 &      B0$-$3& 9.02 $\pm$ 0.11 &  1.51 $\pm$ 0.10 & 1.49 $\pm$ 0.07 & 1.46 $\pm$ 0.04 &  0.11 $\pm$  0.14 & $-$2.83 $\pm$  0.87 &  13.31 $\pm$    0.17& $-$7.11$\pm$  0.89 \\
  8 &      B0$-$3&10.34 $\pm$ 0.00 &  1.67 $\pm$ 0.00 & 1.65 $\pm$ 0.00 & 1.62 $\pm$ 0.00 &  0.11 $\pm$  0.00 & $-$2.83 $\pm$  0.87 &  13.31 $\pm$    0.17& $-$5.80$\pm$  0.89 \\
  9 &  O7$-$8.5f$_K$+& 8.65 $\pm$ 0.11 &  1.75 $\pm$ 0.06 & 1.74 $\pm$ 0.05 & 1.71 $\pm$ 0.16 &  0.11 $\pm$  0.23 & $-$4.05 $\pm$  0.14 &  13.31 $\pm$    0.17& $-$8.71$\pm$  0.25 \\
 14 &    O6$-$7f$_K$+& 8.32 $\pm$ 0.04 &  1.86 $\pm$ 0.04 & 1.82 $\pm$ 0.02 & 1.75 $\pm$ 0.05 &  0.18 $\pm$  0.10 & $-$4.17 $\pm$  0.08 &  13.31 $\pm$    0.17& $-$9.16$\pm$  0.19 \\
 15 &    O6$-$7f$_K$+& 7.93 $\pm$ 0.04 &  1.59 $\pm$ 0.03 & 1.56 $\pm$ 0.02 & 1.50 $\pm$ 0.05 &  0.17 $\pm$  0.09 & $-$4.17 $\pm$  0.08 &  13.31 $\pm$    0.17& $-$9.55$\pm$  0.19 \\
 16 &   O9$-$9.5e& 7.50 $\pm$ 0.04 &  1.68 $\pm$ 0.03 & 1.64 $\pm$ 0.02 & 1.55 $\pm$ 0.05 &  0.18 $\pm$  0.08 & $-$3.62 $\pm$  0.32 &  13.31 $\pm$    0.17& $-$9.43$\pm$  0.36 \\
 17 &  O9$-$9.5f$_K$+& 9.21 $\pm$ 0.04 &  1.37 $\pm$ 0.03 & 1.35 $\pm$ 0.02 & 1.32 $\pm$ 0.06 &  0.11 $\pm$  0.10 & $-$3.80 $\pm$  0.21 &  13.31 $\pm$    0.17& $-$7.90$\pm$  0.27 \\
 18 &    O6$-$7f$_K$+& 8.41 $\pm$ 0.03 &  1.55 $\pm$ 0.03 & 1.46 $\pm$ 0.02 & 1.29 $\pm$ 0.04 &  0.36 $\pm$  0.07 & $-$4.17 $\pm$  0.08 &  13.31 $\pm$    0.17& $-$9.07$\pm$  0.19 \\
 21 &   B7.5$-$A2& 9.64 $\pm$ 0.04 &  1.11 $\pm$ 0.03 & 1.07 $\pm$ 0.02 & 1.00 $\pm$ 0.05 &  0.11 $\pm$  0.08 & $-$0.92 $\pm$  0.92 &  13.31 $\pm$    0.17& $-$4.59$\pm$  0.94 \\
 22 &      cLBV& 6.50 $\pm$ 0.05 &  1.13 $\pm$ 0.04 & 1.15 $\pm$ 0.02 & 1.19 $\pm$ 0.08 & $-$0.04 $\pm$  0.14 & $-$1.09 $\pm$  0.62 &  13.31 $\pm$    0.17& $-$7.90$\pm$  0.64 \\
 23 &    O6$-$7f$_K$+& 9.04 $\pm$ 0.04 &  1.39 $\pm$ 0.03 & 1.36 $\pm$ 0.02 & 1.30 $\pm$ 0.05 &  0.17 $\pm$  0.09 & $-$4.28 $\pm$  0.09 &  13.31 $\pm$    0.17& $-$8.55$\pm$  0.20 \\
 25 &      O4f$_K$+& 8.57 $\pm$ 0.04 &  1.34 $\pm$ 0.03 & 1.32 $\pm$ 0.02 & 1.30 $\pm$ 0.05 &  0.10 $\pm$  0.09 & $-$4.40 $\pm$  0.15 &  13.31 $\pm$    0.17& $-$9.14$\pm$  0.23 \\
 31 &      B0$-$3& 9.05 $\pm$ 0.06 &  1.26 $\pm$ 0.05 & 1.26 $\pm$ 0.03 & 1.24 $\pm$ 0.08 &  0.08 $\pm$  0.15 & $-$2.83 $\pm$  0.87 &  13.31 $\pm$    0.17& $-$7.09$\pm$  0.89 \\
 36 &    O6$-$7f$_K$+& 8.13 $\pm$ 0.04 &  1.93 $\pm$ 0.03 & 1.90 $\pm$ 0.02 & 1.84 $\pm$ 0.05 &  0.16 $\pm$  0.09 & $-$4.17 $\pm$  0.08 &  13.31 $\pm$    0.17& $-$9.35$\pm$  0.19 \\
 38 &      B0$-$3& 9.37 $\pm$ 0.04 &  1.13 $\pm$ 0.03 & 1.11 $\pm$ 0.02 & 1.07 $\pm$ 0.05 &  0.13 $\pm$  0.08 & $-$2.83 $\pm$  0.87 &  13.31 $\pm$    0.17& $-$6.77$\pm$  0.89 \\
  $[$MFD2010$]$ 3 &      B0$-$3& 6.48 $\pm$ 0.30 &  1.48 $\pm$ 0.03 & 1.49 $\pm$ 0.16 & 1.51 $\pm$ 0.45 & $-$0.02 $\pm$  0.54 & $-$2.50 $\pm$  0.80 &  13.31 $\pm$    0.17& $-$9.33$\pm$  0.87 \\
  $[$MFD2010$]$ 4 &      B0$-$3& 7.48 $\pm$ 0.05 & $..$  &$..$  & 1.66 $\pm$ 0.04 & $..$  & $-$2.50 $\pm$  0.80 &  13.31 $\pm$    0.17& $-$8.33$\pm$  0.82 \\
  WR 39 &       WC8& 8.01 $\pm$ 0.03 &  1.35 $\pm$ 0.03 & 1.28 $\pm$ 0.02 & 1.17 $\pm$ 0.05 & $-$0.44 $\pm$  0.08 & $-$3.60 $\pm$  0.50 &  13.31 $\pm$    0.17& $-$8.90$\pm$  0.53 \\
\hline
  1 &       OBe& 8.86 $\pm$ 0.04 &  0.31 $\pm$ 0.03 & 0.33 $\pm$ 0.02 & 0.36 $\pm$ 0.05 & $-$0.02 $\pm$  0.08 & $-$2.87 $\pm$  1.21 &  12.39 $\pm$    1.00& $-$6.40$\pm$  1.57 \\
  2 &   O9$-$9.5e& 6.10 $\pm$ 0.04 &  0.51 $\pm$ 0.03 & 0.48 $\pm$ 0.02 & 0.44 $\pm$ 0.05 &  0.12 $\pm$  0.09 & $-$3.62 $\pm$  0.32 &  12.39 $\pm$    1.00& $-$9.90$\pm$  1.05 \\
  7 &     B4$-$A2&10.54 $\pm$ 0.04 &  0.42 $\pm$ 0.03 & 0.41 $\pm$ 0.02 & 0.39 $\pm$ 0.05 &  0.02 $\pm$  0.09 & $-$0.99 $\pm$  0.96 &  12.39 $\pm$    1.00& $-$2.84$\pm$  1.39 \\
 12 &     B4$-$A2&10.07 $\pm$ 0.05 &  0.32 $\pm$ 0.04 & 0.30 $\pm$ 0.02 & 0.28 $\pm$ 0.07 &  0.02 $\pm$  0.12 & $-$0.99 $\pm$  0.96 &  12.39 $\pm$    1.00& $-$3.31$\pm$  1.39 \\
 13 &     B4$-$A2&10.08 $\pm$ 0.04 &  0.29 $\pm$ 0.03 & 0.26 $\pm$ 0.02 & 0.20 $\pm$ 0.05 &  0.08 $\pm$  0.08 & $-$0.99 $\pm$  0.96 &  12.39 $\pm$    1.00& $-$3.29$\pm$  1.39 \\
 19 &     B4$-$A2& 9.94 $\pm$ 0.04 &  0.42 $\pm$ 0.03 & 0.39 $\pm$ 0.02 & 0.34 $\pm$ 0.05 &  0.07 $\pm$  0.08 & $-$0.99 $\pm$  0.96 &  12.39 $\pm$    1.00& $-$3.44$\pm$  1.39 \\
 20 &     B4$-$A2&11.95 $\pm$ 0.00 &  0.47 $\pm$ 0.00 & 0.35 $\pm$ 0.00 & 0.15 $\pm$ 0.00 &  0.35 $\pm$  0.00 & $-$0.99 $\pm$  0.96 &  12.39 $\pm$    1.00& $-$1.42$\pm$  1.39 \\
 26 &     B4$-$A2&10.42 $\pm$ 0.04 &  0.31 $\pm$ 0.03 & 0.30 $\pm$ 0.02 & 0.27 $\pm$ 0.05 &  0.02 $\pm$  0.08 & $-$0.99 $\pm$  0.96 &  12.39 $\pm$    1.00& $-$2.95$\pm$  1.39 \\
 28 &     B4$-$A2&10.48 $\pm$ 0.04 &  0.30 $\pm$ 0.03 & 0.33 $\pm$ 0.02 & 0.39 $\pm$ 0.06 & $-$0.12 $\pm$  0.10 & $-$0.99 $\pm$  0.96 &  12.39 $\pm$    1.00& $-$2.90$\pm$  1.39 \\
 32 &     B4$-$A2&10.19 $\pm$ 0.05 &  0.37 $\pm$ 0.03 & 0.36 $\pm$ 0.02 & 0.34 $\pm$ 0.07 &  0.02 $\pm$  0.10 & $-$0.99 $\pm$  0.96 &  12.39 $\pm$    1.00& $-$3.18$\pm$  1.39 \\

\hline
\end{tabular}
\begin{list}{}{}
\item[] 
{\bf Notes.} Identification numbers, which are taken from Table  \ref{table.obspectra}, are followed 
by  spectral types,   de-reddened \Ks\ magnitudes,
three estimates of total  extinction (\Aks(JH), \Aks(JK),\Aks(HKs)), $Q1$, \BCKs, DM,
and bolometric magnitudes, \Mbol.~
Errors in \Aks($JH$), \Aks($J$\Ks), and \Aks($H$\Ks) values 
were obtained by propagating the photometric errors of the two considered bands;
missing \Aks\ errors  were filled with 0.2 mag (star \#4). 
Errors in \Mbol\ were derived by propagation of the errors in \Ks\ magnitudes, \Aks, \BCKs, and DM. 
Errors in $Q1$ were derived by propagating the errors in $JH$\Ks; missing errors in \Ks\ (star \#4) 
were filled with 0.80 mag. 
\end{list}
\end{table*}

\begin{table*}\renewcommand{\arraystretch}{0.8}
\caption{\label{table.mbolrsg}  Photometric properties of detected candidate RSGs 
(luminosity L $> 10^4$ \Lsun, masses $> 9$ \Msun).
}
\begin{tabular}{@{\extracolsep{-.09in}}lrrrrrrlllll}
\hline 
 {\rm ID$^a$} & $(J-K_{\rm s})$$_o$ & $(H-K_{\rm s})$$_o$ &\Aks($J$\Ks)$^b$ & \Aks($H$\Ks)$^b$ & $Q1$$^b$ & \Ks$_o$$^b$         & BC$_{\it K _{\rm s}}$  &  DM & \Mbol[mag]$^e$ &  Com. & region\\ 
              & {\rm [mag]} & {\rm [mag]} & {\rm [mag]}      & {\rm [mag]}      & {\rm [mag]}  &    {\rm [mag]}  &     & {\rm [mag]} &  {\rm [mag]} &\\ 
\hline 

 39 & 0.96 & 0.20 & 0.84 $\pm$ 0.02 & 0.87 $\pm$ 0.06 & 0.33 $\pm$ 0.09 & 4.00 $\pm$ 0.03 &  2.64&  13.31 $\pm$ 0.17  & $-$6.67 $\pm$ 0.53  &                 &                \\
 40 & 0.99 & 0.21 & 1.95 $\pm$ 0.02 & 2.04 $\pm$ 0.07 & 0.22 $\pm$ 0.12 & 3.12 $\pm$ 0.03 &  2.70&  13.31 $\pm$ 0.17  & $-$7.49 $\pm$ 0.53  &             RSG$^f$ &          rsgcx1\\
 41 & 0.72 & 0.15 & 1.27 $\pm$ 0.02 & 1.17 $\pm$ 0.07 & 0.47 $\pm$ 0.12 & 5.25 $\pm$ 0.03 &  2.55&  13.31 $\pm$ 0.17  & $-$5.51 $\pm$ 0.53  &                 &          rsgcx1\\
 42 & 0.72 & 0.15 & 1.03 $\pm$ 0.02 & 0.95 $\pm$ 0.06 & 0.44 $\pm$ 0.11 & 5.11 $\pm$ 0.03 &  2.55&  13.31 $\pm$ 0.17  & $-$5.65 $\pm$ 0.53  &                 &          rsgcx1\\
 43 & 0.99 & 0.21 &$..$ & 3.54 $\pm$ 0.06 &$..$ & 4.36 $\pm$ 0.06 &  2.70&  13.31 $\pm$ 0.17  & $-$6.25 $\pm$ 0.53  &             RSG &                \\
 44 & 0.62 & 0.13 & 1.20 $\pm$ 0.02 & 1.13 $\pm$ 0.07 & 0.36 $\pm$ 0.12 & 4.64 $\pm$ 0.03 &  2.50&  13.31 $\pm$ 0.17  & $-$6.17 $\pm$ 0.53  &                 &                \\
 45 & 0.72 & 0.15 & 1.33 $\pm$ 0.02 & 1.33 $\pm$ 0.07 & 0.29 $\pm$ 0.13 & 4.88 $\pm$ 0.03 &  2.55&  13.31 $\pm$ 0.17  & $-$5.88 $\pm$ 0.53  &                 &                \\
 46 & 1.16 & 0.28 & 1.49 $\pm$ 0.02 & 1.53 $\pm$ 0.08 & 0.27 $\pm$ 0.14 & 4.80 $\pm$ 0.03 &  2.84&  13.31 $\pm$ 0.17  & $-$5.67 $\pm$ 0.53  &             RSG &        glimpse9\\
 47 & 1.06 & 0.25 & 1.26 $\pm$ 0.02 & 1.31 $\pm$ 0.06 & 0.24 $\pm$ 0.11 & 4.93 $\pm$ 0.03 &  2.80&  13.31 $\pm$ 0.17  & $-$5.58 $\pm$ 0.53  &             RSG &            reg4\\
 48 & 0.96 & 0.20 & 1.08 $\pm$ 0.02 & 1.10 $\pm$ 0.06 & 0.35 $\pm$ 0.09 & 4.98 $\pm$ 0.03 &  2.64&  13.31 $\pm$ 0.17  & $-$5.69 $\pm$ 0.53  &                 &            reg4\\
\hline
 $[$MFD2010$]$ 5& 1.03 & 0.23 & 1.79 $\pm$ 0.02 & 1.71 $\pm$ 0.06 & 0.50 $\pm$ 0.10 & 5.26 $\pm$ 0.04 &  2.76&  13.31 $\pm$ 0.17  & $-$5.29 $\pm$ 0.53  &             RSG &        glimpse9\\
 BD$-$08 4635 & 1.06 & 0.25 & 0.36 $\pm$ 0.19 & 0.22 $\pm$ 0.51 & 0.63 $\pm$ 0.81 & 2.69 $\pm$ 0.32 &  2.80&  12.39 $\pm$ 0.50  & $-$6.90 $\pm$ 0.78  &                 &            reg2\\
 BD$-$08 4639 & 0.62 & 0.13 & 0.36 $\pm$ 0.17 & 0.24 $\pm$ 0.42 & 0.46 $\pm$ 0.68 & 2.41 $\pm$ 0.27 &  2.50&  12.39 $\pm$ 0.50  & $-$7.48 $\pm$ 0.76  &                 &                \\
 BD$-$08 4645 & 1.06 & 0.25 & 0.31 $\pm$ 0.15 & 0.28 $\pm$ 0.39 & 0.40 $\pm$ 0.65 & 1.99 $\pm$ 0.24 &  2.80&  12.39 $\pm$ 0.50  & $-$7.60 $\pm$ 0.75  &             RSG &                \\

\end{tabular}
\begin{list}{}{}
\item[] {\bf Notes.}
Identification numbers, are followed by  intrinsic $(J-K_{\rm s})$$_o$ , $(H-K_{\rm s})$$_o$ colors, \Aks($J$\Ks) and 
\Aks($H$\Ks),    $Q1$ , de-reddened \Ks, \Ks$_o$, and absolute bolometric magnitudes. ~
($^a$)  Identification numbers are taken from Table \ref{table.crsgspectra}.~
($^b$)  Errors in \Aks, \Ks$_o$, $Q1$, \Mbol\  are calculated by propagation of the photometric errors.~
($^e$) \Mbol\ is obtained with the \BCKs\ of \citet{levesque05}.~
($^f$) The RSG comment denotes known RSGs  \citep[ MFD2010 5 and BD$-$08 4645, e.g.~][]{sylvester98,messineo10},
and stars \#40, \#43, \#46, and \#47, which have EW(CO)s larger than that of a M7III, typical of M0-2I.~
\end{list}
\end{table*}

The distance modulus $DM$ is by definition equal to $$ DM= K_{\mathrm s} - A_{\it K_{\mathrm s}} - M_{\it K},$$ 
where \Aks\ is the extinction  and $M_{\it K}$ is the absolute magnitude
in   $K$-band; \Aks\ and $M_{\it K}$ are function of  spectral types and luminosity 
classes. Early-type stars with known spectral-types yield 
spectro-photometric distances, when assumptions on luminosity classes can be made, 
and erratic behaviors  are not present (e.g. LBVs).
Compilations of absolute $K$ magnitudes and intrinsic colors for O and B types are 
available from \citet{johnson66},  \citet{koornneef83}, \citet{humphreys84}, \citet{wegner94}, 
\citet{lejeune01},    \citet{crowther06},  and \citet{martins06}.
In the near-infrared, spectral classification can be achieved to within a few classes
\citep{hanson96}, and  a range of $M_{\it K}$  must be assumed;  for a O4-6 star, 
for example,  we assumed the average $M_{\it K}$  of those of 
O4 and  O6 stars.  For each  star, $M_{\it K}$ and DM  were estimated for the dwarf, giant, 
and supergiant classes, as summarized in Tables \ref{table.distance} and \ref{table.distsum}. 
We assumed that stars at  similar interstellar extinction were likely to be  at similar distances;
we calculated the DMs of a few detected spectroscopic supergiants; we  assigned
luminosities classes to  fainter stars by comparing
their \Aks\ and \Ks\ to  those of supergiants  of the same spectral type.

O-type stars have $ 1.3\la$ \Aks$\la 1.9$ mag. 
 Stars  \#14, \#16, and \#25  are O-type supergiants, as suggested by  
their emission lines \citep{hanson96,hanson05} --
a strong  \ion{He}{I}  line at 2.058 \um\ appear in emission in the spectrum of star \#16; 
a broad line emission at 2.115 \um\ (typical of f-type and WR stars) and 
 strong \ion{C}{IV} emission lines 
are  seen in star \#25; star \#14 has strong  \ion{He}{I} lines in absorption, but \brg\ in emission. 
There are four O-type stars with   \Ks\ magnitudes  brighter than 
those of stars \#14, \#16, and \#25;  for those we assumed a supergiant class.

Absolute magnitudes of O-types are, however, quite uncertain. The Arches cluster is rich in 
O4--6 stars, and is located at the distance of the Galactic center  \citep{martins08}.
We recalculated an average value of $M_{\it K}=-4.94\pm0.47$ mag for all O4--6I stars in 
the Arches  listed by \citet[][]{martins08}  (hyper-giants F10 and F15 were included),
and of  $-5.16\pm0.13$ mag for those  O4--6I with \ion{He}{I} line at 2.112 \um\ 
in absorption; we used 8.4 kpc, the photometry from \citet{figer02} 
and the extinction law by  \citet[][]{messineo05}.
The O7--9I stars in W33 yield  $M_{\it K}=-5.39\pm0.33$ mag.
All, but one,  Of$_K$+ stars in GMC G23.3$-$0.3 have the \ion{He}{I} line at 2.112 \um\  
in absorption, and mostly weak carbon lines (O6-7). This  empirical comparison implies DM 
from $13.3\pm0.4$ mag to $13.5\pm0.4$ mag for  the newly detected OIf$_K$+ stars. Beside the 
OIf$_K$+ stars, we detected only another OI star (O9-9.5, \#16),  which yields a 
distance of $4.3^{+1.5}_{-1.1}$ kpc (DM=$13.18\pm0.66$ mag) by assuming   $M_{\it K}=-5.68$ mag.

The two  B-type supergiants (\Aks=1.5 mag) yield a spectrophotometric distance of 
4.9 kpc (DM=13.47 mag), when assuming a O9.5-B5 type ($M_{\it K}=-6.49$ mag ), 
or of 4.5 kpc (DM=13.25 mag), when assuming the more frequently observed O9.5-B3  type
($M_{\it K}=-6.27$ mag). 
The results from each group and luminosity class are summarized in Table \ref{table.distsum}.

The derived distance moduli indicate that the OI and BI are  consistent with a unique distance.
By averaging the distance modulus obtained for  BI stars and that for OI stars 
with M$_{\rm K}$ from \citet{martins06},  we obtained DM=13.48$\pm$0.32 mag;
by  using the empirical calibration on the Arches, we obtained DM=$13.35\pm0.14$ mag.
For the remaining O-types, since distances increase with decreasing \Ks$_o$,
a mix of luminosity classes (giants and dwarfs) is inferred by assuming similar distances.

Previous studies of \ion{H}{II} regions  or SNRs (e.g. W41) of this molecular complex
report gasous kinematic distances  from 4 to 5 kpc 
\citep[e.g.~][]{albert06,leahy08}.
Gas measurements in the direction of the GMC are found to peak at a velocity  
from $70.5$ to $82.5$ \kms\ \citep[][and reference therein]{messineo10};
using these  velocities  and the Galactic curve (R$_0=8.4$ kpc and $\Theta_0=254$ \kms) from
\citet{reid09}, we obtained a kinematic distance from   4.35 kpc to 4.78 kpc (DM from 13.19 mag to 13.39 mag);
by using the historical curve of  \citet{brand93} (R$_0=8.5$ kpc and $\Theta_0=220$ \kms),
we obtained a kinematic distance from 4.6 kpc to 5.1 kpc (DM from 13.31 mag to  13.53 mag).
\citet{brunthaler09} provides a parallactic distance of  $4.59^{+0.38}_{-0.33}$ kpc (DM=13.31$\pm$0.17 mag)
for G23.01$-$0.41.

The inferred spectrophotometric distances of O- and B-type supergiants   are within the errors
consistent with those of the GMC G23.3$-$0.3;  these stars are most likely 
associated with the GMC.
Fainter O stars are likely to be giant stars of the same GMC, as supported by their \Aks\ values.

The  spectrophotometric distances agree well within errors with the kinematic distance
of the cloud  and parallactic distance.
In the following,  the photometric properties of stars associated with the GMC
are analyzed by assuming   the parallactic distance by \citet{brunthaler09}.

For the foreground  B4-A2 stars,  we derived  a distance modulus of DM=$10.87\pm0.64$ mag 
by assuming a dwarf class.

\subsection{Luminosities}
\label{seclum}

 Bolometric magnitudes, \Mbol, were derived
using \Ks\ magnitudes, \Aks\  (see Section \ref{section.ext}), 
bolometric corrections, \BCKs, effective temperatures, and distance moduli:

$M_{\mathrm bol} = K_{\mathrm s} - A_{K_{\mathrm s}} + BC_{K_{\mathrm s}} +DM.$

For early-type stars, assumed effective temperatures and \BCKs\
are listed in Tables   \ref{table.obspectra} and  \ref{table.mbolearly} 
\citep[see also Appendix A in ][and references therein]{messineo11};
for late type stars, \BCKs\ and \Teff\ were available from the work of \citet{levesque05}.
Luminosity properties are discussed only for stars with \Aks $> 0.8 $ mag, for which 
a DM of $13.31\pm0.17$ mag is assumed.

The  luminosities of early-types with  emission lines 
(\#14, \#16, \#22, and \#25) range
from  $1.0 \times 10^5$ \Lsun\  to  $4.6 \times 10^5$  \Lsun, 
and are  consistent with those of blue supergiants.
Eleven out of 21  stars with \Aks $> 0.8 $ mag  are  blue supergiants 
(including [MFD2010]3 and [MFD2010]4 and [MVM2011] 39), 
 ten others  are most likely giants;
for the Magellanic clouds, \citet{humphreys84} estimated  7 blue giants
for every 10 blue supergiants in both associations and fields.

We selected as candidate RSGs those observed late-type stars with \Aks $> 0.8$ mag and
with luminosities larger than $>10^4$ \Lsun\ for a distance of 4.6 kpc (stars from \#39 to  \#48
in Table \ref{table.mbolrsg}); 
star (\#46) is  a known RSG \citep{messineo10});
contaminating AGB stars were identified by their strong water absorption, as described in Appendix \ref{agbsel}.
The two RSGs  in the cluster GLIMPSE9  ( $[$MFD2010$]$5 and \#46/$[$MFD2010$]$8  ) 
have an average \Aks=1.6 mag,  an average \Mbol = 5.48 mag (4.6 kpc)
and  M1.5-3 types \citep{messineo10}.
The new RSGs, \#40 and \#47, have  types  M0I and M2I,  \Ak=  2.0 and 1.3 mag, and \Mbol\ 
  $-7.49$ mag and $=-5.58$, respectively; they are consistent with the distance of GLIMPSE9 and the GMC.
The RSG \#43  is a luminous and  distant object with  \Aks= 3.5 mag,
negligible water absorption, and  a large EW(CO).
For completeness, Table \ref{table.mbolrsg} comprises also stars \#39, \#41, \#42, \#44, \#45,  and \#48,  
which, however, have  a slightly lower \Aks\ (1.1 mag) and earlier spectral types (K2-K5).  
Studies of stellar velocities may provide evidence for a cluster of stars.

\subsection{Spatial distribution of massive stars}
\label{sec.spatial}

In Figure \ref{largemap.fig}, the positions  of early-type stars and  candidate RSG stars 
are plotted on a grey scale image of the GMC complex at 3.6 \um\ by GLIMPSE. 
In the following sections,  the properties of the detected massive stars
across several  regions of the cloud  (see Table \ref{regions}) are described.

\subsection{ GLIMPSE9Large }

\begin{figure*}
\begin{center}
\resizebox{0.8\hsize}{!}{\includegraphics[angle=0]{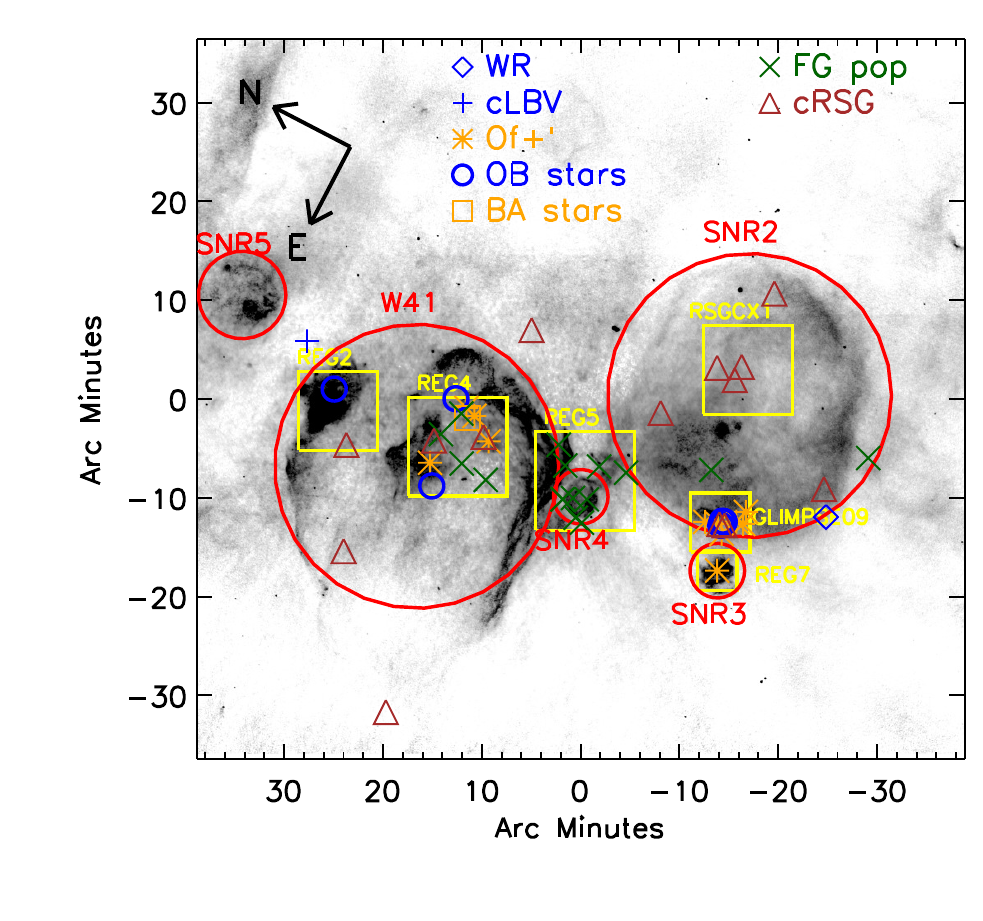}}
\end{center}
\caption{\label{largemap.fig} 
Positions of detected stars  are marked on 
a gray scale, which  is a MAGPIS image   at 20 cm of the whole G$23.3-0.3$ complex \citep{white05,helfand06}.
Positions of Of$_K^{+}$ stars are marked with asterisks,  the cLBV with a plus sign, 
the WR found by \citet{mauerhan11}  with a diamond, late-O and early-B types with circles,  
late-B and early-A  stars with  
squares, and RSGs and cRSGs with triangles. 
Possible foreground early-types ( \Aks$ < 0.8$ mag) are marked with crosses.
Locations  and sizes of supernova remnants (SNRs) are marked by circles.
Squares and labels display the regions  selected  on the GLIMPSE 3.6 \um\ by \citet{messineo10}
with increased nebular emission (\HH\ regions) and apparent
overdensities of bright stars.
} 
\end{figure*}

\begin{figure*}
\centering{
\resizebox{0.49\hsize}{!}{\includegraphics[angle=0]{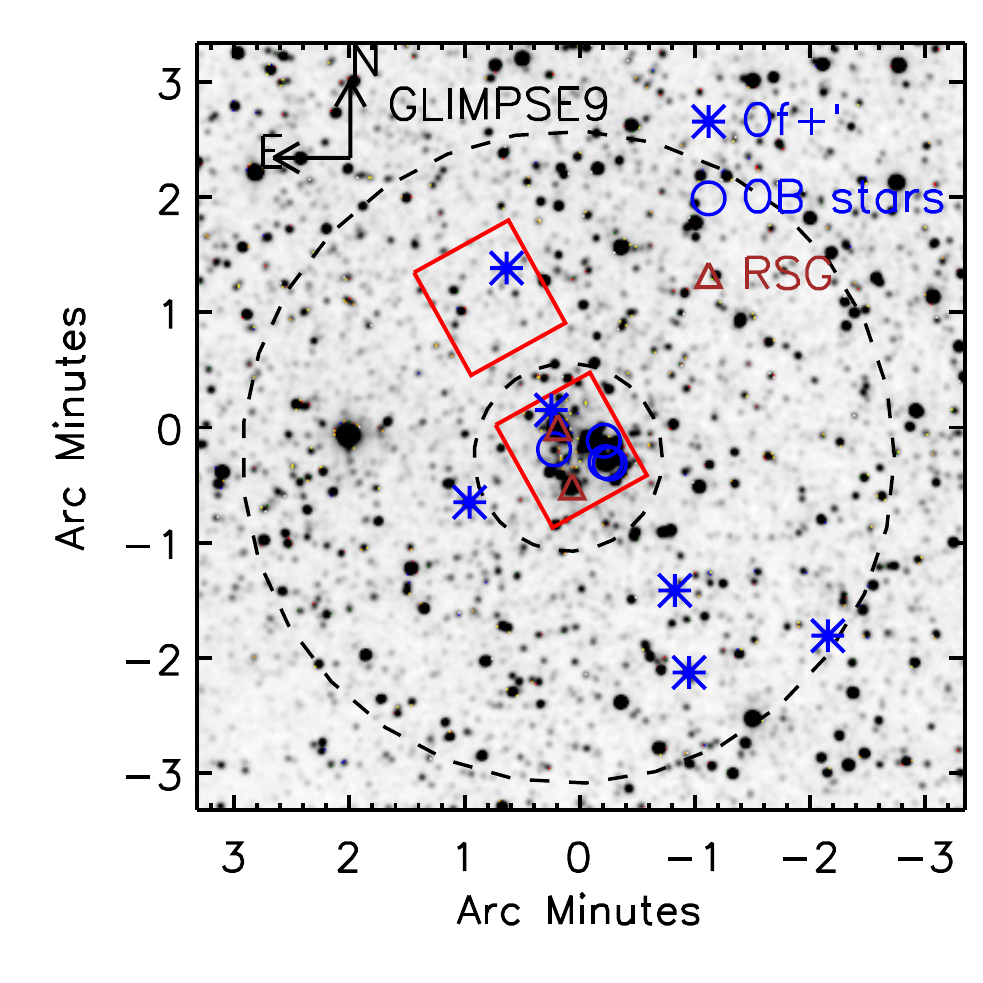}}
\resizebox{0.49\hsize}{!}{\includegraphics[angle=0]{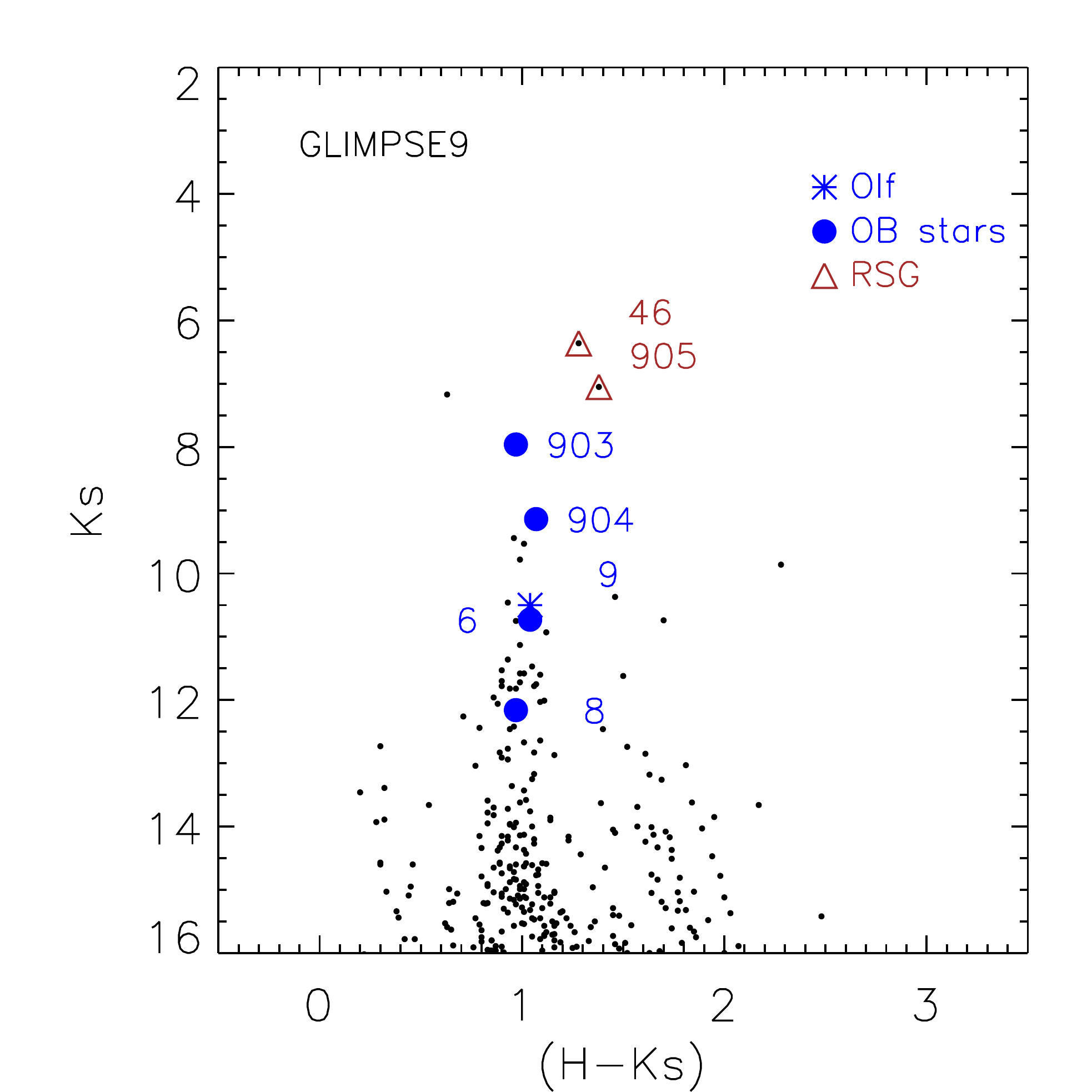}}
}
\centering{
\resizebox{0.49\hsize}{!}{\includegraphics[angle=0]{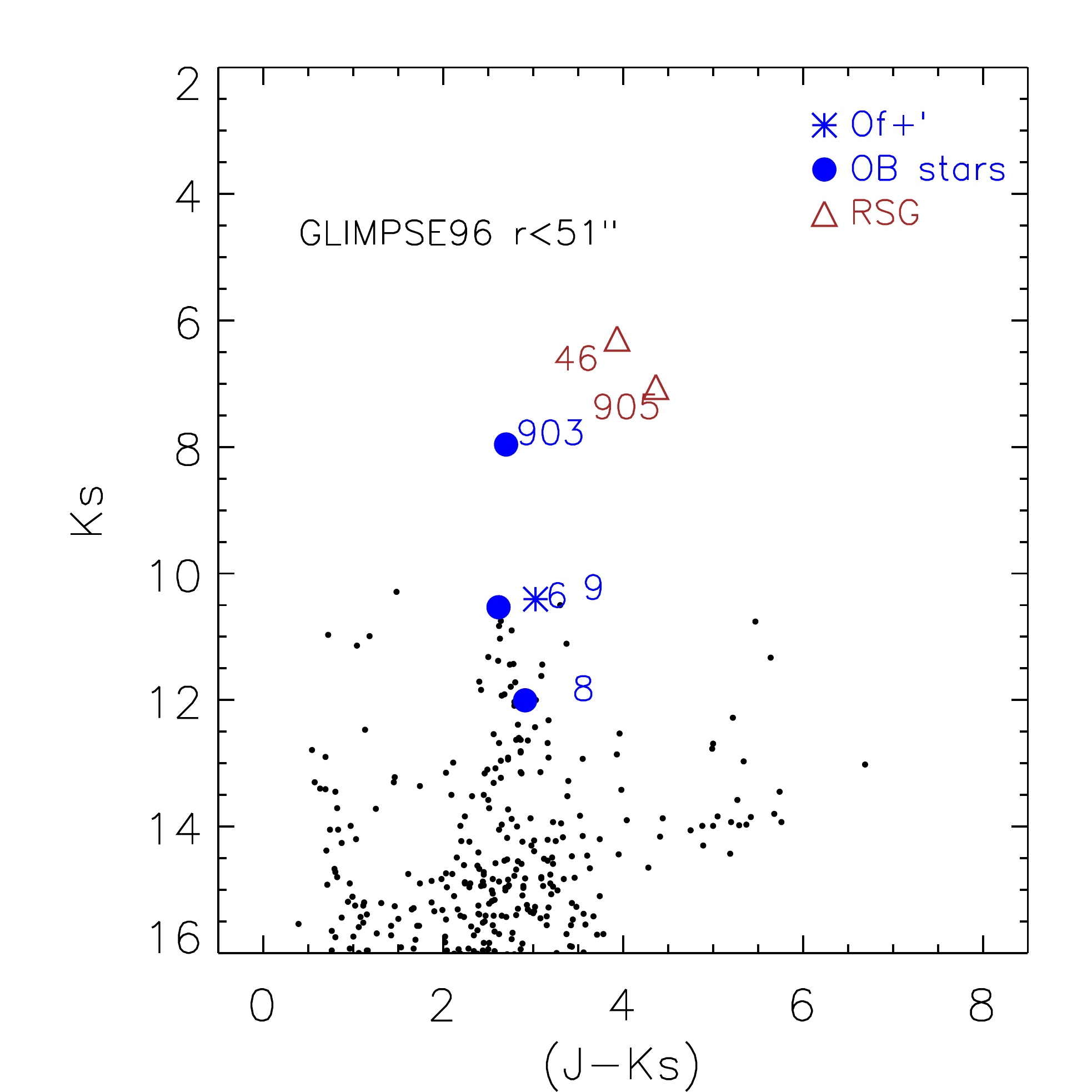}}
\resizebox{0.49\hsize}{!}{\includegraphics[angle=0]{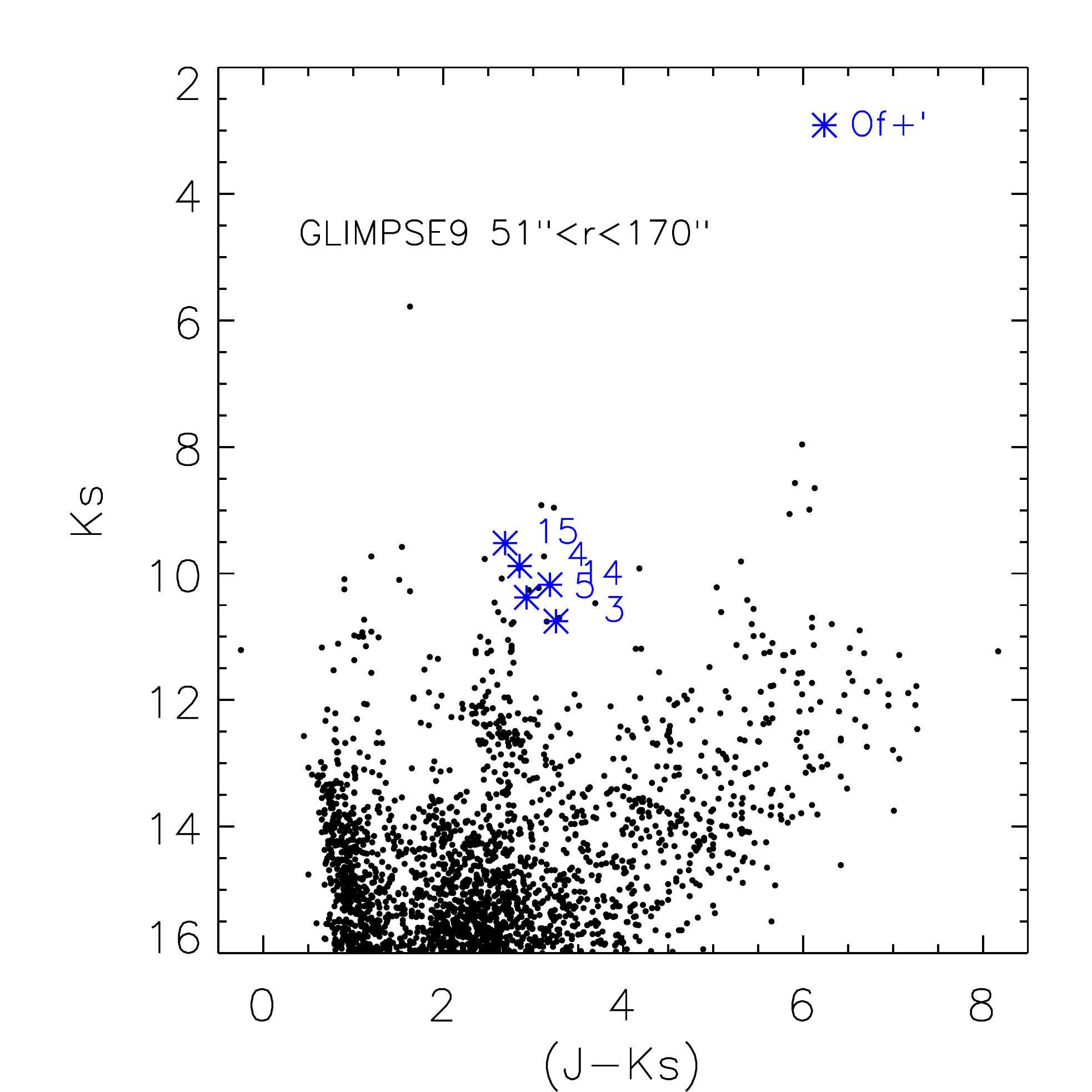}}
}
\caption{\label{cmd.reg}
{\it Left upper panel:}  2MASS \Ks-band image of the observed   region GLIMPSE9Large 
in Table \ref{regions}. The enclosed NICMOS ($51\rlap{.}^{\prime\prime} 5\times 51$\arcsec) 
fields studied by \citet{messineo10}  are shown as  squares; the central NICMOS field covers 
the stellar cluster GLIMPSE9. Two dashed circles indicate the circle and annular regions 
used for the  CMDs. {\it Upper  right panel:}  HST/NICMOS   $H-$\Ks\ vs \Ks\ diagram 
of the  GLIMPSE9 cluster  \citep{messineo10}. {\it Left lower  panel:} UKIDSS-2MASS \Ks\ 
versus $J-$\Ks\ diagram of  the GLIMPSE9 cluster (2MASS data are used above \Ks$\approx 10.5$ mag).
{\it Right lower panel:} UKIDSS-2MASS \Ks\ versus $J-$\Ks\ diagram of  a region
surrounding the GLIMPSE9 cluster. Spectroscopically observed stars are marked
as summarized in the legend.
Of$_K^{+}$ stars are marked with asterisks,  late-O and early-B types 
with filled circles,  RSGs and cRSGs with triangles.   Labels  903, 904, 905 indicate
massive stars  [MFD2010]3, [MFD2010]4, and [MFD2010]5 from \citet{messineo10}.}
\end{figure*}

The surveyed  region  GLIMPSE9Large has a diameter 7 times larger than 
the NICMOS field studied by
\citet{messineo10}, as shown  in Figs.\ \ref{largemap.fig}, and \ref{cmd.reg}.
Only one   Of$_K$+ star lies in the NICMOS field.  A surprisingly large number
of massive Of$_K$+ stars (\#3, \#4, \#5, \#9, \#14, and \#15) are found 
to surround  the GLIMPSE9 cluster.
UKIDSS/2MASS  \Ks\ versus $J-$\Ks\ diagrams of this region are shown in Fig.\ \ref{cmd.reg}.
Most of the bright  stars in the populous diagram of the lower right panel are late-type stars;
indeed, a  sequence made of clump stars 
is recognizable, which runs  from $J-$\Ks$\approx 1$ mag and \Ks=11 mag  to  
$J-$\Ks$\approx 5$ mag and \Ks=14.5 mag; 
there is a  tail of obscured  giants stars  ($J-$\Ks $ > 4$ mag), and
a  blue  main sequence appears at  $J-$\Ks$\approx 1$ mag and \Ks= 12-16 mag. 
Detected massive Of$_K$+ stars have colors similar to those of the GLIMPSE9 cluster, 
$J-$\Ks$\approx 3$ mag, and \Ks\ from 9.52 to 10.75 mag.

The central concentration, i.e. the stellar cluster  GLIMPSE9, hosts
two RSGs and two B0-3 supergiants \citep[][]{messineo10}.
The RSG  members ([MFD2010]5 and \#46/[MFD2010]8) have \Aks\ from 1.49 to 1.79 mag,  
and \Mbol\  from $-$5.67  to $-$5.29 mag (for 4.6 kpc), respectively.

The  Of$_K$+ stars  are not concentrated, but sparse on a
6\arcmin\ radius area (8.0 pc at 4.6 kpc).
Their \Aks\ range from 1.59 mag to 1.90 mag.
The  infrared magnitudes of the Of$_K$+ stars are consistent with 
a distance of 4.6 kpc, and with their  association with  GMC G23.3$-$0.3;
their \Mbol\ range from $-$9.55 to $-$8.70 mag;
stars \#4, \#14,  and \#15 (Of$_K$+)    are   supergiants.

\subsection{ REG4 }
\label{reg4}
 An overdensity of bright stars  on a nebular background, which extends for
about 6\arcmin, was visually detected in REG4 (Figs.\  \ref{largemap0.fig}, \ref{largemap.fig}) 
by \citet{messineo10}. Four  Of$_K$+ stars, 2 B-type stars, 
1 RSG, and 1 cRSG  were detected in region REG4. 
The minimum circle enclosing the four Of$_K$+ has a diameter of 7\arcmin.

\begin{figure*}
\centering{
\resizebox{0.4\hsize}{!}{\includegraphics[angle=0]{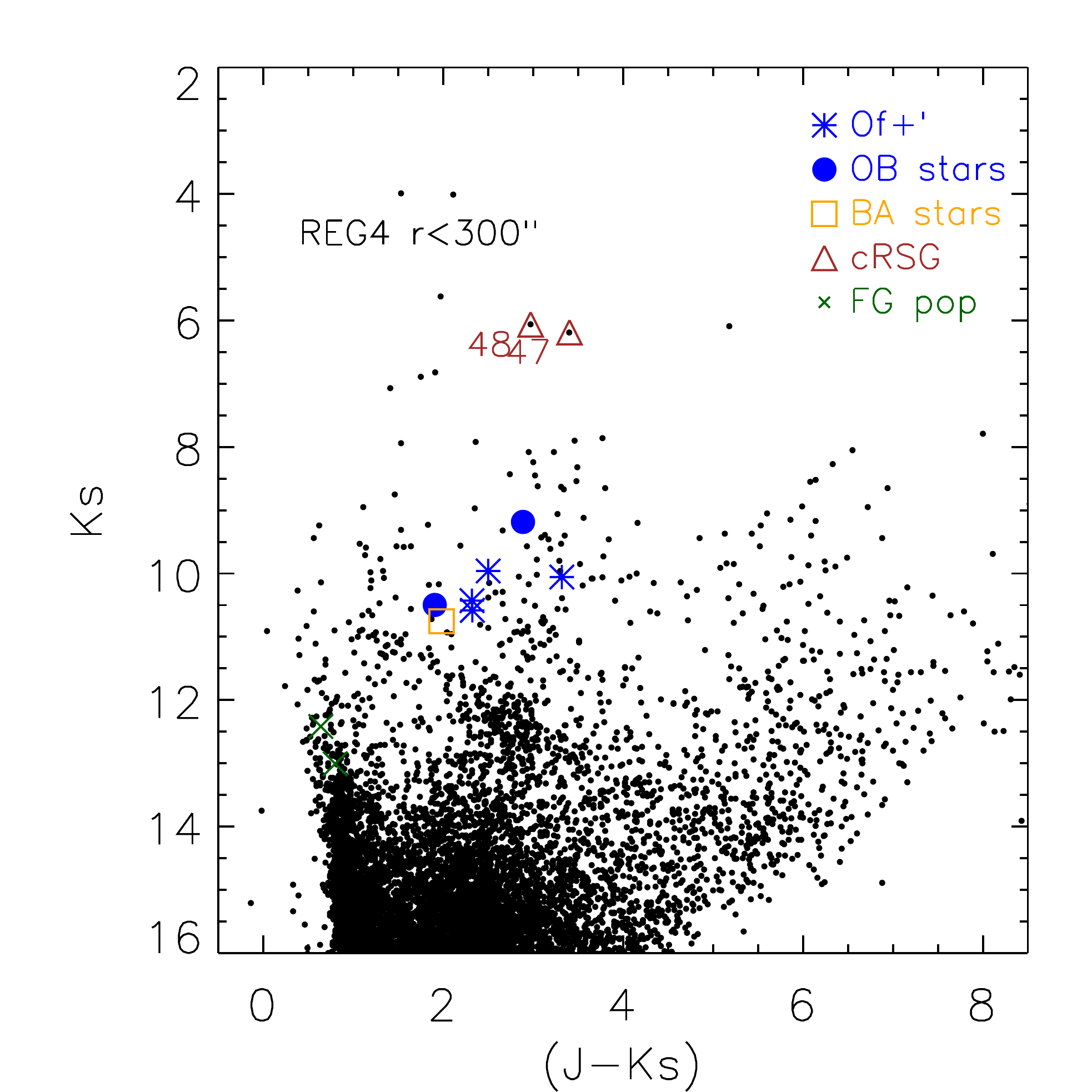}}
\resizebox{0.4\hsize}{!}{\includegraphics[angle=0]{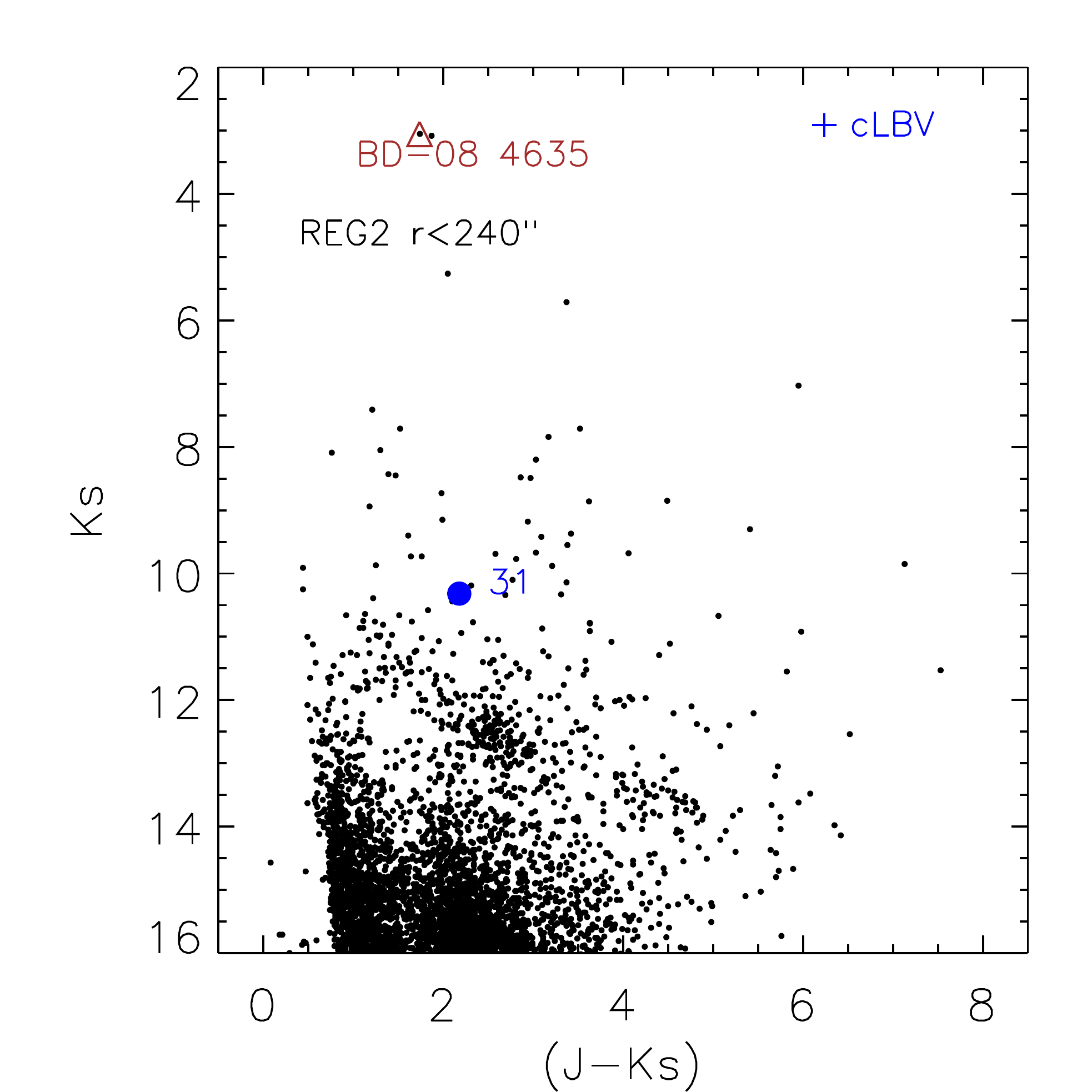}}
\resizebox{0.4\hsize}{!}{\includegraphics[angle=0]{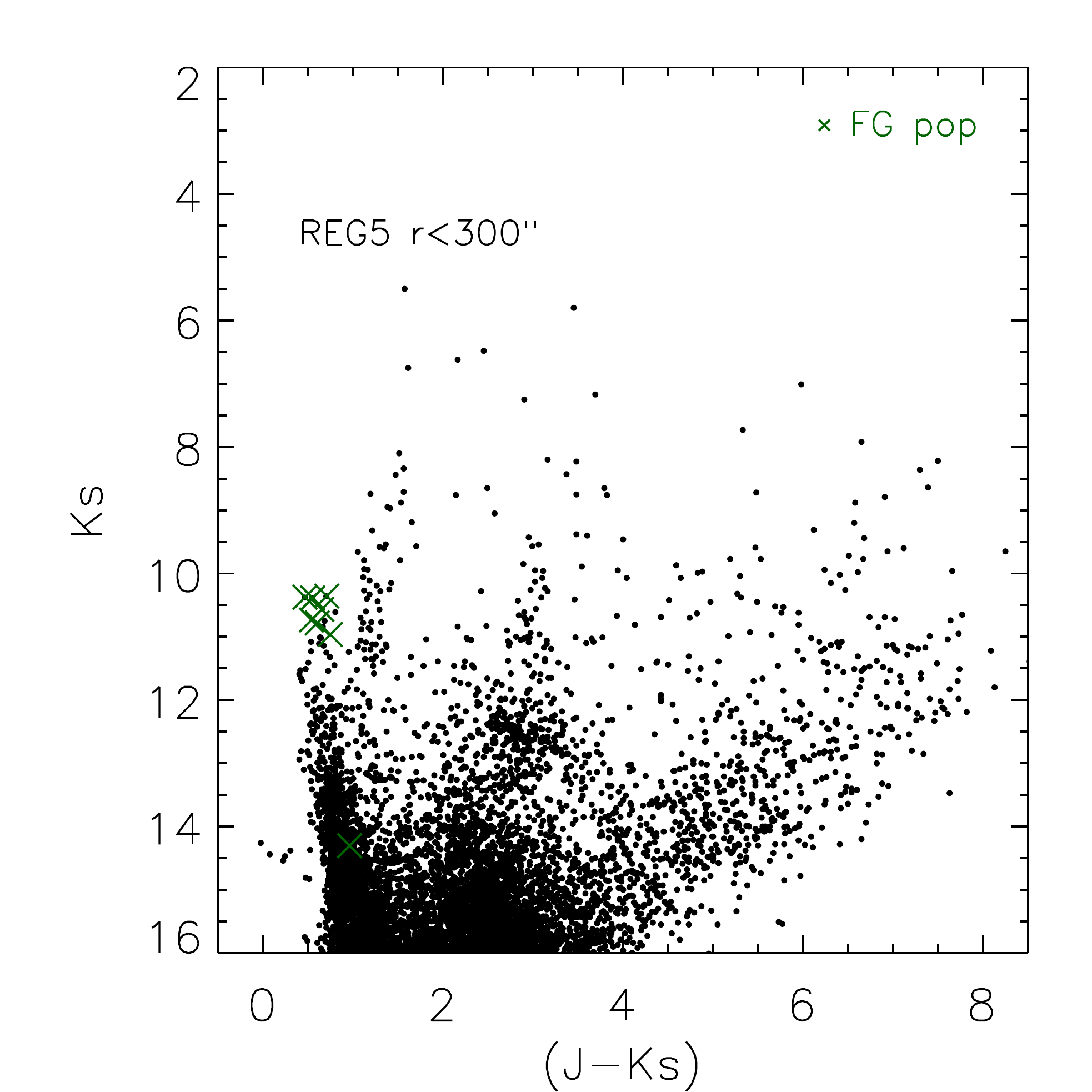}}
}
\centering{
\resizebox{0.4\hsize}{!}{\includegraphics[angle=0]{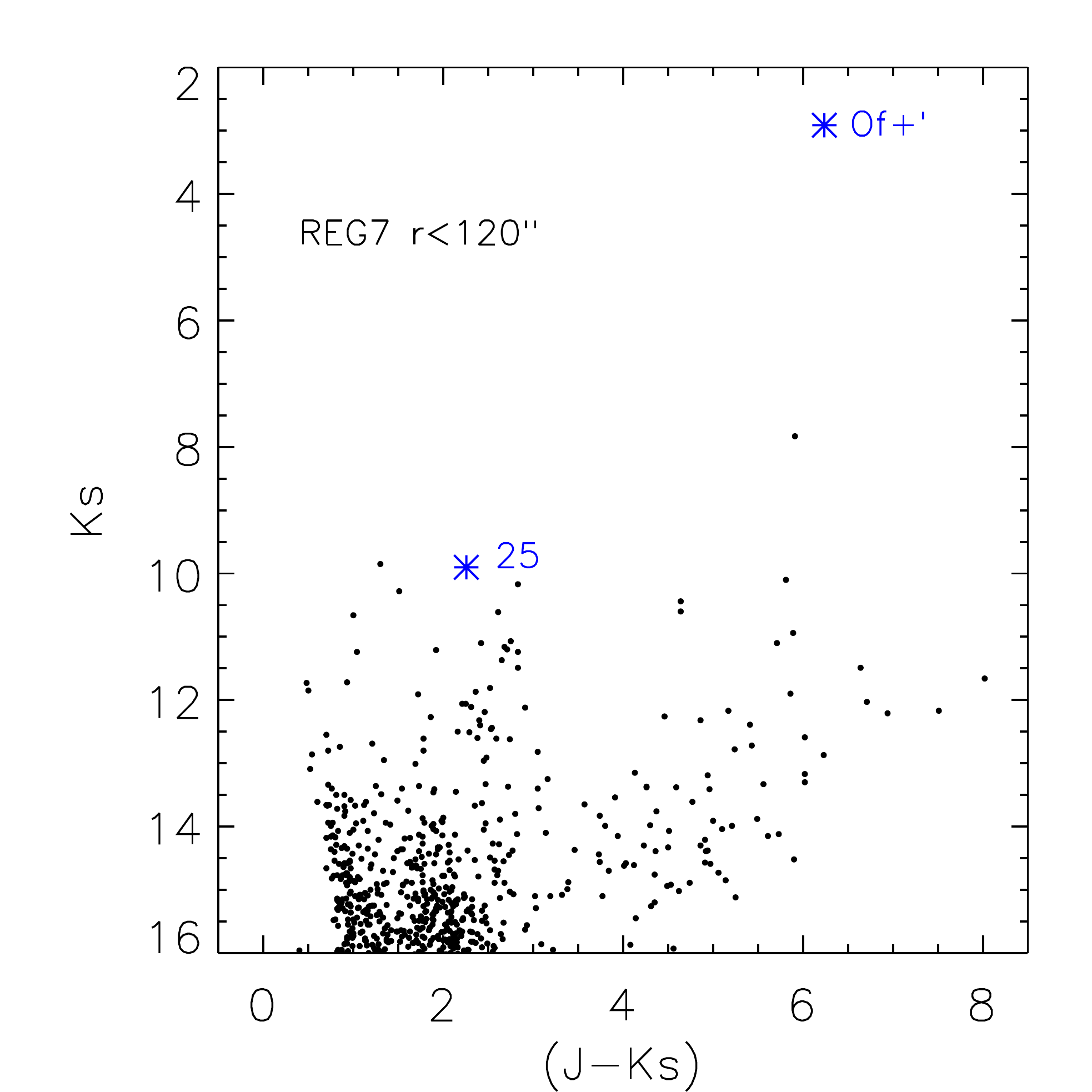}}
\resizebox{0.4\hsize}{!}{\includegraphics[angle=0]{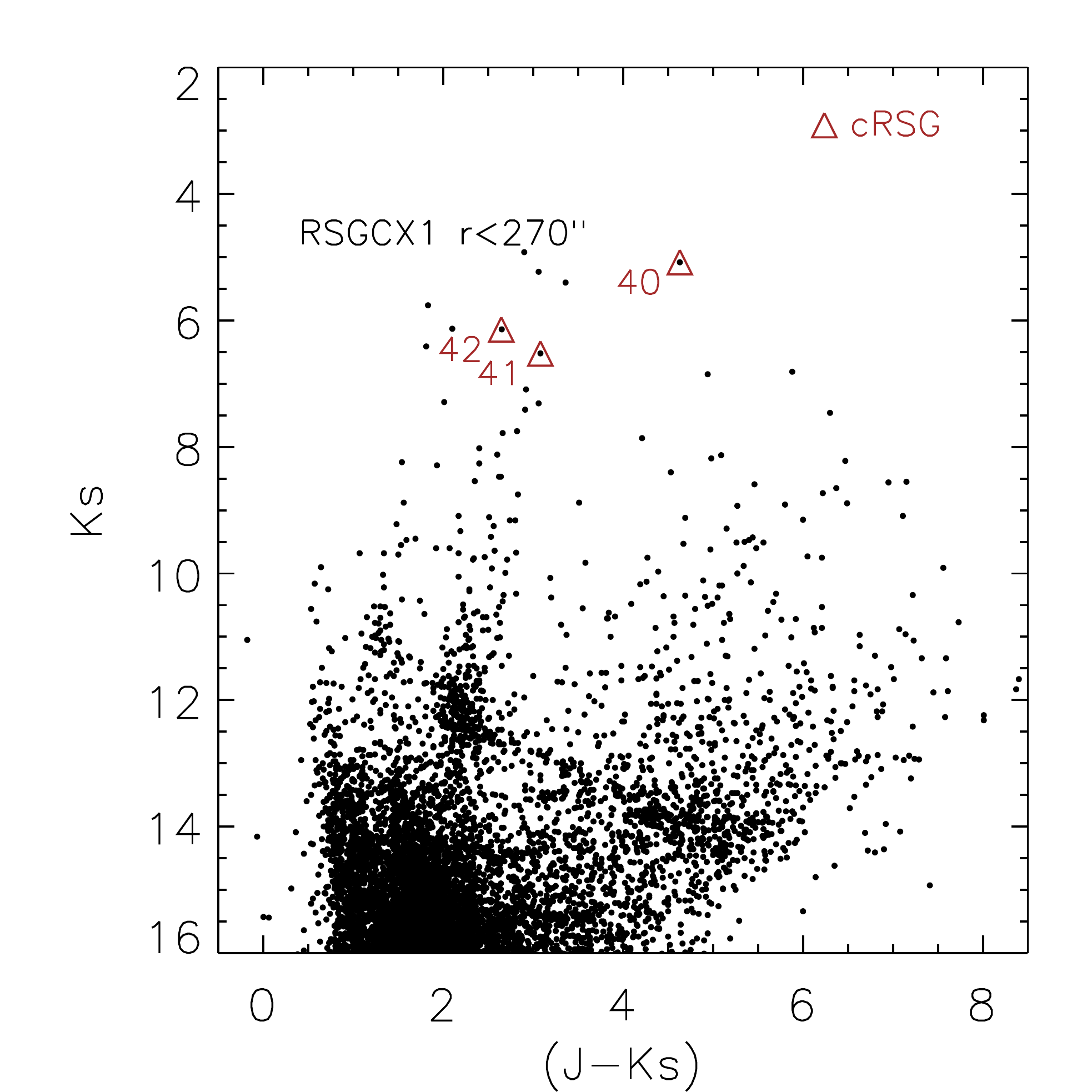}}
\resizebox{0.4\hsize}{!}{\includegraphics[angle=0]{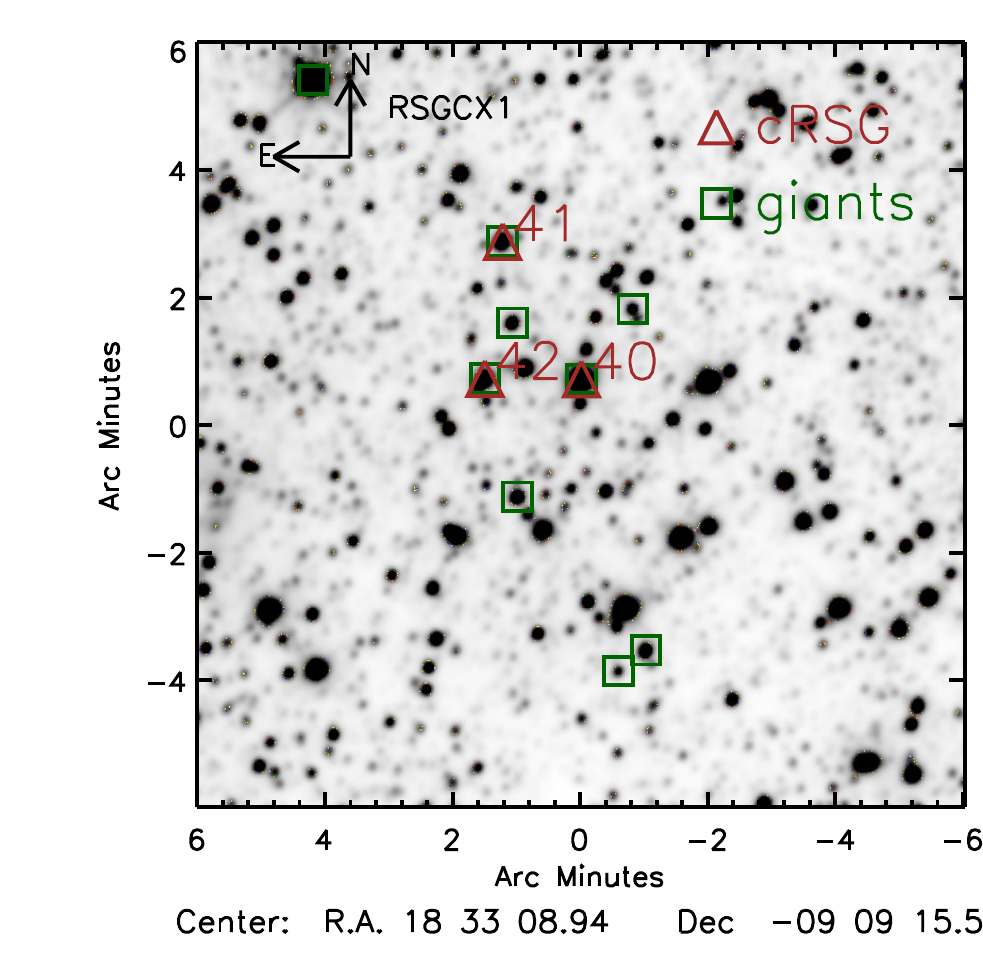}}
}
\caption{ \label{cmdall} UKIDSS-2MASS \Ks\ versus $J-$\Ks\ diagrams 
of regions REG2 ({\it top-right}), REG4 ({\it top-left}), REG5 
({\it middle-left}), REG7 ({\it middle-right}), 
RSGCX1 ({\it bottom-left})  from Table \ref{regions}.
2MASS data are used for \Ks\ brighter than 10.5 mag. 
Spectroscopically observed stars are labeled  as summarized in the legend.
Of$_K^{+}$ stars are marked with asterisks,  late-O and early-B types 
with circles, late-B and early-A stars with  squares. Possible foreground early-types 
( \Aks$ < 0.8$ mag) are marked with crosses.  Triangles  indicate
RSGs/cRSGs.
The {\it bottom-right panel} shows a map  (WISE 3.4 \um) of  RSGCX1,  where squares
 indicate the observed giants,  and triangles  the cRSGs.} 
\end{figure*}

The CMD of region REG4   shows  (see Fig.\ \ref{cmdall}) 
a blue sequence ($J-$\Ks $\approx 0.8$ mag, \Ks
$> 12 $ mag), where we detected a few stars (\#20, \#35, and  \#27);   
a red clump sequence crosses the diagram from $J-$\Ks  $\approx 1.5$ mag, 
\Ks  $\approx 11$ mag to $J-$\Ks  $\approx 3.5$, \Ks $\approx 13$ mag
\citep[e.g.][]{messineo05}.
Detected massive stars have  $J-$\Ks\ color from 2 to 4 mag. 
Their photometric properties  are similar to those seen in region GLIMPSE9Large. 
 The Of$_K$+ types (\#36, \#18,
\#23, and \#17) have \Aks\ from 1.4 to 1.9 mag, and  \Mbol\  from $-7.9$ to $-9.4$ mag. 
Star \#16  is a  blue supergiant (\Ks=9.19 mag).
Stars  \#47 (RSG) and \#48 (cRSG)  are located 3 magnitudes  above the blue supergiants;
they have  extinction  \Aks\ = 1.26 and 1.08  mag, M2 and K5 types, and \Mbol= $-5.58$ and $-5.69$ mag, 
respectively.

\subsection{REG2 and the new candidate LBV}
Region REG2  contains   an \HH\ region (Figs.\  \ref{largemap0.fig}, \ref{largemap.fig}), 
as inferred from the mid-infrared emission and coincident radio continuum emission.  
Star \#31 was detected on  the Western edge of this \HH\ region.
 The CMD of REG2 presents  structures similar to  those  in  REG4
(see \ref{reg4}). The color and magnitude of star \#31 overlap those of the massive 
early-types  in REG4, with \Aks= 1.27 mag and \Ks=10.32 mag.

The cLBV \#22   does  not appear to be part of this \HH\ region, it lies about 
$5\rlap{.}^{\prime}5$
away from star \#31, and is not part of any visible cluster of stars.  
Star \#22  has  \Aks=1.13 mag and \Ks=7.63 mag 
(see Tables \ref{table.obspectra} and \ref{table.mbolearly}).
We assumed a spectral range from B3I to B8I, 
which corresponds to an average effective temperature  of $13200\pm2300$ K. We used an average
\BCKs\ of $-$1.09 mag, and a distance of 4.6 kpc;  we derived  \Mbol $=-7.90$ mag,
 $M_{\it V}= M_{\it K}+V-K=-6.93$ mag, and L $= 1.1 \times 10^5$ \Lsun; intrinsic $V-K$ color is from \citet{koornneef83} 
and \citet{martins06}.  The star would be the faintest known cLBV
\citep[e.g.][]{clark09,messineo12}, but within error consistent 
with the minimum predicted luminosity of L $= 1.6 \times 10^5$  \citep{groh13}.
By assuming a higher temperature \citep[24500 K, similar to that of the peculiar WRA751][]{garcia98},
we would derive an average \BCKs\ of $-2.97$ mag,    
\Mbol  $=-9.84\pm0.64$ mag,  L $= 6.8 \times 10^5$ \Lsun.

\subsection{ REG7, REG5, and RSGCX1}

Region REG7 coincides with  nebular emission (Figs.\  \ref{largemap0.fig}, \ref{largemap.fig}),
without a clear stellar concentration. It also coincides with the candidate cluster [BDS2003]117  \citep{bica03}.
We observed  star \#25, which lies at the center of the nebula,
and identified it as an O4If$_K$+ star with \Aks = $1.34$ mag, and \Mbol = $-9.14$ mag (for 4.6 kpc).

In region REG5, we detected  early-type stars 
(\#7, \#12, \#13, \#19,   \#26, \#28, and \#32 ) from  a blue sequence,
with an average \Aks= $0.35 \pm0.06$ mag, as 
shown in  Fig.\ \ref{cmdall}.  
Their \Ks\ range from 10.36 to 10.96 mag.
They are foreground to the stellar population of the GMC 
(for example, the GLIMPSE9 cluster has an \Aks\ of $1.6\pm0.2$ mag).

Star \#40 (M0I) has a broad EW(CO), and $Q1$=0.22 mag,  which is a typical value for  red supergiants
\citep{clark09,messineo12}. 
It is located, along with stars \#41  and \#42, in direction of the  center of SNR $22.7-0.2$, 
in region RSGCX1 (see Figs.\ \ref{largemap.fig}  and \ref{cmdall}). 
The three stars (\#40,\#41, and \#42) 
have \Aks\ of 1.95, 1.27, and 1.03 mag, which imply  distances larger than 4 kpc
\citep{clark09,drimmel03}.
By assuming that they are at the distance of 4.6 kpc, 
we  derived \Mbol$=-7.49$, $-5.51$, and $-5.65$ mag, respectively, and
their likely association  with the SNR.
The presence of 3 cRSGs implies also the presence of a candidate massive cluster of stars
\citep[$> 10000$ M$_\odot$,][]{clark09}. 

\begin{figure}
\resizebox{1\hsize}{!}{\includegraphics[angle=0]{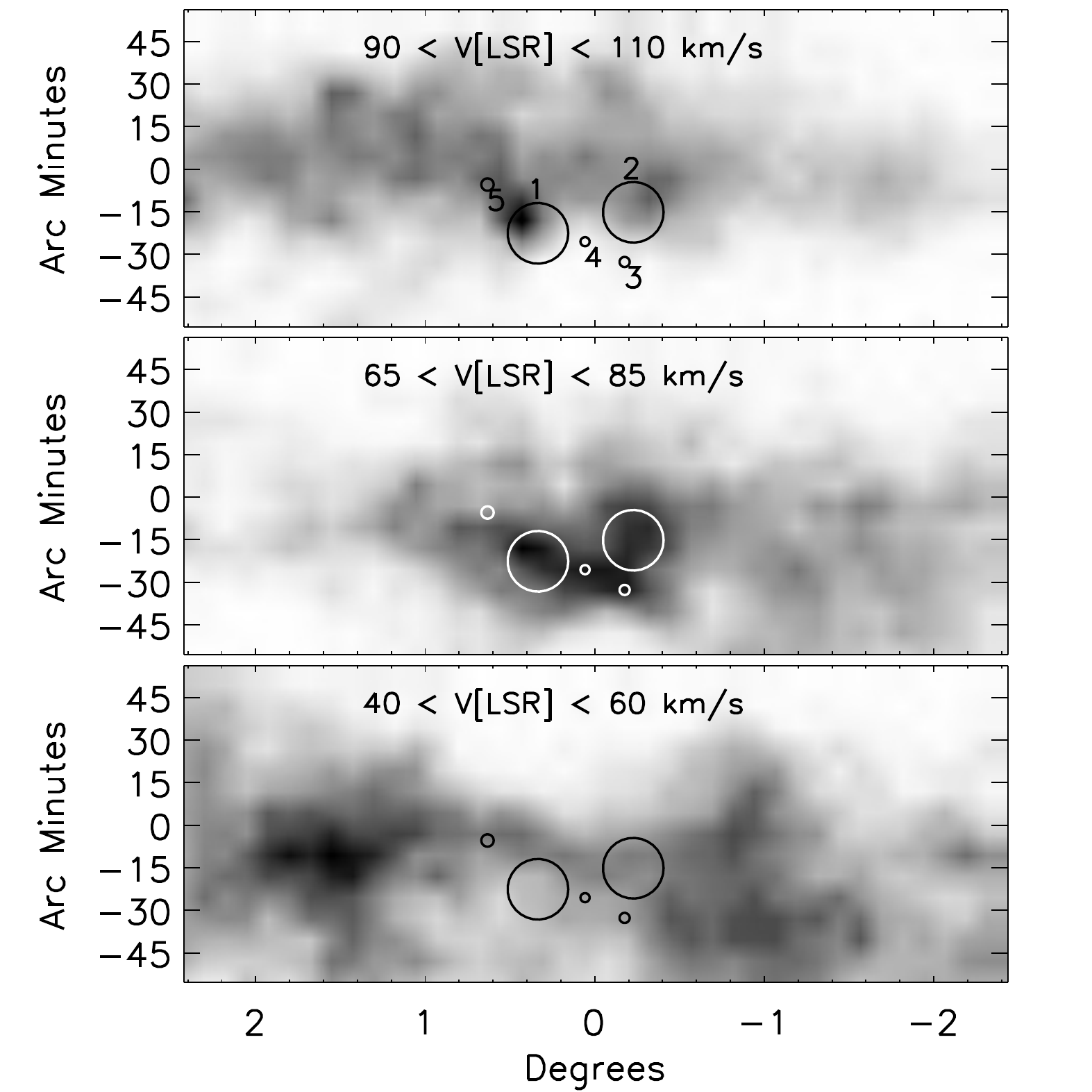}}
\caption{\label{fig:comap} $^{12}$CO integrated maps in direction of the giant molecular complex $G23.3-0.3$ 
\citep{dame01},  from 40 \kms\ to  60 \kms\ ({\it bottom panel}), from  65 \kms\ to  85 \kms\ 
({\it middle panel}), and from 
90 \kms\ to 110 \kms\ ( {\it top panel}). Labels refer to SNR1, 2, 3, 4, and 5 in Table 
\ref{tablexray}. Center is at a longitude of $22\rlap{.}^{\circ}92$ and a latitude of $+0\rlap{.}^{\circ}07$.}
\end{figure}

\subsection{High-energy sources in the GMC and progenitor masses}

Four SNRs are projected over the wide giant molecular cloud  G23.3$-$0.3 \citep{messineo10}.
In Figure \ref{fig:comap}, the SNRs  are superimposed on a  $^{12}$CO map of 
the giant molecular complex, with data-cubes from 
\citet{dame01}. 
Several peaks of CO emission are seen, for example at velocity (in the local standard of rest system) of
\vlsr $\approx 55$ \kms, $77-82$ \kms,  and $100$ \kms;
there is a similar velocity structure in the CO emission detected towards GLIMPSE09/SNR2, 
REG7/SNR3, REG5/SNR4, and REG4/W41.
The prominent emission  has a  maximum peak at \vlsr = 77--82 \kms\ (middle panel of Fig.\ \ref{fig:comap}); 
this is the cloud GMC G23.3$-$0.3, which is described by 
\citet{albert06} with a mass of about $2\times10^6$ \Msun\ , and an
extent of  two degrees of longitude  from $l\approx 22$\degr\ to $l\approx 24\rlap{.}^{\circ}25$, with a peak at 
$l \approx 23\rlap{.}^{\circ}3$ and $b \approx -0 \rlap{.}^{\circ}  3$;
a strong velocity component at \vlsr  $\approx 100$  \kms\  (upper panel of Fig.\ \ref{fig:comap}) 
appears only in the two higher latitude 
regions (SNR1/W41 border, as measured by  \citet[][]{brunthaler09}, and SNR2/SNR22.7$-$0.2).

Two  SNRs with  apparent diameters of $\sim$30\arcmin\
are listed  in the catalogue of \citet{green09},  G022.7$-$00.2 (SNR2) and  G023.3$-$00.3 (W41);   
two other  highly probable shell SNRs with an angular diameter of $4\rlap{.}^{\prime} 7$
and  $4\rlap{.}^{\prime} 5$, G$22.7583-0.4917$ (SNR3) and G$22.9917-0.3583$ (SNR4), were 
identified by \citet{helfand06}  with MAGPIS data;
their negative spectral indexes are also confirmed by \citet[][]{messineo10}.  
There is an extraordinary symmetry in the CO gas distribution of the giant cloud and  
locations (and even sizes) of the SNRs, which suggests  their physical association with the cloud. 
\citet{leahy08} concluded that  W41 is associated with the GMC G23.3$-$0.3.
G$22.7583-0.4917$ (SNR3)  and G$22.9917-0.3583$ (SNR4) can similarly be associated 
with the  GMC \citep{messineo10}; the  SNR G23.5667$-$0.0333/SNR5 and G22.7$-$0.2 are 
at a slightly higher latitude, where the 77 \kms\ and the $\sim$100 \kms\ clouds overlap;
however, at the position of   G22.7$-$0.2 the 77 \kms\ cloud has the strongest
CO intensity \citep{messineo10}. The SNR G23.5667$-$0.0333/SNR5  
\citep{helfand06,messineo10} is located at  $l=23\rlap{.}^{\circ} 57$ 
and $b=-0\rlap{.}^{\circ} 03$, outside the bulk of infrared emission of 
the main complex.

\begin{figure}
\resizebox{0.99\hsize}{!}{\includegraphics[angle=0]{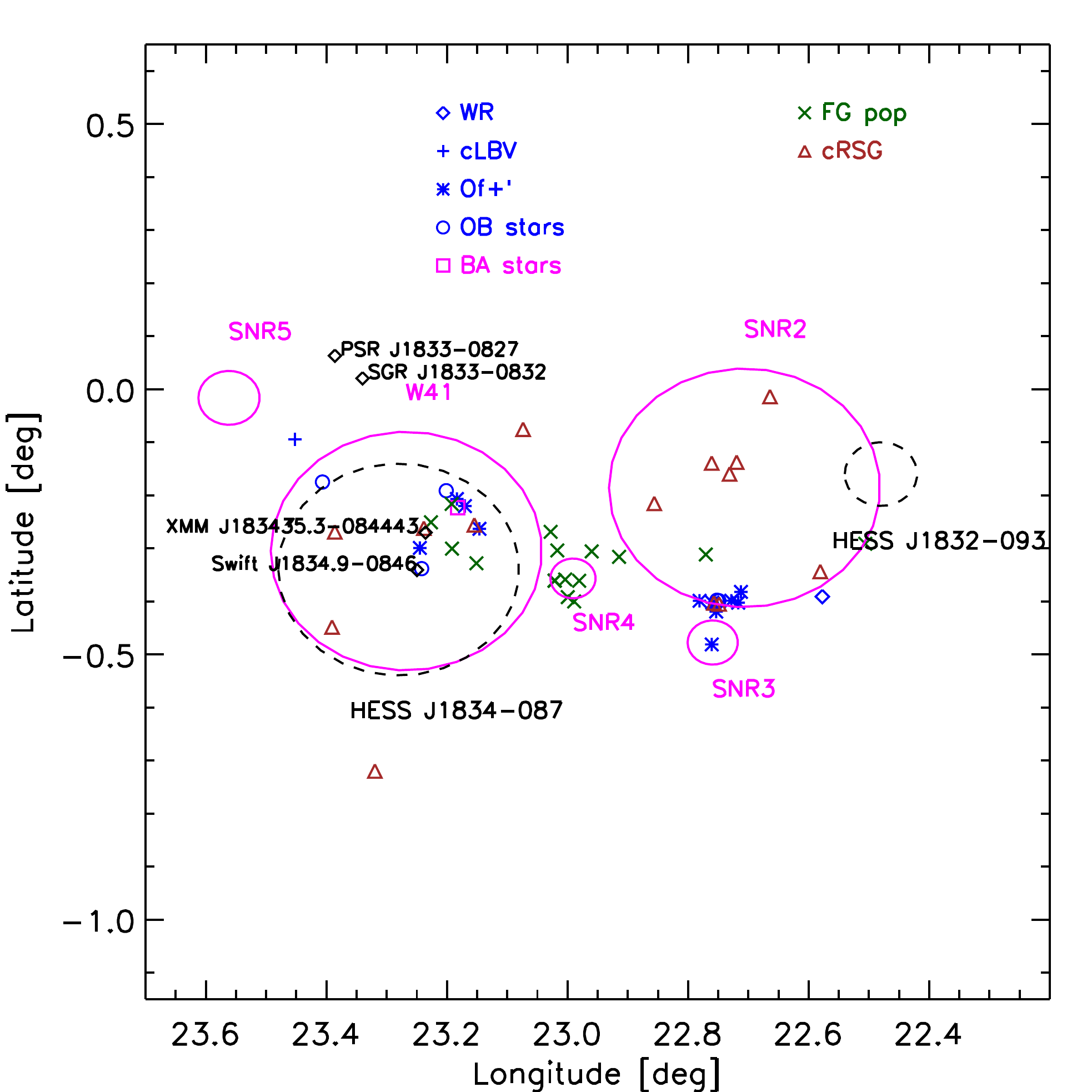}}
\caption{\label{fig:draw} Locations and angular sizes of  the supernova remnants (SNRs) 
are indicated with circles.
Two dashed circles mark the positional uncertainties of  HESS J1834$-$087 and HESS J1832$-$093.
Dark diamonds indicates the location of PWN XMM-J183435.3$-$084443,   Swift-J1834.9$-$0846,
 PSR J1833$-$0827, and  SGR J1833$-$0832. 
Symbols used for stars are as in Fig.\ \ref{largemap.fig}.}
\end{figure}

A large number of X-ray and TeV emitters have been reported in the direction 
of  the two largest SNRs (W41/SNR1 and G22.7$-$0.2/SNR2). A schematic of the giant molecular 
cloud with  the location 
of the SNRs,  high-energy emitters, and the newly discovered massive stars is shown 
in  Fig.\ \ref{fig:draw} (see also Table \ref{tablexray}).

%%%%%%%%%%%%%%%%%%%%%%%%%%%%%%%%%%%%%%%%%%%%%%%%%%%%%%%%%%%%%%%%%%%%%%%%%%%%%%%%%%%%%%%%%%%%%%

The TeV source HESS $J1834-087$ is located at the center of the  shell-type remnant 
W41 (SNR1) \citep[e.g.][]{aharonian05,tian07,leahy08}. 
For the majority of   extended TeV detections, 
young pulsars have been proposed as counterparts (young pulsar wind nebulae). 
\citet{misanovic11} identified the faint X-ray point-source  XMM J183435.3$-$084443 
(CXOU J183417.2$-$084901) \citep[number 7 in Table 1 ][]{mukherjee09} as a  pulsar wind nebula (PWN). 
Swift observations   unveiled another possible TeV emitter,
the magnetar Swift J1834.9$-$0846 \citep{gogus11,kargaltsev12}.
So far, distances of 4-5 kpc have been assumed for both candidate TeV emitters  by associating them  with 
W41  \citep{leahy08}.

HESS J1834$-$087, XMM J183435.3$-$084443, and Swift J1834.9$-$0846  fall  in the  center  of the W41  shell, 
and in our region REG4 (see Fig.\ \ref{largemap.fig}).
In region REG4, we detected several rare O-type supergiants 
(from 28 to 45 \Msun\   at a spectro-photometric distance of 4.6 kpc)
and two cRSGs.
Swift J1834.9$-$0846   is one of the few Galactic magnetars  associated with  massive stars
\citep[.e.g.][]{figer05,bibby08,muno06,davies09,mori13}.

%%%%%%%%%%%%%%%%%%%%%%%%%%%%%%%%%%%%%%%%%%%%%%%%%%%%%%%%%%%%%%%%%%%%%%%%%%%%%%%%%%%%%%%%%%%%%%

SNR G22.7$-$0.2 has a size similar to that of W41  \citep[40 pc at 4.6 kpc,][]{green09}.
The presence of a candidate cluster of RSGs (RSGCX1)  with three cRSG stars  toward the  center
of this SNR  suggests  that the progenitor
of the supernova was from this population.
HESS J1832$-$093 overlaps with SNR G22.7$-$0.2 (SNR2)
\citep{laffon11}.

%%%%%%%%%%%%%%%%%%%%%%%%%%%%%%%%%%%%%%%%%%%%%%%%%%%%%%%%%%%%%%%%%%%%%%%%%%%%%%%%%%%%%%%%%%%%%%

G22.7583-04917 (SNR3) has a diameter
of about 5\arcmin, or 6.7 pc at the distance of 4.6 kpc.
The 90 cm shell-type emission is centered on the massive O4f$_K$+, star \#25.
This suggests that the SN progenitor had a mass similar to that of star \#25 
($28-36$ \Msun\ at 4.6 kpc). 

%%%%%%%%%%%%%%%%%%%%%%%%%%%%%%%%%%%%%%%%%%%%%%%%%%%%%%%%%%%%%%%%%%%%%%%%%%%%%%%%%%%%%%%%%%%%%%

G22.9917$-$0.3583 (SNR4)  has a size of about $4\rlap{.}^{\prime} 5$, or 6.0 pc at the distance of 4.6 kpc,
and falls in region REG5  of Table \ref{regions}.
We detected only "foreground stars", which are  unrelated to the GMC.

%%%%%%%%%%%%%%%%%%%%%%%%%%%%%%%%%%%%%%%%%%%%%%%%%%%%%%%%%%%%%%%%%%%%%%%%%%%%%%%%%%%%%%%%%%%%%%

\begin{table*}
\caption{ \label{tablexray} List of associated high energy objects per supernovae remnants.}
\begin{tabular}{@{\extracolsep{-.06in}}lllllllll}
\hline
OBJECT                      & RA[J2000]    & DEC[J2000]   & diam & diam  & Vel   & Comment  \\               
                            &{\rm [hh mm ss]}  & {\rm [deg mm ss]}  &[\arcmin]&[pc]&[\kms]& \\
\hline
SNR1 / W41                  & 18 34 46.42  & $-$08 44 00    &  30&40.1 &  $77 \pm 5$ & 3,6 \\
HESS J1834-087              & 18 34 55.31  & $-$08 44 17.64 & 12 &&     	 & 1, 11, 13 \\
PWN XMM J183435.3-084443    & 18 34 35.32  & $-$08 44 43.80 &    &&     	 & 12 \\
PSR J1833-0827              & 18 33 40.35  & $-$08 27 30.44 &    &&     	 & 13 \\
SGR J1833-0832              & 18 33 44.38  & $-$08 31 07.71 &    &&     	 & 4  \\
Magnetar Swift J1834.9-0846 & 18 34 52.12  & $-$08 45 55.97 &    &&     	 & 5, 7\\

\hline
SNR2 / G22.7-0.2            &18 33 17.86   & $-$09 10 35    &  30&40.1 &    82.5& 3, 6,  8\\
HESS J1832-093              &18 32 46.85   & $-$09 21 54.49 & 0.6 &&        & 10  \\

SNR3 -G22.7583-0.4917       &18 34 26.70 & $-$09 15 50 & 5.0&6.7 & 75.5        & 2, 3, 8, 9     \\

SNR4 -G22.9917-0.3583       &18 34 26.59 & $-$09 00 09 & 4.5&6.0 & 70.9        & 3, 8, 9 \\

SNR5 -G23.5667-0.0333       &18 34 17.09 & $-$08 20 21 & 6.1&    & 91.3        & 3, 8, 9 \\
\hline
\end{tabular}

\begin{list}{}{}
\item[] 
{\bf References.}
  (1) \citet{aharonian05};~
  (2) \citet{bronfman96};~
  (3) \citet{dame86};~
  (4)  \citet{gogus10};~
 (5) \citet{gogus11};~
 (6) \citet{green09};~
 (7) \citet{kargaltsev12};~
 (8) \citet{kuchar97};~
 (9)\citet{helfand06};~
 (10)\citet{laffon11};~
 (11) \citet{leahy08};~
 (12)\citet{mukherjee09};~  
 (13) \citet{tian07}.~
\end{list}

\end{table*}

\section{Discussion and summary}
\label{discussion}
\subsection{Massive stars}
Analysis of the spectroscopic data presented in this paper has revealed  of a rich  population of 
evolved massive stars associated with GMC G23.3$-$0.3, yielding  38 new early-type stars, 3 new RSGs, and 6 new cRSGs.

Complementary photometric data indicate a bi-modality in the distribution of \Aks\ of early-type stars.
A component  with \Aks\ from 0.9 to 2.0 mag contains a large  variety  of massive stars 
from O-types to late B-types, and  a large fraction of those are  associated with 
the  GMC. The nine O- and B-type  supergiants have average \Aks=1.63 mag with  $\sigma=0.18$ mag.
Despite the uncertain absolute calibration of O-type stars,
we obtained  average  spectro-photometric distance moduli from  
$13.18\pm0.66$ mag (O9-9.5I) to  $13.4\pm0.4$ mag (Of$_K$+ stars).  
This range is consistent with that derived from B supergiants and with 
the distance to the GMC G23.3$-$0.3.
We adopted a  DM=$13.31\pm0.17$ mag  to characterize the luminosity and mass properties 
of obscured stars (\Aks $> 0.8$ mag), with the  parallactic distance modulus of \citet{brunthaler09} 
in good agreement with the  spectro-photometric distance.

\begin{figure}
\resizebox{0.99\hsize}{!}{\includegraphics[angle=0]{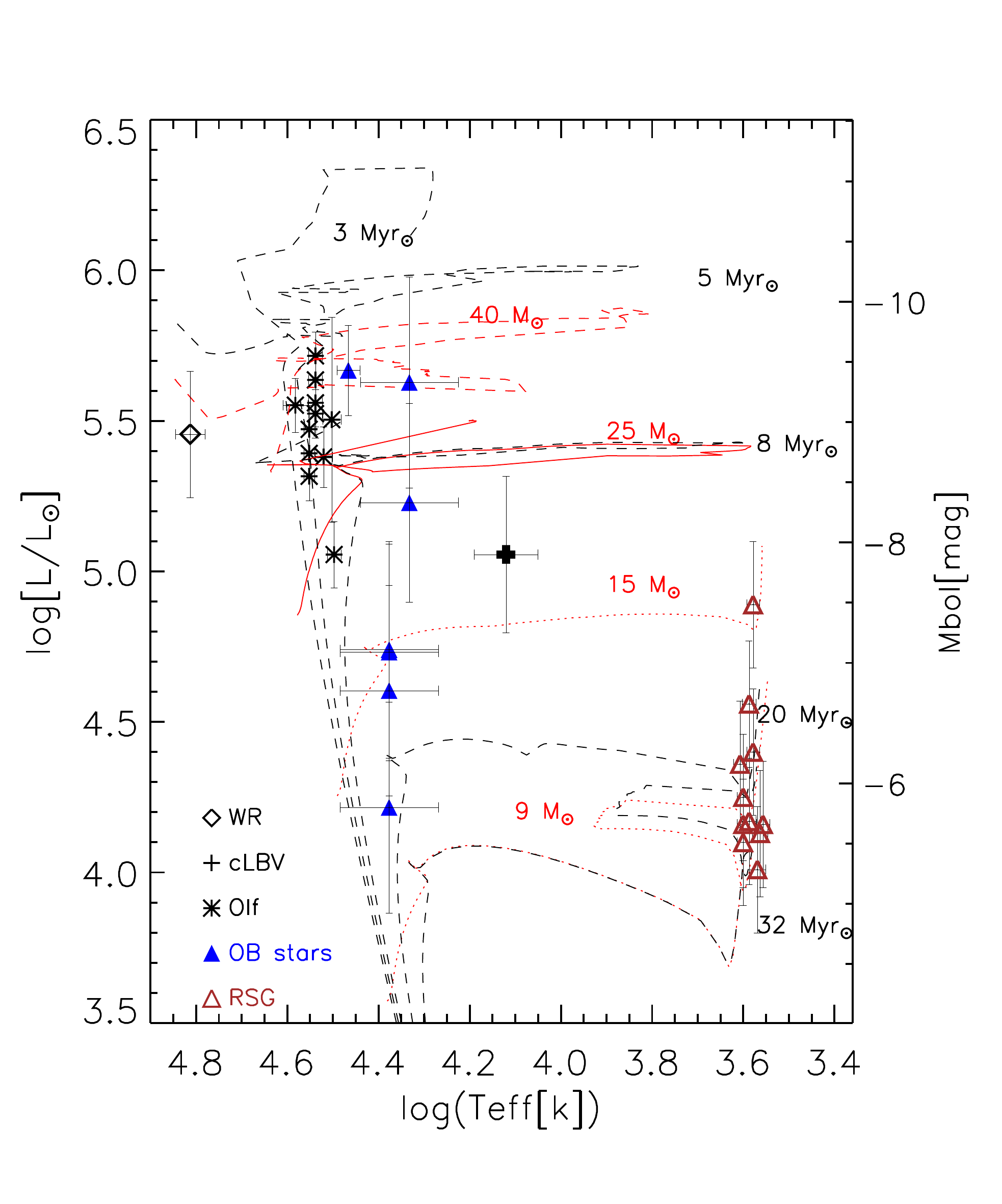}}
\caption{\label{fig:luminosity} Luminosities of massive stars with  \Aks$>$0.8 mag 
are plotted versus their effective temperatures. 
Stellar tracks for stars of   9,   15, 25, and 40 \Msun,
based on the new rotating Geneva models with a solar metallicity, 
are  shown with dotted and  dashed lines; darker curves show
the corresponding isochrones at 32, 20,  8,  5, and 3 Myr \citep{ekstrom12}.
The positions of Of$_K$+ stars  are marked by  asterisks, the cLBV by a plus
sign, the WR number 39 in \citet{mauerhan11} by a diamond symbol, other OB stars by filled triangles, and
RSGs and cRSGc  by empty triangles. }
\end{figure}

Concerning the massive stellar cohort, a cLBV was  detected in region REG2
and 10  massive Of$_K$+ stars in   REG4 and in the vicinity of GLIMPSE9. 
The Of$_K$+ stars 
have  \Ks$_o$\ from 7.9 to 9.2 mag, 
and in Fig. \ref{fig:luminosity} we plot their position on an HR diagram; comparison to 
theoretical predictions  for rotating massive stars  suggests
masses from 25 to 45 \Msun, and ages from 5 to 8 Myr \citep{ekstrom12}. 
This finding would suggest the likely presence of 
more evolved WRs in the complex; indeed, one  WC8 
is reported by \citet{mauerhan11}.

RSGs have a large span of magnitudes even for an almost coeval population \citep[for example, the RSGs in RSGC1, ][]{figer06},
and are not suitable as distance indicators.
By assuming a  distance of 4.6 kpc, we found  3 new RSGs (\#40, \#43, and \#47), \i.e.\ stars with
luminosities larger than $>10^4$ \Msun\  and \Aks\  $>1.3$ mag.
Their spectral types (from M0 to M2) closely align with 
the distribution of  spectral types of Galactic RSGs, which peaks at M2-M3 \citep{davies07,elias85}.
As shown in Figure  \ref{fig:luminosity}, the  new RSGs   are much older than the detected O-stars;
we estimated masses from 9 to 15 \Msun\ and ages from 20 to 30 \Msun.

\subsection{Distribution over the cloud}

The location of the massive stars  provides  insights on the star formation history of GMC G23.3$-$0.3.
The same mix  of massive stars (RSGs, Of$_K$+ stars, and B stars) 
at similar \Aks, spectral types, and magnitudes 
was detected in REG4 and GLIMPSE9Large. The two regions are separated  by 27\arcmin\ (36 pc at 4.6 kpc).
This  provides evidence for  repeated  multi-seeded  bursts of star formation across 
the complex, which appears  to form a unique extended structure at a distance of about 4.6 kpc.
Two main generations of massive stars were located;  RSGs and cRSGs   have ages of 20-30 Myr;
massive Of$_K$+ stars trace star formation occurred  5-8 Myr ago.  

The luminosities of  the 4  RSGs in GMC G23.3$-$0.3  
\citep[two new ones, plus the  two RSGs in][]{messineo10} 
are consistent with  ages from 18 to 30 Myr. It is difficult to   accurately infer the mass of the natal 
stellar aggregate of RSGs, because of  their short lifetimes leading to a small population potentially affected by 
  stochastic effects. Following the analysis  by \citet{clark09},
we might reasonably expect them to be associated with a population of stars of $>>10^4M_{\odot}$.
Under the assumption of a Salpeter  initial mass function \citep{salpeter55}, we 
determine that  additional  stellar populations of total mass $\sim2200$  and $\sim1500$ \Msun\  
were necessary to
account for the presence of the  six Of$_K$+ stars in region GLIMPSE9Large, 
and four in region REG4, respectively.

The  G$23.3-0.3$ complex contains only one  stellar cluster, GLIMPSE9, with a mass of $\sim3000$ \Msun\ and an age of 15-27 Myr.
The younger Of$_K$+ stars  are not part of a stellar cluster and are distributed sparsely over  two regions with radii of about 8.0 pc, with 
six surrounding  the GLIMPSE9 cluster. A few examples of isolated massive star formation are reported in literature. 
For example, one O2If*/WN6 star with a mass of 40-80 \Msun\ (0.6 Myr old)
and an O2If*/WN6 with mass  $> 100$ \Msun\ were detected in the \HH\ region  surrounding the 
Galactic cluster NGC3603 \citep{roman-lopes13a,roman-lopes13b,roman-lopes12}. Further observations will be required to understand the 
origin of this population.

We, therefore, infer a substantial difference in ages between the young massive stars (about 5 Myr) 
and older  RSGs (18-30 Myr)  in  GMC G$23.3-0.3$. An age spread is
common seen in giant molecular complexes, such as G305   \citep{clark04,davies12}, W51
\citep{clark09}  and 30 Dor. 
The latter region is of particular interest with regard to GMC G$23.3-0.3$, with
 star formation apparently
 commencing $\sim$25Myr ago and continuing to the present day  
\citep{walborn97,grebel00,walborn13}.

 Finally, the G$23.3-0.3$ complex is located at 23\degr, at a Galactocentric distance of about 4.6 kpc. 
The existence of a  number of massive clusters/complexes rich in RSGs 
from $l\approx23$\degr\ to $l\approx35$\degr\  seems a peculiar feature of the Galactic barred potential 
\citep[e.g.][]{deguchi06,habing06, clark09,rsgchem09}.

\subsection{Progenitor masses of SNRs}

Adopting the initial mass function  of \citet{salpeter55} and employing the  
isochrones of  \citet{ekstrom12}, 
at an age of 5 Myr
a representative stellar population will have lost 
$\sim 2$ \textperthousand\ of stars with masses $> 1$ \Msun\ and 3.5\% of stars 
with masses $> 8$ \Msun\ {\bf as} SNe;  
at 
 30 Myr these fractions increase to 3\%\ and  45\%,  respectively. As such, we would expect multiple SNe to have occurred within 
G23.3$-$0.3; 
indeed, four of the 274 known SNR \citep{green13} are found to reside in it.
 We detected massive evolved stars towards the  centers of 
SNR W41, G22.7$-$0.2(SNR2), and G22.7583$-$0.4917 (SNR3).
Massive O stars (5-8 Myr) and RSGs (20-30 Myr) were detected in the center of W41;
several candidate RSG stars were found in the  center  of G22.7$-$0.3/SNR2;
a O4 supergiant in isolation was found at  the center of  G22.7583$-$0.4917
(SNR3).

\begin{appendix}

\section{$Q1$ parameter}
\label{q1q2sec}

\begin{figure}
\resizebox{0.99\hsize}{!}{\includegraphics[angle=0]{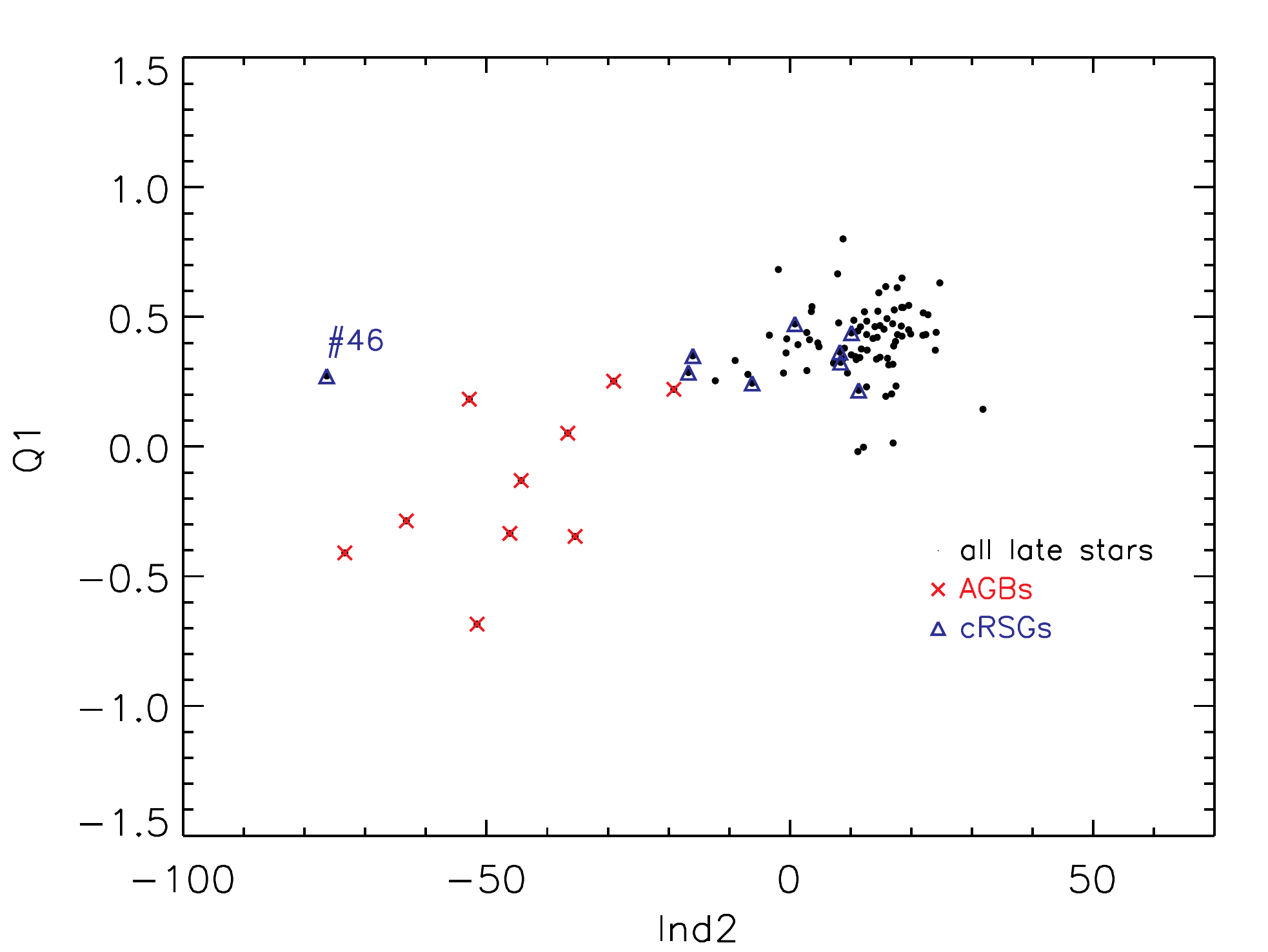}  }
\caption{\label{ind2q1-ewlum} 
 $Q1$ values of late-type stars (dots) versus the H$_2$O water index.
We marked   the AGB stars with crosses, and  the cRSGs with triangles.} 
\end{figure}

$Q1$ is defined as a combination of $J,H$ and \Ks\ magnitudes, and it is proportional
to the distance of a point-source from the interstellar reddening vector passing trough 
the origin in the $J-H$ versus $H-$\Ks\ plane \citep{messineo12};
positive values are for point-sources  to the left of the reddening vector, negative to the right.
The reddening vector is defined with a power law and an index of $-1.9$ \citep{messineo05}.

$Q1$ values are plotted against  H$_2$O water indexes in
Fig.\  \ref{ind2q1-ewlum}.
Gaseous water absorption in the envelopes of late-type stars (for example Mira-type AGB stars) causes 
a  dimming of the $H$ magnitude, and results in a weaker $Q1$ value. 
The average  and standard deviation of the $Q1$ values of  RSGs and cRSGs 
(L $> 1 \times 10^4$ \Lsun\ for a distance of 4.6 kpc) are
0.38 and 0.11 mag, respectively;  those of AGBs are $-0.15$ and 0.31 mag.

\section{ Giant stars and  selection of AGB stars}
\label{agbsel}

\begin{figure}
\resizebox{1.0\hsize}{!}{\includegraphics[angle=0]{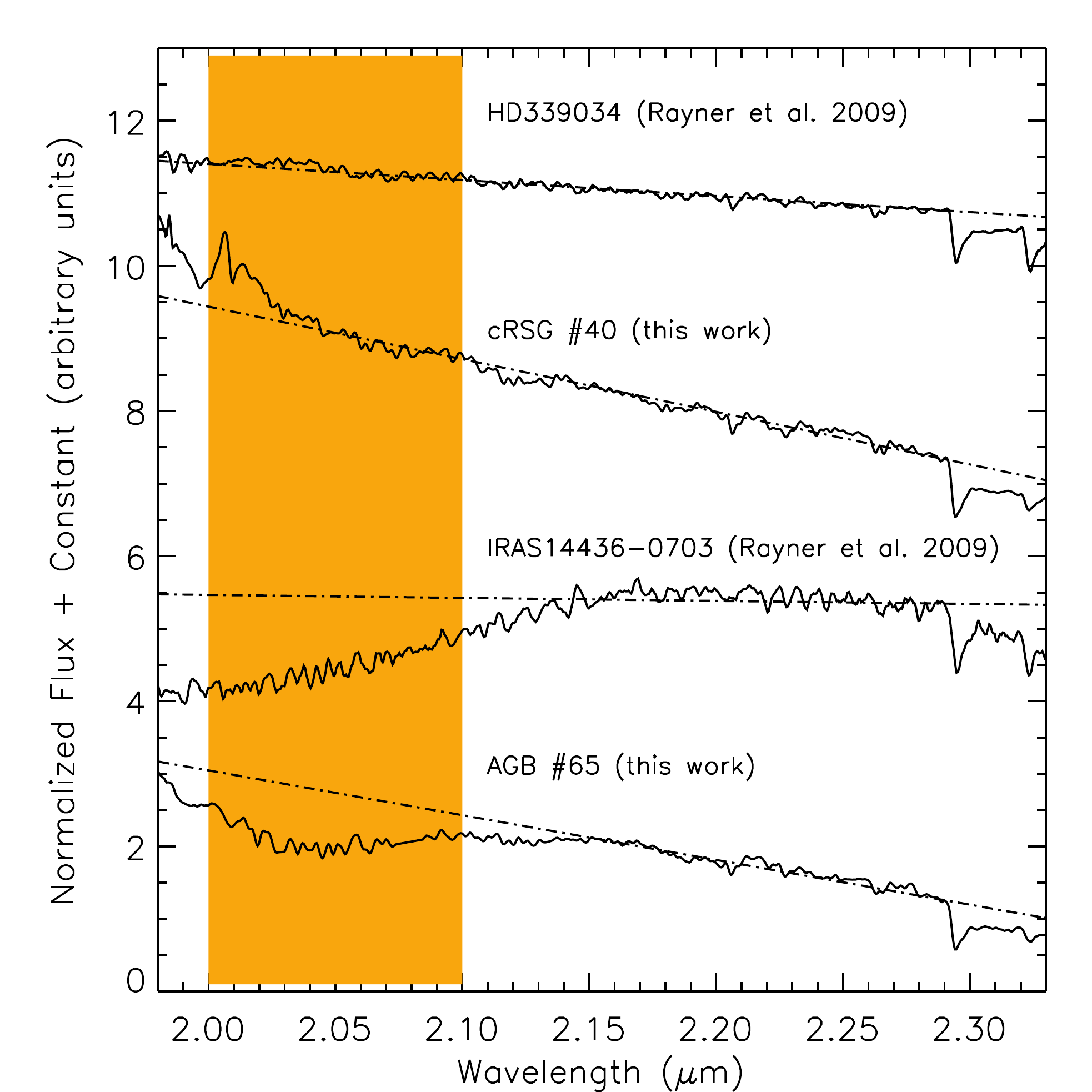}}
\caption{ \label{agb.fig} Normalized reddened spectra (arbitrarily shifted for clarity)
of late-type stars. As an example, the spectrum of a new 
cRSG, \#40, is compared to the IRTF spectrum of HD339034 \citep{rayner09};
the spectrum of the AGB \#65 is compared to the IRTF spectrum of 
IRAS 14436-0703 \citep{rayner09}.
The dotted-dashed lines are  linear fits to the stellar continuum in the range 2.15-2.29 \um.
The darker region is used for measuring an $H_2O$ index. 
} 
\end{figure}

In $K-$band spectra of AGBs,  absorption by water is visible as a change in shape of the stellar 
continuum short-ward of 2.1 \um\  \citep{blum03,alvarez00,rayner09}. 
Some examples of $K$-band spectra of AGBs and RSGs are shown in Fig.\  
\ref{agb.fig}. 

We linearly interpolated the de-reddened spectra from  2.15 \um\ to 2.29 \um, extrapolated
this fit to 2.0 \um, and calculated the difference 
of the linear fit and the observed spectrum  from 2.0 \um\ to 2.1 \um;
we defined the sum of this difference vector as the H$_2$O index.
The distribution of the H$_2$O values  resembles  a gaussian with an additional
tail of negative values. We classified as AGB stars those stars with a H$_2$O index
deviating more than 6 $\sigma$ from the central mean.
The same classification is  obtained in the region 2.025-2.100 \um.
A variation of 10\% in the \Aks\ results in a typical variation of the H$_2$O index within 20\%.
This criterium  reproduces the "visual selection" of highly curved spectra.
Stars \#55, \#56, \#58, \#60, \#61, \#62, \#63, \#65, \#104, \#134,  
\#147, and \#149
were classified as  AGB stars, 
i.e. 12 out of 113 observed late-type  stars (11\%). 
AGB stars are listed in Tables \ref{table.crsgspectra} and \ref{table.giantspectra}.

Star \#46  \citep[\#8 star in ][]{messineo10}  is the brightest star of the 
GLIMPSE9 cluster in \Ks-band. Despite its curved stellar continuum, 
it was classified as a likely RSG by 
comparison of its luminosity and extinction to those of other cluster members;
its spectrum resembles   My CEP  \citep[a rare M7I,][]{rayner09}.

\section{Finding charts and giant stars}
\label{charts}
\begin{figure*}
\begin{center}
\resizebox{0.7\hsize}{!}{\includegraphics[angle=0]{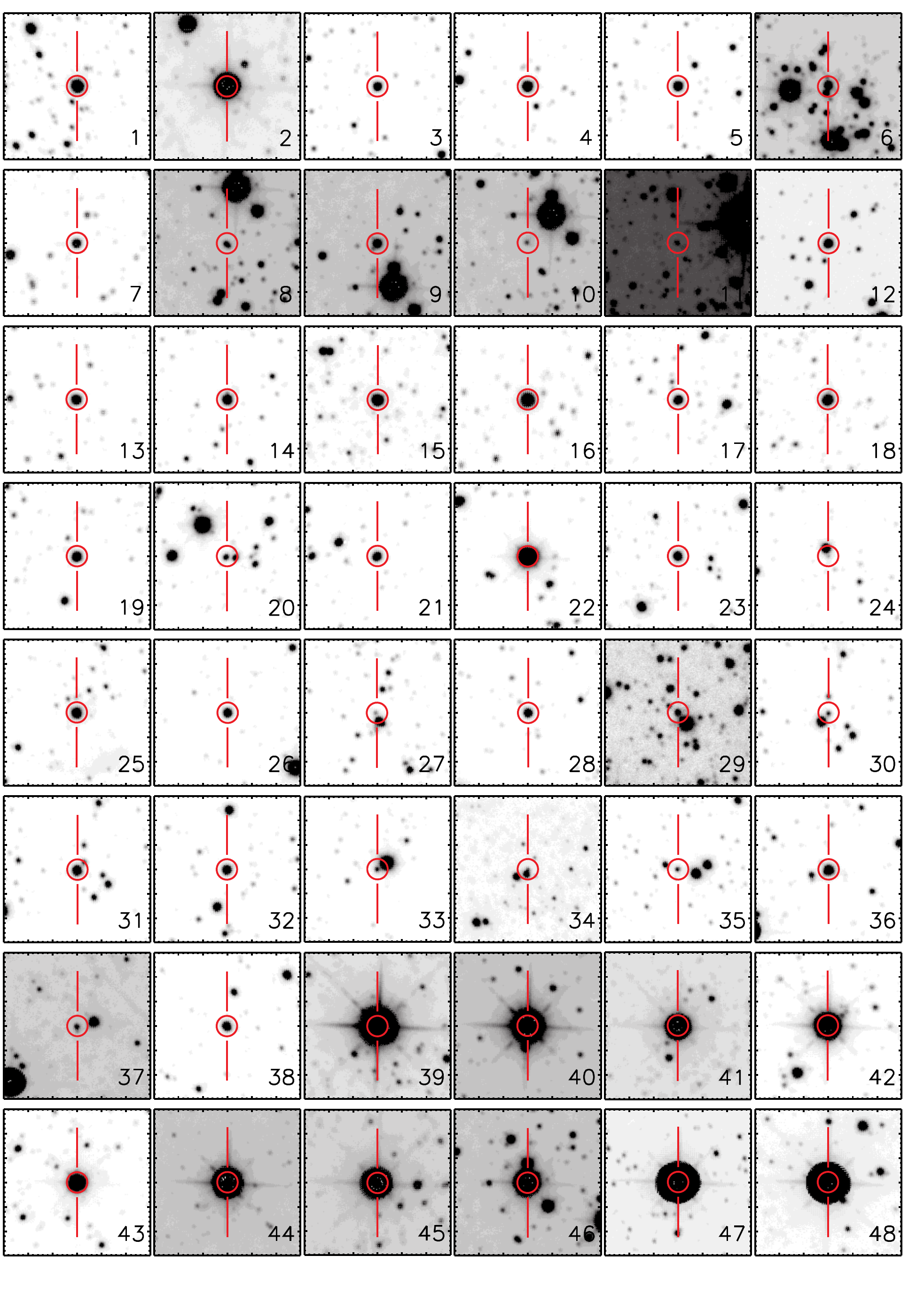}  }
\end{center}
\caption{\label{chartes-late}  UKIDSS \K-band images ( 30\arcsec $ \times $ 30\arcsec ) 
of the detected  stars. Targets are indicated with 2 line-pointers.
Identification numbers are from Tables  
\ref{table.obspectra},\ref{table.crsgspectra}, and \ref{table.giantspectra}.  
North is up and Est to the left.}
\end{figure*}

Finding charts for the detected stars are given in  Figs.\ \ref{chartes-late}
and \ref{chartessinfo-early}.

A list of detected red giant stars is provided in Table \ref{table.giantspectra}.

\addtocounter{figure}{-1}
\begin{figure*}
\begin{center}
\resizebox{0.7\hsize}{!}{\includegraphics[angle=0]{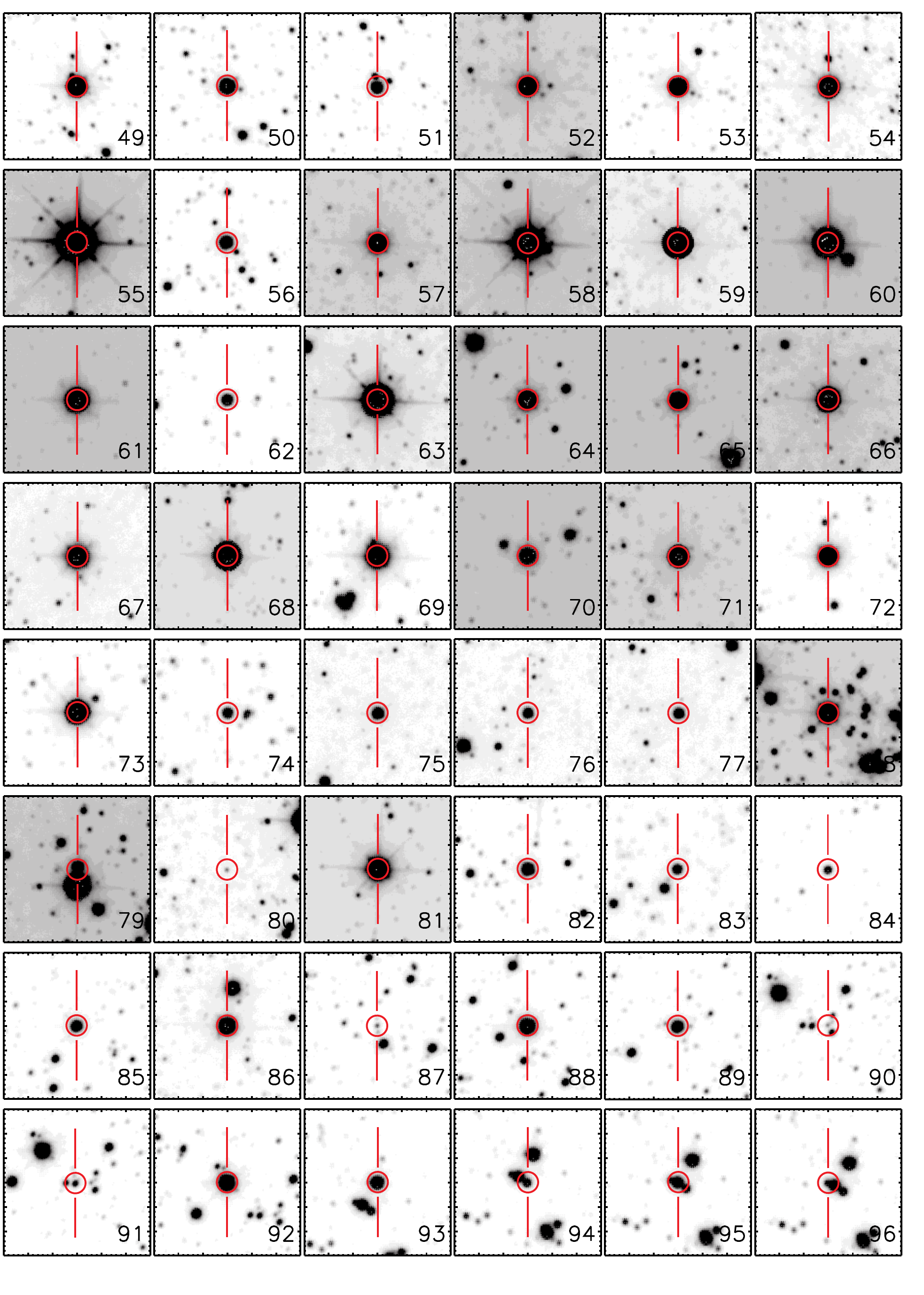}  }
\end{center}
\caption{Figure \ref{chartes-late} continued.} 
\end{figure*}

\addtocounter{figure}{-1} 
\begin{figure*}
\begin{center}
\resizebox{0.7\hsize}{!}{\includegraphics[angle=0]{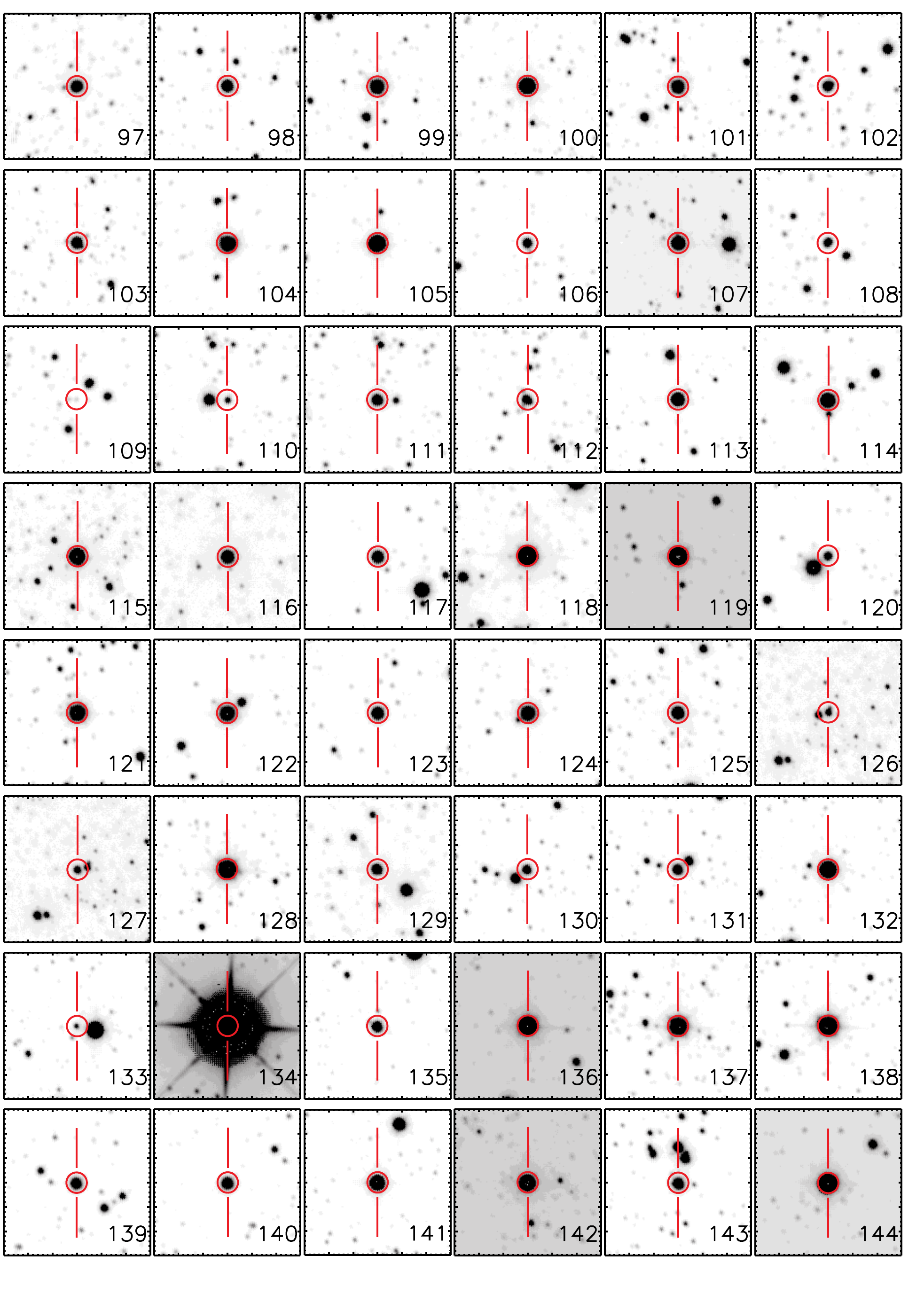}  }
\end{center}
\caption{Figure \ref{chartes-late} continued.} 
\end{figure*}

\addtocounter{figure}{-1}
\begin{figure*}
\begin{center}
\resizebox{0.7\hsize}{!}{\includegraphics[angle=0]{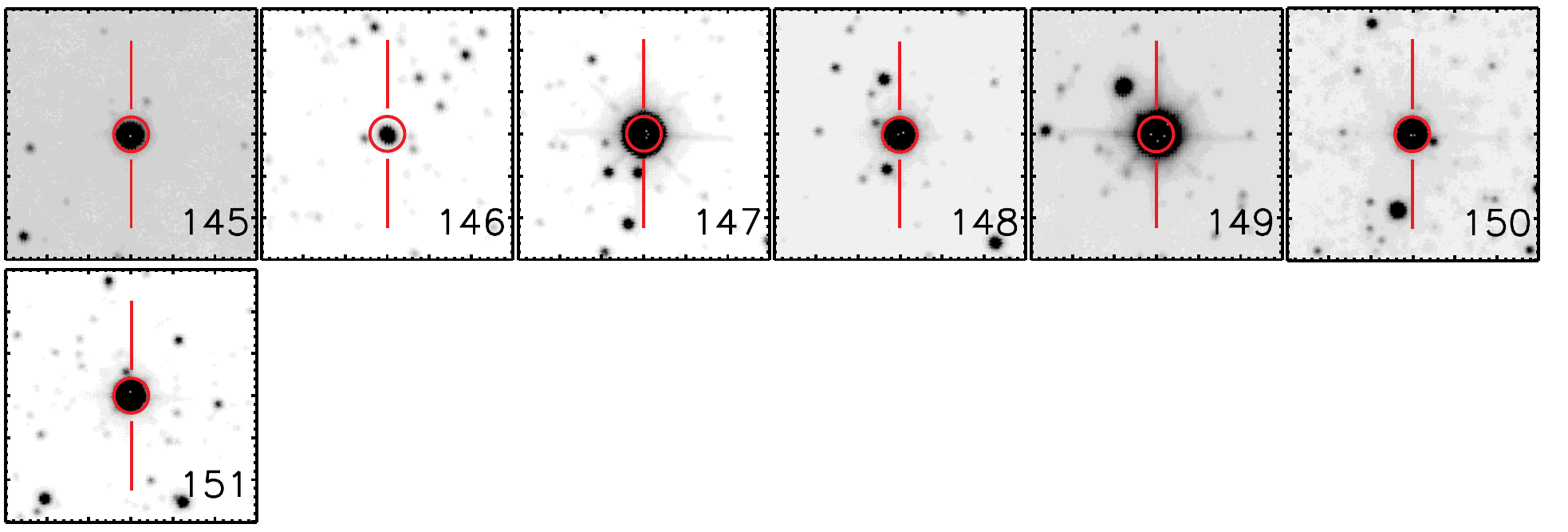}  }
\end{center}
\caption{ Figure \ref{chartes-late} continued.} 
\end{figure*}

\begin{figure*}
\begin{center}
\resizebox{0.7\hsize}{!}{\includegraphics[angle=0]{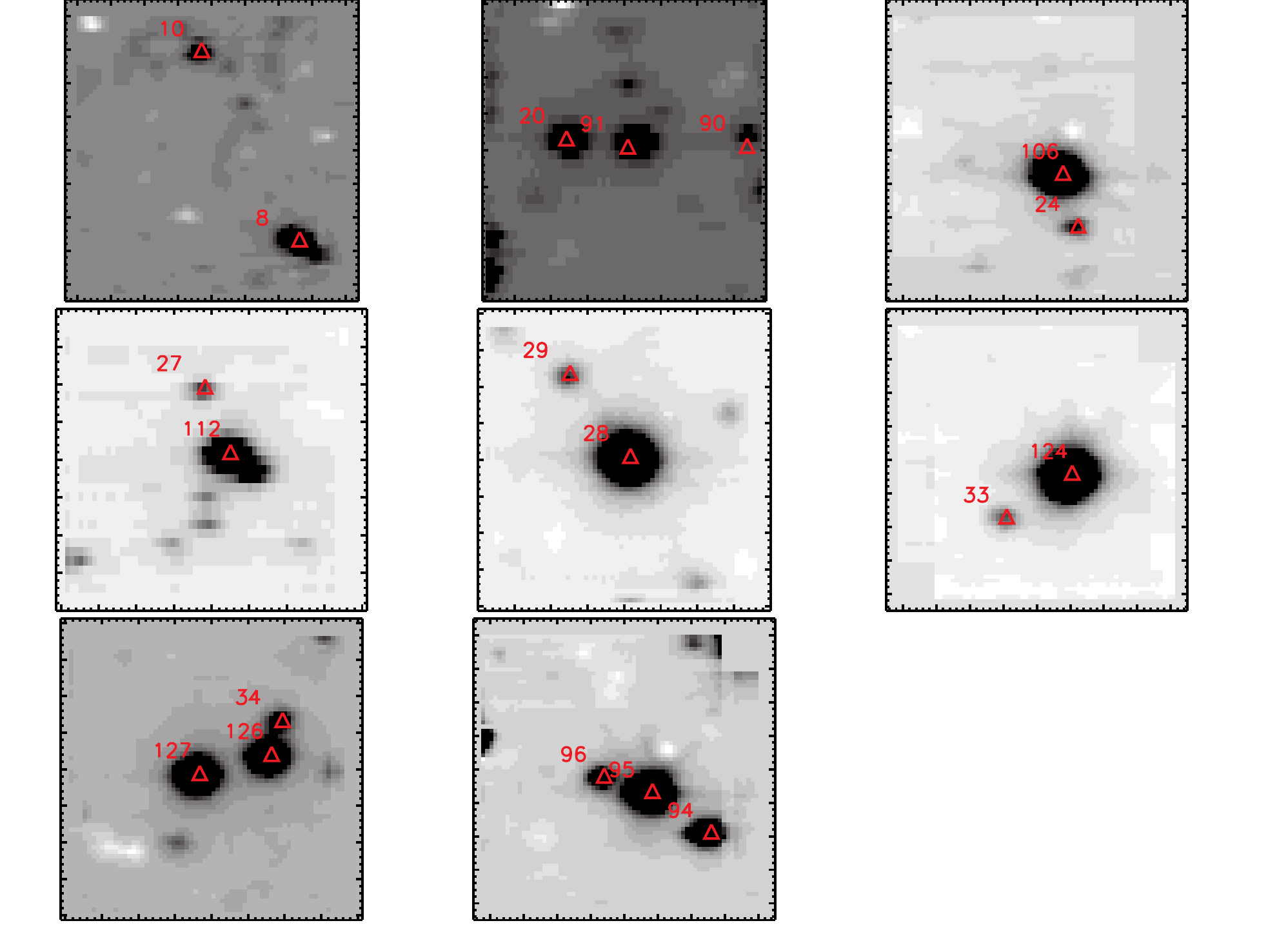}  }
\end{center}
\caption{\label{chartessinfo-early} 
Average SINFONI cubes of faint early-type stars,  which are  difficult to identify in the UKIDSS 
images due to confusion. The SINFONI field of view is 8\arcsec $\times$ 8\arcsec, 
two cubes with a positional shift of $1\rlap{.}^{\prime\prime}5$   were taken per observation. 
 North is up and Est to the left of the image.} 
\end{figure*}

\begin{table*}
\caption{\label{table.giantspectra} List of observed giant stars
($L < 4 \times 10^4 $\Lsun\  for a distance of 4.6 kpc, or AGB stars). 
}
{\tiny
\begin{tabular}{@{\extracolsep{-.03in}}rrr|llrrr|lr|rrrr|rr| l}
\hline 
 {\rm ID}   & {\rm RA(J2000)} & {\rm DEC(J2000)}  & \multicolumn{5}{c}{\rm Spectral Type} & Comment\\ 
\hline 
 &                 &                      &{\rm Instr.}  & {\rm EW(CO)}  & {\rm Sp[giant]}  & {\rm \Teff[giant]$^+$}  & {\rm H$_2$O}& \\ 
 &{\rm [hh mm ss]}  & {\rm [deg mm ss]}    &             &   [$AA$]      &                  & {\rm [K]}               & [\%]        & \\ 
\hline 

 49 &  18 33 04.80 &  $-$9~12~47.7 &            SofI &   25&       M3 &    3605$\pm$  120&    8&                                \\
 50 &  18 33 05.64 &  $-$9~07~26.6 &            SofI &   17&       K3 &    3985$\pm$  121&   10&                                \\
 51 &  18 33 06.54 &  $-$9~13~07.4 &            SofI &   28&       M5 &    3450$\pm$  203&   16&                              BG\\
 52 &  18 33 12.95 &  $-$9~10~23.3 &            SofI &   23&       M1 &    3745$\pm$  130&    1&                                \\
 53 &  18 33 13.27 &  $-$9~07~39.7 &            SofI &   26&       M4 &    3540$\pm$  155&   $-$3&                                \\
 54 &  18 33 13.82 &  $-$9~23~37.6 &            SofI &   25&       M3 &    3605$\pm$  120&    2&                                \\
 55 &  18 33 25.94 &  $-$9~03~51.2 &            SofI &   27&     $..$ &      $..$ &  $-$63&                             AGB\\
 56 &  18 33 29.37 &  $-$8~51~22.9 &            SofI &   35&     $..$ &      $..$ &  $-$52&                             AGB\\
 57 &  18 33 30.98 &  $-$8~50~27.0 &            SofI &   24&       M2 &    3660$\pm$  140&   $-$6&                                \\
 58 &  18 33 33.16 &  $-$8~48~15.8 &            SofI &   28&     $..$ &      $..$ &  $-$44&             AGB IRAS 18307$-$0850\\
 59 &  18 33 33.64 &  $-$9~13~51.5 &            SofI &   26&       M3 &    3605$\pm$  120&   13&                                \\
 60 &  18 33 38.69 &  $-$9~10~06.3 &            SofI &   25&     $..$ &      $..$ &  $-$46&                             AGB\\
 61 &  18 33 40.24 &  $-$9~09~07.3 &            SofI &   19&     $..$ &      $..$ &  $-$51&              AGB OH22.77$$-$$0.26$^a$\\
 62 &  18 33 40.97 &  $-$9~02~13.4 &            SofI &   27&     $..$ &      $..$ &  $-$25&                             AGB\\
 63 &  18 33 41.07 &  $-$9~22~53.2 &            SofI &   22&     $..$ &      $..$ &  $-$73&                             AGB\\
 64 &  18 33 44.64 &  $-$8~48~09.7 &            SofI &   23&       M1 &    3745$\pm$  130&    0&                                \\
 65 &  18 33 45.38 &  $-$8~47~57.7 &            SofI &   21&     $..$ &      $..$ &  $-$35&                             AGB\\
 66 &  18 33 46.45 &  $-$9~21~36.1 &            SofI &   25&       M2 &    3660$\pm$  140&   10&                                \\
 67 &  18 33 46.75 &  $-$8~33~00.2 &            SofI &   22&       M0 &    3790$\pm$  124&    4&                                \\
 68 &  18 33 48.54 &  $-$9~12~36.2 &            SofI &   24&       M2 &    3660$\pm$  140&    8&                                \\
 69 &  18 33 48.82 &  $-$8~43~01.5 &            SofI &   23&       M1 &    3745$\pm$  130&   11&                                \\
 70 &  18 33 49.67 &  $-$8~33~05.9 &            SofI &   25&       M3 &    3605$\pm$  120&    2&                                \\
 71 &  18 33 49.92 &  $-$9~11~45.9 &            SofI &   22&       M1 &    3745$\pm$  130&    0&                                \\
 72 &  18 33 50.48 &  $-$8~42~42.1 &            SofI &   23&       M1 &    3745$\pm$  130&   10&                                \\
 73 &  18 33 53.33 &  $-$9~09~40.3 &            SofI &   22&       M1 &    3745$\pm$  130&    7&                                \\
 74 &  18 33 55.10 &  $-$9~08~12.3 &            SofI &   22&       M1 &    3745$\pm$  130&   10&                              BG\\
 75 &  18 34 05.34 &  $-$8~56~56.5 &         SINFONI &   42&       M4 &    3540$\pm$  155&   18&                                \\
 76 &  18 34 05.47 &  $-$8~57~58.9 &         SINFONI &   36&       M1 &    3745$\pm$  130&   14&                                \\
 77 &  18 34 06.07 &  $-$8~57~10.7 &         SINFONI &   37&       M1 &    3745$\pm$  130&   24&                                \\
 78 &  18 34 09.28 &  $-$9~14~00.7 &         SINFONI &   43&       M4 &    3540$\pm$  155&   12&                     [MFD2010] 1$^b$\\
 79 &  18 34 10.37 &  $-$9~13~49.5 &         SINFONI &   38&       M2 &    3660$\pm$  140&    9&                     [MFD2010] 6$^b$\\
 80 &  18 34 11.46 &  $-$9~14~03.0 &         SINFONI &   15&      $<$  K0 &   $>$  4185$\pm$  204&   $-$1&                                \\
 81 &  18 34 11.70 &  $-$8~57~09.3 &         SINFONI &   48&       M7 &    3223$\pm$  226&  $-$12&                                \\
 82 &  18 34 11.87 &  $-$8~57~29.4 &         SINFONI &   39&       M3 &    3605$\pm$  120&   11&                                \\
 83 &  18 34 12.10 &  $-$9~04~02.4 &         SINFONI &   12&      $<$  K0 &   $>$  4185$\pm$  204&   15&                                \\
 84 &  18 34 14.70 &  $-$8~35~01.1 &         SINFONI &   10&      $<$  K0 &   $>$  4185$\pm$  204&   $-$1&                      XMM$-$5 $^c$\\
 85 &  18 34 15.22 &  $-$8~47~41.3 &         SINFONI &   38&       M2 &    3660$\pm$  140&   17&                                \\
 86 &  18 34 15.79 &  $-$8~48~33.3 &         SINFONI &   41&       M4 &    3540$\pm$  155&   19&                                \\
 87 &  18 34 15.97 &  $-$8~45~41.6 &         SINFONI &   34&       M0 &    3790$\pm$  124&    3&                              BG$^d$\\
 88 &  18 34 16.03 &  $-$8~45~20.1 &         SINFONI &   38&       M2 &    3660$\pm$  140&   21&                                \\
 89 &  18 34 16.38 &  $-$8~46~19.0 &         SINFONI &   35&       M1 &    3745$\pm$  130&   23&                                \\
 90 &  18 34 18.52 &  $-$8~45~33.2 &         SINFONI &   28&       K3 &    3985$\pm$  121&  $..$ &                                \\
 91 &  18 34 18.72 &  $-$8~45~33.0 &         SINFONI &   25&       K2 &    4049$\pm$  131&    7&                                \\
 92 &  18 34 19.19 &  $-$8~45~26.4 &         SINFONI &   43&       M5 &    3450$\pm$  203&    4&                                \\
 93 &  18 34 19.50 &  $-$9~04~35.5 &         SINFONI &   27&       K3 &    3985$\pm$  121&   16&                                \\
 94 &  18 34 19.58 &  $-$9~04~41.4 &         SINFONI &   45&       M5 &    3450$\pm$  203&  $..$ &                                \\
 95 &  18 34 19.70 &  $-$9~04~40.1 &         SINFONI &   26&       K2 &    4049$\pm$  131&   17&                                \\
 96 &  18 34 19.79 &  $-$9~04~39.7 &         SINFONI &   25&       K2 &    4049$\pm$  131&  $..$ &                                \\
 97 &  18 34 20.73 &  $-$8~48~49.0 &         SINFONI &   37&       M2 &    3660$\pm$  140&    3&                                \\
 98 &  18 34 21.42 &  $-$8~50~31.0 &         SINFONI &   38&       M2 &    3660$\pm$  140&   14&                                \\
 99 &  18 34 22.16 &  $-$9~14~16.0 &         SINFONI &   38&       M2 &    3660$\pm$  140&   14&                                \\
100 &  18 34 23.13 &  $-$8~48~04.1 &         SINFONI &   42&       M4 &    3540$\pm$  155&   10&                                \\
 \hline
\end{tabular}
\begin{list}{}{}
\item[] {\bf Notes.}
The identification numbers are followed by celestial coordinates, instrument, 
EW(CO)s,   spectral types, \Teff, H$_2$O indexes, and comments.~
($^+$) Temperature erros account for accuracy in spectral types of $\pm2$.~
($^a$) This Asymptotic giant branch stars (AGB)  coincides with  the maser
OH22.77$-$0.26 \citep{blommaert94}. The  stellar velocity (LSR) is 93.1 \kms.~
($^b$) \citet{messineo10}.~
($^c$) XMM point source number 5 in Table 1 of \citet{mukherjee09}.~
($^d$) BG= object in the background of the cloud.
\end{list}
}
\end{table*}

\addtocounter{table}{-1}
\begin{table*} 
\caption{  continuation  of Table \ref{table.giantspectra}.} 
{\tiny
\begin{tabular}{@{\extracolsep{-.03in}}rrr|llrrr|lr|rrrr|rr| l}
\hline 
 {\rm ID}   & {\rm RA(J2000)} & {\rm DEC(J2000)}  & \multicolumn{3}{c}{\rm Spectral Type} & Comment\\ 
\hline 
 &                 &                      &{\rm Instr.}   & {\rm \Teff[giant]}  & {\rm Sp[giant]}  &  \\ 
 &{\rm [hh mm ss]}  & {\rm [deg mm ss]}    &               &	                  & {\rm [K]}         &                              \\ 
\hline 

101 &  18 34 24.18 &  $-$9~14~34.0 &         SINFONI &   34&       M0 &    3790$\pm$  124&   18&                                \\
102 &  18 34 24.26 &  $-$9~02~44.5 &         SINFONI &   23&       K1 &    4117$\pm$  136&   16&                                \\
103 &  18 34 24.31 &  $-$8~47~38.9 &         SINFONI &   34&       M0 &    3790$\pm$  124&   17&                                \\
104 &  18 34 24.39 &  $-$8~29~10.0 &            SofI &   31&     $..$ &      $..$ &  $-$32&                             AGB\\
105 &  18 34 25.86 &  $-$8~35~32.9 &         SINFONI &   40&       M3 &    3605$\pm$  120&   15&                                \\
106 &  18 34 26.41 &  $-$9~00~47.6 &         SINFONI &   20&      $<$  K0 &   $>$  4185$\pm$  204&   31&                                \\
107 &  18 34 26.42 &  $-$8~47~18.4 &         SINFONI &   36&       M1 &    3745$\pm$  130&   22&                                \\
108 &  18 34 27.65 &  $-$9~00~49.6 &         SINFONI &   34&       M0 &    3790$\pm$  124&   12&                                \\
109 &  18 34 27.82 &  $-$9~00~52.8 &         SINFONI &   33&       M0 &    3790$\pm$  124&  $..$ &\\
110 &  18 34 29.48 &  $-$8~45~03.8 &         SINFONI &   33&       K5 &    3869$\pm$  137&  $..$ &                                \\
111 &  18 34 29.74 &  $-$8~45~03.7 &         SINFONI &   40&       M3 &    3605$\pm$  120&   12&                                \\
112 &  18 34 30.10 &  $-$8~44~42.4 &         SINFONI &   24&       K1 &    4117$\pm$  136&   11&                                \\
113 &  18 34 31.05 &  $-$8~51~41.2 &         SINFONI &   39&       M2 &    3660$\pm$  140&   21&                                \\
114 &  18 34 31.70 &  $-$8~34~09.9 &         SINFONI &   41&       M3 &    3605$\pm$  120&   19&                                \\
115 &  18 34 31.87 &  $-$8~47~14.8 &         SINFONI &   37&       M1 &    3745$\pm$  130&   14&                                \\
116 &  18 34 32.31 &  $-$8~33~06.6 &         SINFONI &   33&       M0 &    3790$\pm$  124&   16&                                \\
117 &  18 34 32.32 &  $-$8~34~03.0 &         SINFONI &   38&       M2 &    3660$\pm$  140&   17&                                \\
118 &  18 34 32.48 &  $-$8~44~05.2 &         SINFONI &   39&       M2 &    3660$\pm$  140&   14&                                \\
119 &  18 34 33.69 &  $-$8~32~39.8 &         SINFONI &   39&       M2 &    3660$\pm$  140&   19&                                \\
120 &  18 34 33.73 &  $-$9~01~32.2 &         SINFONI &   38&       M2 &    3660$\pm$  140&    8&                                \\
121 &  18 34 33.83 &  $-$9~17~56.9 &            SofI &   22&       M0 &    3790$\pm$  124&   17&                                \\
122 &  18 34 33.93 &  $-$9~01~34.6 &         SINFONI &   36&       M1 &    3745$\pm$  130&   14&                                \\
123 &  18 34 34.98 &  $-$8~33~08.0 &         SINFONI &   34&       M0 &    3790$\pm$  124&   16&                                \\
124 &  18 34 35.61 &  $-$9~01~26.3 &         SINFONI &   13&      $<$  K0 &   $>$  4185$\pm$  204&   12&                                \\
125 &  18 34 36.73 &  $-$8~51~19.1 &         SINFONI &   42&       M4 &    3540$\pm$  155&   16&                                \\
126 &  18 34 36.96 &  $-$8~47~55.6 &         SINFONI &   28&       K3 &    3985$\pm$  121&  $..$ &                                \\
127 &  18 34 37.10 &  $-$8~47~56.2 &         SINFONI &   27&       K3 &    3985$\pm$  121&   24&                                \\
128 &  18 34 37.25 &  $-$9~17~44.5 &            SofI &   20&       K5 &    3869$\pm$  137&    8&                                \\
129 &  18 34 37.67 &  $-$8~30~53.4 &         SINFONI &   31&       K5 &    3869$\pm$  137&   11&                                \\
130 &  18 34 37.95 &  $-$8~50~48.7 &         SINFONI &   48&       M7 &    3223$\pm$  226&    2&                              BG\\
131 &  18 34 38.11 &  $-$8~50~50.5 &         SINFONI &   31&       K5 &    3869$\pm$  137&   17&                                \\
132 &  18 34 38.72 &  $-$8~48~53.3 &         SINFONI &   41&       M3 &    3605$\pm$  120&   15&                                \\
133 &  18 34 38.97 &  $-$8~48~52.5 &         SINFONI &   47&       M6 &    3336$\pm$  226&  $-$23&                              BG\\
134 &  18 34 39.18 &  $-$8~34~47.4 &            SofI &   20&     $..$ &      $..$ &  $-$36&              AGB IRAS18318$-$0837\\
135 &  18 34 39.23 &  $-$8~49~08.8 &         SINFONI &   32&       K5 &    3869$\pm$  137&   15&                                \\
136 &  18 34 39.55 &  $-$8~30~48.1 &         SINFONI &   51&       M7 &    3223$\pm$  226&   $-$9&                                \\
137 &  18 34 39.58 &  $-$9~16~44.4 &            SofI &   23&       M1 &    3745$\pm$  130&    8&                                \\
138 &  18 34 40.60 &  $-$9~14~45.9 &            SofI &   22&       M1 &    3745$\pm$  130&   14&                                \\
139 &  18 34 41.35 &  $-$8~35~30.3 &         SINFONI &   33&       M0 &    3790$\pm$  124&   18&                                \\
140 &  18 34 42.51 &  $-$8~35~34.7 &         SINFONI &   16&      $<$  K0 &   $>$  4185$\pm$  204&   12&                                \\
141 &  18 34 45.02 &  $-$8~47~23.5 &         SINFONI &   33&       M0 &    3790$\pm$  124&   18&                                \\
142 &  18 34 47.45 &  $-$8~48~09.7 &         SINFONI &   39&       M2 &    3660$\pm$  140&   22&                                \\
143 &  18 34 47.50 &  $-$8~47~33.9 &         SINFONI &   38&       M2 &    3660$\pm$  140&   17&                                \\
144 &  18 34 48.09 &  $-$8~34~50.4 &         SINFONI &   43&       M4 &    3540$\pm$  155&   12&                                \\
145 &  18 34 49.20 &  $-$8~34~22.4 &         SINFONI &   38&       M2 &    3660$\pm$  140&   18&                                \\
146 &  18 35 14.66 &  $-$8~37~40.6 &         SINFONI &   35&       M0 &    3790$\pm$  124&    3&      XMM$-$15 $^a$\\
147 &  18 35 17.93 &  $-$8~29~54.1 &            SofI &   26&     $..$ &      $..$ &  $-$29&                             AGB\\
148 &  18 35 18.53 &  $-$8~29~10.7 &            SofI &   23&       M1 &    3745$\pm$  130&    3&                                \\
149 &  18 35 21.28 &  $-$8~59~34.2 &            SofI &   27&     $..$ &      $..$ &  $-$19&                             AGB\\
150 &  18 35 27.04 &  $-$8~29~06.2 &            SofI &   21&       M0 &    3790$\pm$  124&   17&                                \\
151 &  18 35 28.17 &  $-$8~28~34.7 &            SofI &   19&       K5 &    3869$\pm$  137&   11&                                \\

\hline
\end{tabular}
\begin{list}{}{}
\item[{\bf Notes.}] ($^a$) XMM point source number 15 in Table 1 of \citet{mukherjee09}.
\end{list}
}
\end{table*}

\end{appendix}

\begin{acknowledgements}

This work was partially funded by the ERC Advanced Investigator Grant GLOSTAR (247078).
This work was partly supported by NASA under award NNG 05-GC37G, through 
the Long-Term Space Astrophysics program. 
This research
was partly performed in the Rochester Imaging Detector Laboratory
with support from a NYSTAR Faculty Development Program
grant.
This publication makes use of data products from the Two Micron All Sky Survey, which is a joint project 
of the University of Massachusetts and the Infrared Processing and Analysis Center/California Institute 
of Technology, funded by the National Aeronautics and Space Administration and the National Science Foundation.
This work is based  on observations made with the Spitzer Space Telescope, 
which is operated by the Jet Propulsion Laboratory, California Institute of Technology under a contract with NASA.
DENIS is  a joint effort of several Institutes mostly located in Europe. It has
    been supported mainly by the French Institut National des
    Sciences de l'Univers, CNRS, and French Education Ministry, the
    European Southern Observatory, the State of Baden-Wuerttemberg, and
    the European Commission under networks of the SCIENCE and Human
    Capital and Mobility programs, the Landessternwarte, Heidelberg and
    Institut d'Astrophysique de Paris.
This research made use of data products from the
Midcourse Space Experiment, the processing of which was funded by the Ballistic Missile Defence Organization with additional
support from the NASA office of Space Science. This research has made use of the SIMBAD data base, 
operated at CDS, Strasbourg, France. This publication makes use of data products from
WISE, which is a joint project of the University of California, Los
Angeles, and the Jet Propulsion Laboratory/California Insti-
tute of Technology, funded by the National Aeronautics and
Space Administration. 
Based on observations with the NASA/ESA Hubble Space Telescope (GO program 11545), obtained at the Space Telescope 
Science Institute, which is operated by the Association of Universities for Research in Astronomy (AURA), 
Inc., under NASA contract NAS5-26555. 
This research made use of Montage, funded by the National Aeronautics and Space Administration's 
Earth Science Technology Office, Computational Technnologies Project, under Cooperative 
Agreement Number NCC5-626 between NASA and the California Institute of Technology. 
The code is maintained by the NASA/IPAC Infrared Science Archive.
%We acknowledge use of the Atomic Line List ($http://www.pa.uky.edu/~peter/atomic/$).
The authors thank Professor Dame for providing CO data.
A special thank goes to the great support offered by  the European Southern Observatory.
This work is seeded on the speculations presented by Messineo et al. (2010).
MM does not have enough words to thank all co-authors of that paper -- 
Don Figer, Ben  Davies, Rolf Kudritzki, Mike Rich, 
John MacKenty, and Christine Trombley -- for their support and enthusiasm on 
this region that  was becoming every day bigger and full of rings.
Free speculations are  the beauty and potentiality of science.
MM thanks the Jos de Bruine and  Timo Prusti for 
useful discussions and support while at ESA.
We thank the anonymous referee for his constructive comments.

\end{acknowledgements}

\end{document}